\documentclass[pdflscape,lscape,iop,apj,numberedappendix,appendixfloats,stfloats,url,microtype]{emulateapj}

\pdfoutput=1
\usepackage{paralist,amsmath}
\usepackage{apjfonts}
\usepackage{textcomp}
\makeatletter
  \newcommand\tinyv{\@setfontsize\tinyv{7.0pt}{7.0}}
\makeatother
\slugcomment{Submitted to The Astrophysical Journal Supplement Series.}
\shorttitle{A Broadband Polarization Catalog of Extragalactic Radio Sources}
\shortauthors{Farnes et al.}
\usepackage{color}
\newcounter{editno}
\newcommand{\edits}[1]{\stepcounter{editno}
  \textcolor{blue}\bgroup[\arabic{editno}] #1\egroup}
\begin{document}

\title{A Broadband Polarization Catalog of Extragalactic Radio Sources}

\author{J.~S. Farnes\altaffilmark{1,2}, B.~M. Gaensler\altaffilmark{1,2}, E.~Carretti\altaffilmark{3}}
\altaffiltext{1}{Sydney Institute for Astronomy, School of Physics, The University of Sydney, NSW 2006, Australia.}
\altaffiltext{2}{ARC Centre of Excellence for All-sky Astrophysics (CAASTRO).}
\altaffiltext{3}{CSIRO Astronomy and Space Science, PO Box 76, Epping, NSW 1710, Australia.}
\email{jamie.farnes@sydney.edu.au}

\begin{abstract}
An understanding of cosmic magnetism requires converting the polarization properties of extragalactic radio sources into the rest-frame in which the corresponding polarized emission or Faraday rotation is produced. Motivated by this requirement, we present a catalog of multiwavelength linear polarization and total intensity radio data for polarized sources from the NRAO VLA Sky Survey (NVSS). We cross-match these sources with a number of complementary measurements -- combining data from major radio polarization and total intensity surveys such as AT20G, B3-VLA, GB6, NORTH6CM, Texas, and WENSS, together with other polarization data published over the last 50 years. For 951 sources, we present spectral energy distributions (SEDs) in both fractional polarization and total intensity, each containing between 3 and 56 independent measurements from 400~MHz to 100~GHz. We physically model these SEDs, and where available provide the redshift of the optical counterpart. For a superset of 25,649 sources we provide the total intensity spectral index, $\alpha$. Objects with steep versus flat $\alpha$ generally have different polarization SEDs: steep-spectrum sources exhibit depolarization, while flat-spectrum sources maintain constant polarized fractions over large ranges in wavelength. This suggests the run of polarized fraction with wavelength is predominantly affected by the local source environment, rather than by unrelated foreground magnetoionic material. In addition, a significant fraction (21\%) of sources exhibit `repolarization', which further suggests that polarized SEDs are affected by different emitting regions within the source, rather than by a particular depolarization law. This has implications for the physical interpretation of future broadband polarimetric surveys.
\end{abstract}


\section{Introduction}
The combination of cosmic magnetic fields and charged particles, both of which are ubiquitous in the universe, results in the emission of synchrotron radiation from radio sources \citep[e.g.][]{2011hea..book.....L}. This radiation is fundamentally linearly polarized, and both the fractional polarization and electric vector polarization angle (EVPA) of this emission show significant frequency-dependence \citep[e.g.][]{1966MNRAS.133...67B}. This frequency-dependence is due to Faraday rotation and depolarization which, for extragalactic sources, are typically considered to be caused by magnetoionic material that either intervenes between us and the observed emitting region, or is intermixed with the emitting region itself \citep[e.g.][]{1966MNRAS.133...67B,1991MNRAS.250..726T,1998MNRAS.299..189S,2009A&A...502...61M}.

Understanding of Faraday rotation and astrophysical depolarization requires broadband radio measurements, and a large number of facilities are currently available or planned that will have the necessary bandwidth to study cosmic magnetic fields. For example, the Australian Square Kilometre Array Pathfinder (ASKAP) will observe at frequencies between 700~MHz and 1.8~GHz, \citep{2008ExA....22..151J}, the Australia Telescope Compact Array (ATCA) at $\ge1.1$~GHz \citep[e.g.][]{2011MNRAS.416..832W}, the Giant Metrewave Radio Telescope (GMRT) at frequencies $<$1.4~GHz \citep[e.g.][]{FARNESETAL}, the GALFA Continuum Transit Survey (GALFACTS) with Arecibo between 1.2 and 1.5~GHz \citep{2010ASPC..438..402T}, the Karl~G.~Jansky Very Large Array (JVLA) at $>$1.2~GHz \citep[e.g.][]{2009IEEEP..97.1448P}, the Low-Frequency Array (LOFAR) at $<$230~MHz \citep{2013A&A...556A...2V}, and the Murchison Widefield Array (MWA) between 80 and 300~MHz \citep{2013PASA...30....7T}.

Many of these facilities are pathfinders towards the Square Kilometre Array (SKA) Cosmic Magnetism Project \citep{2004NewAR..48.1003G,2011arXiv1111.5802B}. The SKA will detect up to $\approx10^{7}$ polarized extragalactic sources on the sky at a mean spacing of $\sim90$~arcsec \citep{2004NewAR..48.1003G}, and provide broadband measurements of both the polarized fraction and Faraday rotation. The polarimetric measurements that result will be used to construct a densely-sampled `Rotation Measure grid' \citep{2004NewAR..48.1289B,2005AAS...20713703G} that allow for analysis and reconstruction of the Galactic magnetic field \citep[e.g.][]{2009ApJ...702.1230T,2012ApJ...757...14J,2012A&A...542A..93O}, for investigation of the evolution of magnetic fields over cosmic time \citep[e.g.][]{2008ApJ...676...70K,2012arXiv1209.1438H,2013MNRAS.435.3575B}, and can reveal physical properties of the central engines in radio sources \citep[e.g.][]{2012MNRAS.421.3300O}.

Making sense of such data will require two capabilities: fast cross-matching algorithms for identification of counterpart sources with known redshifts at complementary wavelengths, and more importantly `$k$-corrections' that will allow for the polarization properties to be determined in the emitting frame of a source. These $k$-corrections will be necessary for the polarized fractions, the rotation measures, and possibly also for Faraday rotators that have a non-linear relationship between the EVPA and $\lambda^2$. In order to $k$-correct the polarized fraction to the emitting frame, we will require well-defined polarized spectral energy distributions (SEDs). Such SEDs can also be used to investigate the predominant causes of depolarization, and to classify individual sources.

As continuous broadband polarization data for a large number of sources will not be available until the next generation of radio telescopes, we can instead attempt to construct SEDs from existing data, to enable $k$-corrections of polarized sources. Polarized SEDs can be reconstructed using the considerable amount of archival polarimetric radio data, but require careful consideration of numerous systematics including resolution effects, beam depolarization, multiple source components, time-variability, Rician bias, differing uncertainties, outliers, and the use of either an interferometer or a single dish for data collection. The polarized fractions and rotation measures of 37,543 sources have been determined previously at 1.4~GHz using the NRAO VLA Sky Survey (NVSS) \citep{1998AJ....115.1693C,2009ApJ...702.1230T}. Polarized fractions and EVPAs have also been presented in a large number of other radio catalogs \citep[e.g.][]{1980A&AS...40..319S,1980A&AS...39..379T,1981A&AS...43...19S,1982A&AS...48..137S,1999A&AS..135..571Z,2003A&A...406..579K,2003PASJ...55..351T,2010MNRAS.402.2403M}. In addition, surveys that provide total intensity data allow for measurement of the spectral index and any spectral curvature \citep[e.g.][]{1991ApJS...75....1B,1996AJ....111.1945D,1996ApJS..103..427G,1997A&AS..124..259R}. Furthermore, the NVSS Rotation Measures (RMs) have recently been matched against optical catalogs, providing spectroscopic redshifts for more than 4,000 polarized radio sources \citep{2012arXiv1209.1438H}. This conglomeration of data not only provides the EVPAs, polarized fractions, and redshifts needed to construct and then $k$-correct a polarized SED, but also accumulates a large number of complementary parameters that are relevant to efforts for understanding cosmic magnetism.

In this paper, we cross-correlate the aforementioned data using a K-Dimensional Tree \citep[as described by][]{bentley1975}. Such a technique is able to rapidly eliminate large numbers of sources from a list of possible cross-matches. Consequently, the algorithm allows for a computationally inexpensive nearest-neighbor search -- reducing the problem from $\mathcal{O}(N^2)$ to $\mathcal{O}(N \log N)$. By accumulating data obtained at different frequencies, the algorithm allows us to construct a polarized radio SED with measurements between 0.4~GHz to 100~GHz. We then present a catalog of model fits to the polarized SEDs. The catalog also contains narrow and broadband RMs, total intensity spectral indices, estimates of the depolarization, and spectroscopic redshifts. We carefully detail the systematics and limitations of the catalog, including resolution effects, and the effect of multiple source components. This catalog constitutes the most comprehensive database of linear radio polarization currently publicly available, and increases the number of well-defined polarized SEDs by over an order of magnitude \citep[e.g.][]{2003A&A...406..579K,2004A&A...427..465F,2008A&A...487..865R,2009A&A...502...61M}. In this paper, we present the catalog of measurements of compact polarized radio sources, and define the possible uses and limits of our sample. Thorough scientific exploitation of this catalog is beyond the scope of this paper and will be the subject of future studies.

Throughout this paper, we characterize Faraday rotation as a rotation of the observed EVPA as a function of wavelength, so that
\begin{equation}
\Theta_{\textrm{EVPA}} = \Theta_{\textrm{0}} + \text{RM}\lambda^2 \,,
\label{rotationmeasureequation}
\end{equation}
where $\lambda$ is the observing wavelength, $\Theta_{\textrm{EVPA}}$ and $\Theta_{\textrm{0}}$ are the measured and intrinsic EVPA respectively, and the factor of proportionality RM, the rotation measure, is the gradient of the EVPA with $\lambda^2$. A more generalised quantity to parameterize Faraday rotation is the fundamental physical quantity, the Faraday depth. As a special case, the Faraday depth at which all polarized emission is produced is equal to the RM if there is only one emitting source along the line of sight, which has no internal Faraday rotation, and is not affected by beam depolarization, and there are only Faraday screens along the sight line \citep[for further detail, please see][]{2005A&A...441.1217B}. In such a case
\begin{equation}
\label{RM2}
\text{RM} = \frac{e^3}{2 \pi m_{\rm e}^2 c^4} \int_{d}^0 n_{\rm e} {\bf B} \cdot \text{d}{\bf s} \,,
\end{equation}
where \(n_{\rm e}\) is generally the electron number density of the plasma and \(\bf B\) is the magnetic field strength. The constants \(e\), \(m_{\rm e}\), and \(c\) are the electronic charge, the mass of the electron, and the speed of electromagnetic radiation in a vacuum respectively. The integral is performed along the line of sight from the source (at distance \(d\)) to the observer.

The paper is structured as follows: Section~\ref{accumulating} details the collection and preparation of the data used from various radio facilities. The cross-matching and fitting of each SED are explained in Sections~\ref{crossmatching} and \ref{sourcefitting} respectively. The catalog itself is presented in Section~\ref{catalog}, and possible systematic effects on the catalog are considered in Section~\ref{systematics}. The results are presented in Section~\ref{results}. Possible future applications and a general discussion are presented in Section~\ref{discussion}. The total intensity spectral index, $\alpha$, is defined such that $S_{\nu}\propto\nu^{+\alpha}$, where $S_{\nu}$ is the radio flux density and $\nu$ is the observing frequency. The polarized SEDs are fit using a number of applicable models, which include a polarization spectral index; $\beta$ is defined such that $\Pi \propto \lambda^{\beta}$, where $\Pi$ is the polarized fraction and $\lambda$ is the observing wavelength. Note that $\beta$ is defined in the opposite sense to the total intensity spectral index, $\alpha$, in that it is the exponent of observing frequency rather than wavelength. We refer to `polarization' on multiple occasions, in all cases we are referring to linear radio polarization -- circular polarization is beyond the scope of this work.

\section{Collection and Preparation of Data}
\label{accumulating}

We have collected data from throughout the literature at many different observing frequencies. The various parameters of the data incorporated into the catalog are displayed in Table \ref{tab:surveyparameters}, along with the abbreviations with which we will refer to them. Many of these data are available in an electronic format, while for others we manually converted them into a digitally accessible form.

\begin{deluxetable*}{ccccccc}
\tablewidth{0pt}
\tablecaption{Parameters of the archival data incorporated into our catalog. Data are listed in the order they were cross-matched.}
\tablehead{
\colhead{Abbrev.} & \colhead{$N_{\textrm{sources}}$} & \colhead{Resolution} & \colhead{Frequency}  & \colhead{Polarization}  & \colhead{Redshift}    & \colhead{Ref.}      \\
 \colhead{} & \colhead{}  &  \colhead{} & \colhead{/MHz}    & \colhead{}    & \colhead{} &  \colhead{}       }
\startdata
NVSS RM & 37543   & 45$^{\prime\prime}$                             & 1400    & Yes & No  & 1\\
NVSS & $1.8\times10^6$   & 45$^{\prime\prime}$                  & 1400    & Not used\tablenote{\small{The original NVSS contains polarization measurements. However, the NVSS RM catalog also provides the RMs and takes the effects of bandwidth depolarization into account.}} & No  & 2\\
H12 & 4003  & 45$^{\prime\prime}$                                       & 1400    & Not used\tablenote{\small{The catalog of H12 contains identical polarization information to that of the NVSS RM data.}} & Yes  & 3\\
TI80 & 1510    & Various (coarsest of $\approx6.5^{\prime}$)\tablenote{\small{TI80 accumulates data published prior to 1978, and does not state the angular resolution -- we therefore assume the observing beam to be large throughout our analysis.}}      & 400--100000 & Yes & Not used\tablenote{\small{The catalog of TI80 contains redshift data, but due to the better angular resolution of the Sloan Digital Sky Survey (SDSS) we consider the counterpart sources identified in \citet{2012arXiv1209.1438H} to be superior.}} & 4  \\
AT20G & 5890    &  $8.3^{\prime\prime}\times21.1^{\prime\prime}$\tablenote{\small{The AT20G has a beam that changes with declination. The values listed are the typical values for the survey.}}  & 4860    & Yes & No & 5 \\
\nodata      &  \nodata   & $4.6^{\prime\prime}\times11.7^{\prime\prime}$   & 8640    & Yes & No &  \nodata \\
 \nodata     &  \nodata   & 10.8$^{\prime\prime}$                           & 20000   & Yes & No &  \nodata\\
Z99  & 154     & 3$^{\prime}$                                  & 4700    & Yes & No & 6  \\
T03  & 185     & 7$^{\prime\prime}$                              & 1400    & Yes & No & 7  \\
 \nodata     & \nodata    & 4.5$^{\prime\prime}$                            & 2496    & Yes & No & \nodata  \\
 \nodata     &  \nodata  & 2$^{\prime\prime}$                              & 4800    & Yes & No &  \nodata \\
 \nodata      &  \nodata  & 1$^{\prime\prime}$                              & 8640    & Yes & No  & \nodata \\
B3-VLA\tablenote{\small{The B3-VLA catalog includes data from the NVSS at 1.4~GHz; we do not include this data a second time.}} & 192   & 261$^{\prime\prime}$                            & 2695    & Yes & No & 8  \\
 \nodata      &  \nodata  & 147$^{\prime\prime}$                            & 4850    & Yes & No & \nodata  \\
 \nodata      &  \nodata  & 69$^{\prime\prime}$                             & 10500   & Yes & No & \nodata  \\
SN80  &  103  & 6.5$^{\prime}$\tablenote{\small{The stated observing beams of SN80, SN81, SN82 are approximate values based on the dish diameter.}}                                  & 1580    & Yes & No & 9  \\
 \nodata     &  \nodata   & 6.0$^{\prime}$                                  & 1720    & Yes & No &  \nodata  \\
SN81  &  185  & 6.5$^{\prime}$                                  & 1580    & Yes & No & 10   \\
 \nodata    &  \nodata    & 5.9$^{\prime}$                                  & 1760    & Yes & No &  \nodata   \\
 \nodata    &  \nodata   & 0.7$^{\prime}$                                  & 14750   & Yes & No &  \nodata \\
SN82  & 68    & 2.1$^{\prime}$                                  & 10500   & Yes & No & 11  \\
GB6  & 75162     & 3.6$^{\prime}$                                  & 4850    & No  & No & 12   \\
WENSS  & 229420   & 54$^{\prime\prime}\times54^{\prime\prime}\text{cosec}\delta$  & 326 & No & No & 13  \\
Texas & 66841    & $\approx3.4^{\prime}$                                  & 365     & No  & No & 14 \\
NORTH6CM & 53522 & 3.5$^{\prime}$                                  & 4850    & No  & No & 15
\enddata
\tablerefs{\small{(1) \citet{2009ApJ...702.1230T}; (2) \citet{1998AJ....115.1693C}; (3) \citet{2012arXiv1209.1438H}; (4) \citet{1980A&AS...39..379T}; (5) \citet{2010MNRAS.402.2403M}; (6) \citet{1999A&AS..135..571Z}; (7) \citet{2003PASJ...55..351T}; (8) \citet{2003A&A...406..579K}; (9) \citet{1980A&AS...40..319S}; (10) \citet{1981A&AS...43...19S}; (11) \citet{1982A&AS...48..137S}; (12) \citet{1996ApJS..103..427G}; (13) \citet{1997A&AS..124..259R}; (14) \citet{1996AJ....111.1945D}; (15) \citet{1991ApJS...75....1B}.}}
\label{tab:surveyparameters}
\end{deluxetable*}

In all of the various catalogs, sources flagged as extended or unreliable in the original data were removed. Many source fluxes and polarized fractions in the literature were stated as an upper limit based on the detection threshold of a particular set of observations -- these measurements were also excluded. On occasion, some data displayed unusual values, with a flux density of zero janskys, negative values for positive-definite quantities, and polarized fractions of $>100$\% -- such values were also removed. Furthermore, the EVPAs were defined using differing conventions, with values between either $0^{\circ}$ to $180^{\circ}$ or between $-90^{\circ}$ to $+90^{\circ}$. All input data were adjusted so that the EVPA in the final catalog is consistently defined between $0^{\circ}$ to $+180^{\circ}$ from North through East. The coordinates of each input catalog were all converted to equatorial J2000 (in decimal degrees). Note that not all polarization catalogs also contain the corresponding total intensity data.

The final catalog contains data measured with a variety of different instruments, including that obtained during many large-scale surveys. The NVSS \citep{1998AJ....115.1693C} was a 1.4~GHz survey with the Very Large Array (VLA) covering the entire sky north of $-40^{\circ}$ declination at a resolution of 45~arcsec. The rms brightness fluctuations are approximately uniform across the sky at $\sim0.45$~mJy~beam$^{-1}$ in Stokes $I$ and $\sim0.29$~mJy~beam$^{-1}$ in Stokes $Q$ and $U$. The astrometry is accurate to within $<1$~arcsec for point sources with flux densities $>15$~mJy, and to $<7$~arcsec for the faintest ($\sim2.3$~mJy) detectable sources. The survey has a completeness limit of 2.5~mJy, and resulted in a catalog of over 1.8~million discrete sources in Stokes $I$. \citet{2009ApJ...702.1230T} reanalyzed the NVSS in order to derive RMs towards the radio sources with signal-to-noise (s/n) greater than $8\sigma$ in polarized intensity, leading to RMs for 37,543 radio sources -- an average density of approximately one RM per square degree. The RMs are based on just two narrow adjacent frequency bands, resulting in the possibility of significant uncertainty in each measurement. An attempt to overcome ambiguities in the RM was made using the depolarization of each radio source. The overwhelming majority of these sources are presumed to be extragalactic.

We now detail other previous observations that we have included into our catalog. We complement our catalog with measurements from the B3-VLA sample \citep[see e.g.][for further details]{2003A&A...406..579K}, which consists of 1,049 radio sources in five flux-limited subsamples stronger than 0.1~Jy at 408~MHz. Follow-up measurements with the Effelsberg 100~m telescope extended the frequency range to 151~MHz through to 10.5~GHz, which allowed detailed spectral studies of a sample exceeding 1,000 sources in total intensity. For the 192 sources with detected polarization at 10.5~GHz, additional measurements were taken at 2.695~GHz and 4.85~GHz.

We also include data from the AT20G survey, which was a blind 20~GHz survey of the entire southern sky at Galactic latitudes, $|b|$$>1.5^{\circ}$, carried out using the Australia Telescope Compact Array (ATCA) \citep{2010MNRAS.402.2403M}. The AT20G consisted of 5,890 sources above a flux density limit of $40$~mJy, and included near-simultaneous observations at 4.8~GHz and 8.6~GHz for most sources south of $-15^{\circ}$ in declination. The AT20G includes linear polarization measurements at all frequencies.

Our catalog also includes additional linear polarization measurements obtained at 4.7~GHz for 154 extragalactic radio sources using the Effelsberg 100~m telescope \citep{1999A&AS..135..571Z}, at 1.4~GHz, 2.5~GHz, 4.8~GHz, and 8.6~GHz for 202 sources using the ATCA \citep{2003PASJ...55..351T}, between 1.58~GHz and 1.76~GHz and at 14.75~GHz for 141 sources, and at 1.58~GHz and 1.72~GHz for 91 sources using the 100~m telescope at the Max-Planck-Institut f\"{u}r Radioastronomie \citep{1980A&AS...40..319S,1981A&AS...43...19S}, and at 10.5~GHz for 68 sources using the Algonquin Radio Observatory (ARO) 46~m telescope \citep{1982A&AS...48..137S}. We also include the accumulated data for 1,510 radio sources published prior to 1978 \citep{1980A&AS...39..379T}.

Although containing no linear polarization information, there are also a significant number of total intensity catalogs. We include data into our catalog from the GB6 survey at 4.85~GHz, which was made with the NRAO seven-beam receiver on the Green Bank 91~m telescope and detected 75,162 discrete sources \citep{1996ApJS..103..427G}. We also include the NORTH6CM survey at 4.85~GHz, which detected 53,522 sources \citep{1991ApJS...75....1B}. Furthermore, we include data at lower frequencies: from the Westerbork Northern Sky Survey (WENSS) at 326~MHz which detected 11,299 sources to a limiting flux density of $\approx18$~mJy \citep{1997A&AS..124..259R}, and from the Texas Interferometer at the University of Texas Radio Astronomy Observatory which detected 66,841 sources at 365~MHz and is 80\% complete at $\approx0.25$~Jy \citep{1996AJ....111.1945D}.

Efforts have previously been made to identify complementary sources at other wavelengths in total intensity \citep[e.g.][]{2008AJ....136..684K,ricci2013}. The NVSS has also been used to identify optical counterparts and spectroscopic redshifts for 4,003 linearly polarized radio sources, using various resources including the Sloan Digital Sky Survey (SDSS) \citep{2009ApJS..182..543A,2012arXiv1209.1438H}. The varying sky coverage, observational frequencies, and other systematics have complicated the interpretation of these previous catalogs, as it is not particularly clear which source population is being studied. A sample selected at a single observational frequency, and with as uniform sky coverage as possible, is therefore beneficial in minimizing the selection effects that are intrinsically abundant when producing such a catalog. As shall be detailed in Section~\ref{crossmatching}, our catalog constitutes a polarized 1.4~GHz flux-limited sample, based on the NVSS.

\section{Cross-Matching}
\label{crossmatching}

\subsection{Foundations of Cross-Matching}
\label{thebasics}
Cross-matching is the process of identifying counterpart sources that are coincident in position on the sky. This process is typically used for total intensity measurements, but we seek to expand the method for cases with a corresponding polarization measurement. We wish to use such a technique to group measurements at different observing frequencies that correspond to the same physical source. Various techniques have been previously implemented to cross-match radio catalogs, and any such method must provide quantitative limits on the probability of false catalog associations. Previously used techniques have included nearest-neighbor searches \citep[e.g.][]{2002AJ....124.2364I,2013MNRAS.429.2080W} and Bayesian approaches \citep[e.g.][]{2008ApJ...679..301B}. Such techniques constitute a form of coincidence-assessment that can be used to determine the probability of finding false matches, i.e.\ the false-detection rate (FDR).

A cross-matching algorithm needs to be computationally efficient. We are limited here by our largest data set, the NVSS, which consists of 1.8~million sources across 75\% of the sky. We therefore implement the use of a K-Dimensional Tree \citep{bentley1975}. Such an algorithm is a computationally inexpensive nearest-neighbor search -- reducing the problem from $\mathcal{O}(N^2)$ to $\mathcal{O}(N \log N)$ -- and returns all of the matches within a pre-defined spatial radius around a reference source. The K-Dimensional Tree is thereby able to rapidly eliminate large numbers of sources from a list of possible cross-matches. 

It is necessary to choose a \emph{reference catalog} to which source associations can be concatenated. We define this reference catalog to be the NVSS RM data at 1.4~GHz \citep{2009ApJ...702.1230T}. Each \emph{secondary catalog} was then cross-matched with the reference catalog. There are two possible outcomes to this process: (i) a cross-match exists and the relevant secondary catalog data are attached to the corresponding NVSS RM source, or (ii) a cross-match does not exist and the secondary catalog data are discarded for that NVSS RM source. In this manner, we produce a flux-limited sample that is selected on the basis of being polarized at 1.4~GHz. 

Further constraints are also required in order to control the FDR, as the various secondary catalogs all have different resolutions, sensitivities, and unevenly-spaced coverage on the sky. The FDR is further affected by the combined astrometric errors of both the reference and secondary catalogs, which limit the accuracy of any given source position. The various sensitivities and sky coverage will only affect the completeness of our final catalog. On the other hand, the differing resolutions can introduce a number of systematics (see Section~\ref{systematics} for further details on catalog systematics and limitations). However, we only consider sources that are unresolved at all wavelengths: ensuring that any measured beam depolarization is not affected by the differing parameter-space of each input catalog. The presence of beam depolarization, or otherwise, is one of the many physical inferences that may be drawn from a resulting catalog. Furthermore, while the effects of multiple source components or internal source structure could likely only be overcome by a broadband and high-resolution all-sky polarization survey, such as that to be performed by ASKAP or the SKA -- other measures of a source (such as the total intensity spectral index and optical counterpart properties) can also allow for such effects to be studied. In this way, differing resolutions do not affect a resulting catalog; each source being consistently unresolved is rather an aspect that will be useful for future investigations. These effects are therefore beyond the scope of this current paper. Importantly however, the differing resolutions could lead to potentially cross-matching independent sources that are unresolved in a given individual catalog -- thereby increasing the FDR. We therefore need a matching criterion that defines a suitable astrometric matching radius, and that considers whether each source is appropriately isolated within the observing beam. We shall consider matching criteria in Section~\ref{thematchingradii}. 

As the NVSS RM data do not contain the angular size, EVPA, or redshift of each source, we add this information to the catalog by cross-matching to the original NVSS \citep{1998AJ....115.1693C} and to a catalog of NVSS redshifts \citep{2012arXiv1209.1438H}. As all three catalogs use the NVSS source positions, an arbitrarily small cross-matching radius was used to combine the data.

\subsection{Defining the Matching Radii}
\label{thematchingradii}
As the FDR is a measure of the probability of coincidental associations, this must be primarily related to the density of sources on the sky. Following \citet{1998AJ....115.1693C}, such a relation is given by
\begin{equation}
P(<r) = 1-\exp{(-\pi\rho r^2)} \,,
\label{probFDR}
\end{equation}
where the probability of the nearest unrelated source, $P(<r)$, lying within an area $\pi r^2$ is dependent on the number of sources per square degree, $\rho$. By adjusting the maximum threshold radius at which cross-matches are confirmed or rejected, it is therefore possible to control the FDR to be below some subjective limit that we deem to be acceptable.

Such methods are useful for checking if optical sources emit in the radio, as the coordinates to optical sources typically have good positional accuracy. The situation becomes more complicated when comparing two sets of radio data, in which there may be significant inaccuracies in both surveys to be matched. Furthermore, while such cross-matching evaluates the positional accuracy (or astrometry) of each measurement, it does not take resolution effects into account. It is possible that unresolved sources as seen with a larger observing beam are resolved into separate components in a high resolution survey. Consequently, we define two radii for our matching criterion  -- an astrometric radius, $r_{a}$, that accommodates the accuracy of a measured source position, and an isolation radius, $r_{i}$, that handles resolution effects.

In choosing an astrometric radius, we need to properly consider that the astrometric errors are not dominated by those in a single survey, and that there are typically significant contributions from both the reference and secondary catalogs. We assume that the effects of s/n on the astrometry are negligible; while the positional uncertainty typically depends on the s/n, all of our sources are of sufficient s/n to be detected in linear polarization -- which tends to select the sources of highest total intensity flux density within a given survey. We approximate the rms uncertainty in the source positions to be given by the combined rms of both catalogs, such that $\sigma=\sqrt{\sigma_{\textrm{1}}^2+\sigma_{\textrm{2}}^2}$, where $\sigma_{\textrm{1}}$ and $\sigma_{\textrm{2}}$ are the contributions from the reference and secondary catalogs respectively. We estimate that a suitable $r_{a}$ could be $\approx3\sigma$, such that we would allow a match if a source is within the $\sim99.7$\% confidence interval. Nevertheless, values of $r_{a}$ smaller than $3\sigma$ would also be acceptable, as this only serves to place a tighter constraint on the identified cross-matches. Moreover, due to the low source density of our catalogs, coincidental matches can be expected to be low even for substantially larger $r_{a}$ (see equation \ref{probFDR}). Of course, the most reliable method would be to independently measure the positional errors in each catalog. To balance these considerations, the exact chosen values of $r_{a}$ are determined empirically from an assessment of the FDR in Section \ref{section:FDR}. The estimated positional errors for each survey are stated in Table \ref{tab:astrometry}. In cases where the astrometry is not provided for a secondary catalog, we make the conservative estimate that the rms uncertainty in each position is given by $<\theta/3$, where $\theta$ is the FWHM of the resolution element.

In terms of an isolation radius, we always take $r_{i}$ to be equal to the FWHM of the catalog with the larger observing beam -- regardless of whether the secondary catalog to be cross-matched has an observing beam that is smaller or larger than the NVSS beam of FWHM $\theta_{\textrm{NVSS}}=45$~arcsec. As $r_{i}$ is a radial distance, the source will always be isolated within $2\times$FWHM of the catalog with the larger observing beam. The isolation radius used is listed in Table \ref{tab:astrometry}. This provides a limited number of possible isolation scenarios while cross-matching an individual source, as follows:
\begin{compactenum}[i]
\item No sources are detected within $r_{i}$.
\item Only one source is detected within $r_{i}$.
\item More than one source is detected within $r_{i}$.
\end{compactenum}
In case (i), nothing is done and we simply iterate to the next source. In case (ii), the detected source is appended to the reference catalog if the astrometric radius does not denote this as a positional coincidence. In case (iii), the source consists of multiple components and neither of these components are appended to the final catalog. Consequently, the catalog selects isolated sources, which are unresolved at all frequencies.

\begin{deluxetable*}{ccccccc}
\tablewidth{0pt}
\tablecaption{The Astrometry, Astrometric and Isolation Radii, and False Detection Rate for Cross-Matched Data}
\tablehead{
\colhead{Abbrev.} & \colhead{Estimated positional uncertainty ($1\sigma$)}  & \colhead{Isolation radius ($r_{i}$)}  & \colhead{Astrometric radius ($r_{a}$)}  & \colhead{False Matches}  & \colhead{Real Matches} & \colhead{FDR}     \\
\colhead{} & \colhead{}         & \colhead{/arcsec} & \colhead{/arcsec}         & \colhead{} & \colhead{} & \colhead{/\%} }
\startdata
NVSS      & $<1^{\prime\prime}$ for $S>15$~mJy              & N/A   & N/A    & N/A    &  N/A    &  N/A \\
 \nodata     & $\sim7^{\prime\prime}$ for $S\approx2.3$~mJy  &  \nodata  & \nodata & \nodata & \nodata  & \nodata   \\
 TI80      & $<2.2^{\prime}$                                 & 390 & 150                                  & 38    &  767     &  5.0   \\
 AT20G     & 0.9$^{\prime\prime}$ in right ascension         & 45  &  7                                   & 4     &  1272    &  0.3  \\
\nodata           & 1.0$^{\prime\prime}$ in declination             & \nodata & \nodata & \nodata & \nodata  & \nodata     \\
 Z99       & $<1.0^{\prime}$                                 & 180 & 25                                   & 0     &  54      &  0.0  \\
 T03       & $<1^{\prime\prime}$                             & 45  & 5                                    & 1     &  103     &  1.0   \\
 B3-VLA    & $\sim1^{\prime\prime}$                          & 69 & 10                                   & 1     &  132     &  0.8   \\
 SN80      & $<2.2^{\prime}$                                 & 390 & 140                                  & 7     &  144     &  4.9  \\
 SN81      & $<2.2^{\prime}$                                 & 390 & \nodata & \nodata & \nodata  & \nodata   \\
 SN82      & $<2.2^{\prime}$                                 & 390 & \nodata & \nodata & \nodata  & \nodata   \\
 GB6       & $\approx7.5^{\prime\prime}$ in right ascension  & 216 & 90                                   & 715   &  16700   &  4.3 \\
 \nodata          & $\approx8.5^{\prime\prime}$ in declination      & \nodata  & \nodata & \nodata & \nodata  & \nodata   \\
 WENSS     & $1.5^{\prime\prime}$                            & 54  & 17                                   & 133   &  11893   &  1.1 \\
 Texas     & $\approx1^{\prime\prime}$                       & 204 & 10                                   & 363   &  5789    &  6.3 \\
 NORTH6CM  & $25^{\prime\prime}$                             & 210 & 90                                   & 463   &  14166   &  3.3
\enddata
\label{tab:astrometry}
\end{deluxetable*}

\subsection{Assessing the False Detection Rate}
\label{section:FDR}

\begin{figure*}
\centering
\includegraphics[clip=true, trim=1.5cm 0cm 0cm 0cm, width=19.5cm]{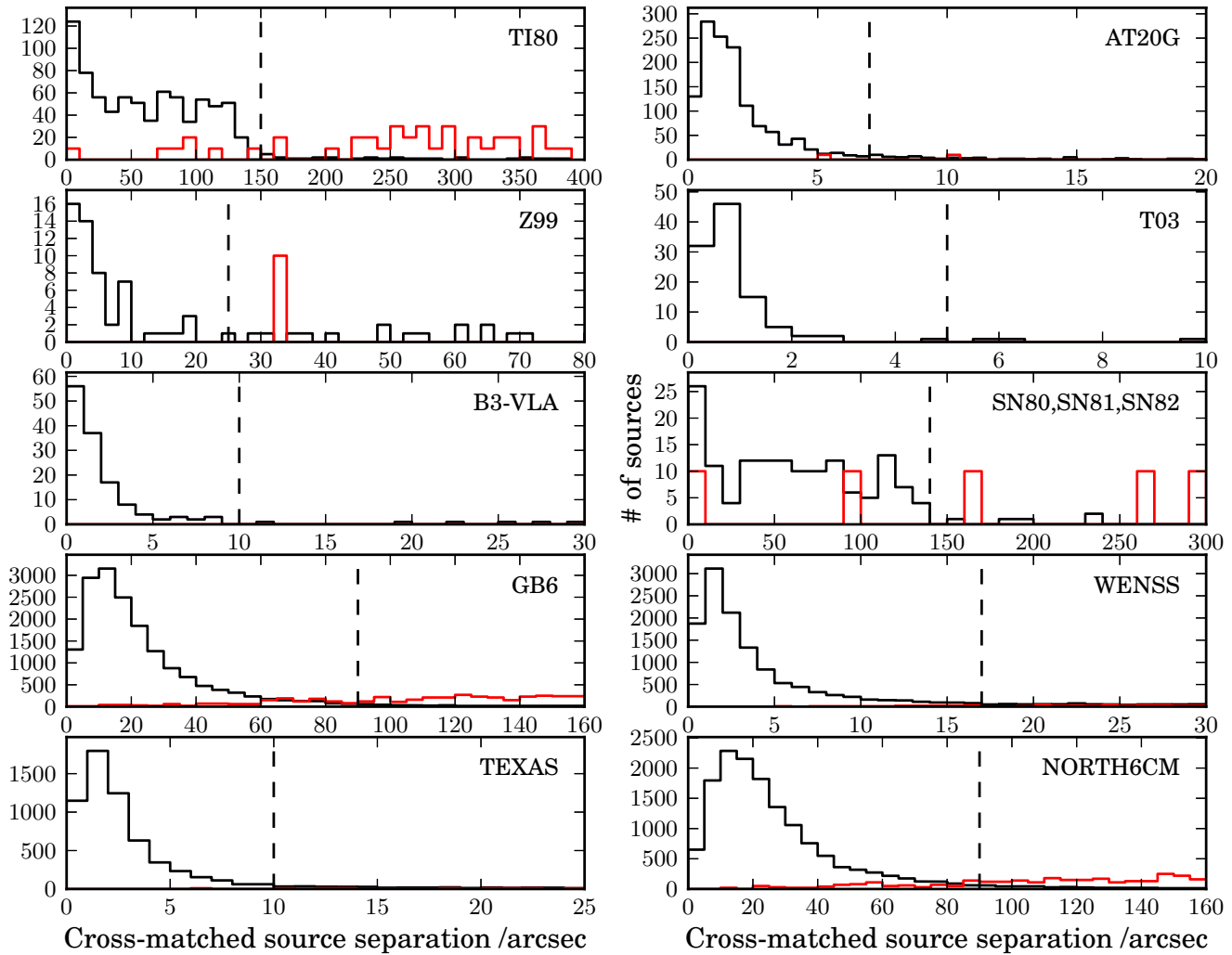}
\caption{Histograms of the number of sources from the NVSS RM catalog \citep{2009ApJ...702.1230T} with identified counterparts, as a function of the measured angular separation between the NVSS RM and counterpart source in arcsec. Each histogram shows the output for every secondary catalog that was added to our sample, relative to the NVSS. In an idealised scenario, with no positional errors in either the NVSS or secondary catalog, we would expect to obtain a delta function at an angular separation of zero with a height equal to the total number of sources. The black solid lines show the results for the true source distribution, and the red solid lines show the results for a quasi-random source distribution with an added positional offset. Note that the red source distribution is scaled in the plot to appear larger by a factor of 10. The number of positional coincidences for the quasi-random source distribution is typically low, and for many plots are not visible on the axes despite the scaling. The estimated astrometric cross-matching radii are shown as dashed vertical black lines, and were estimated based on the angular separation at which the number of true and randomized sources is approximately equal.}
\label{histograms}
\end{figure*}

In order to assess the reliability of a particular cross-match, we need to measure the combined positional errors of each of our catalogs relative to the NVSS RM reference catalog. This also allows us to measure the most suitable astrometric cross-matching radius, $r_{a}$. In addition, we would also like to estimate the FDR by placing quantitative limits on the number of false cross-matches that would arise if a given survey was cross-matched to a collection of sources that were distributed randomly across the sky. Following the methodology of \citet{2013MNRAS.429.2080W}, we run the cross-matching as normal, and then a second time with the position of each source being offset by $10/\sqrt{2}$~arcmin in both right ascension and declination. For both runs, we record the radial distance between every source and its identified counterpart. The histogram of the number of matches as a function of the radial distance between matches can then be created for both the true and offset sky positions. The separation between matched sources is a measure of the combined positional errors in both catalogs, and allows us to estimate the positional errors and the number of coincidental matches. These histograms are shown in Fig.~\ref{histograms} for every cross-matched catalog. The number of matches for the true and offset sources becomes approximately equal at some specific separation, and this distance was estimated and chosen as the astrometric matching radius. If this astrometric radius is underestimated, a substantial number of real matches will be rejected. Conversely, if overestimated, we will be accepting an increasing number of false matches with minimal return in terms of an increased sample size. Our estimates therefore provide a balance between maximizing our sample size and maintaining a low FDR. The estimated values for $r_{a}$ are typically proportional to the estimated uncertainties in the source positions as shown in Table \ref{tab:astrometry}. For cross-matching, the set of secondary measurements were only appended to the reference catalog if the listed source positions agreed to within $r_{a}$.

\begin{figure}[hpt]
\centering
\includegraphics[clip=true, trim=1cm 0cm 0cm 0cm, width=9.5cm]{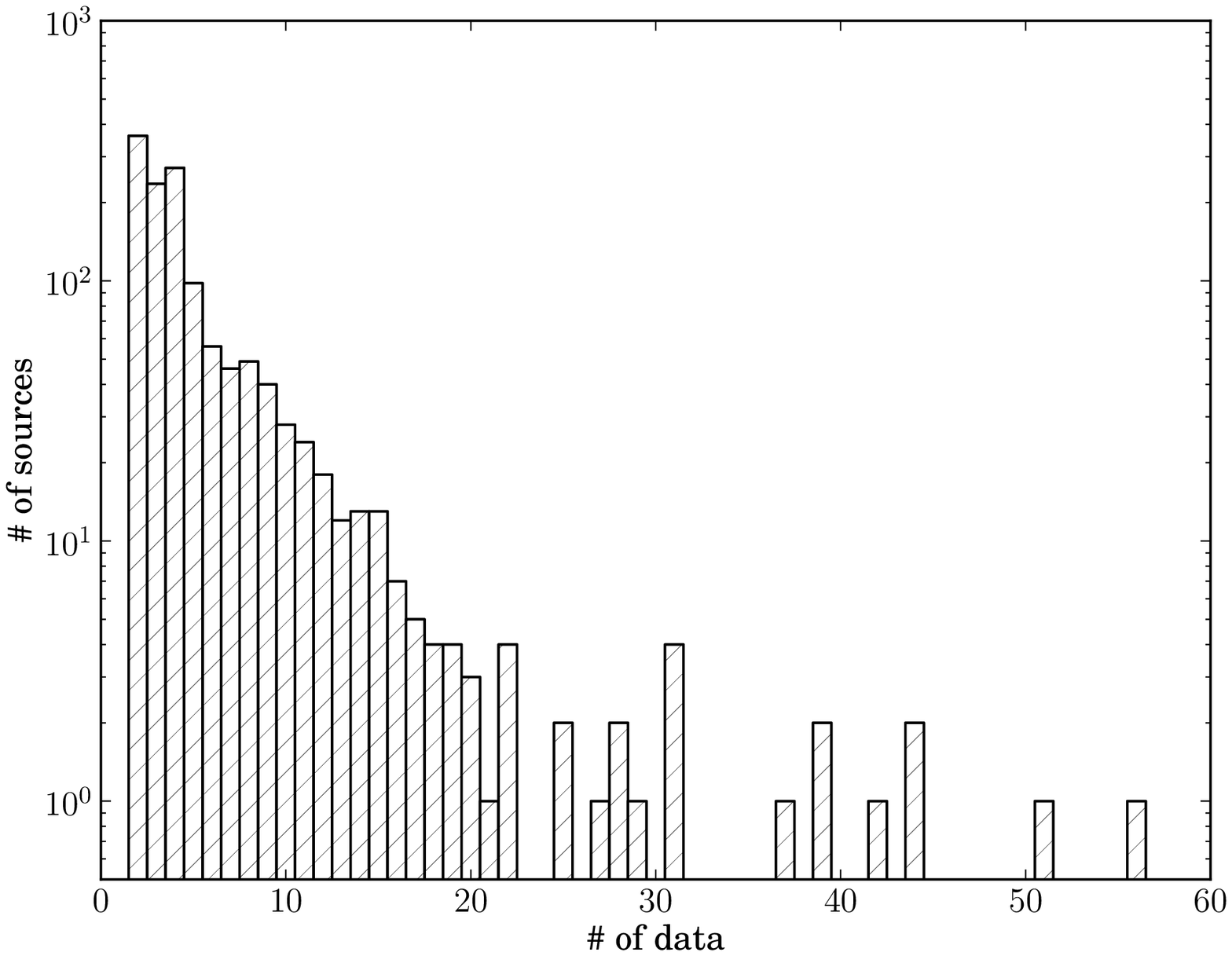}
\bigskip
\includegraphics[clip=true, trim=1cm 0cm 0cm 0cm, width=9.5cm]{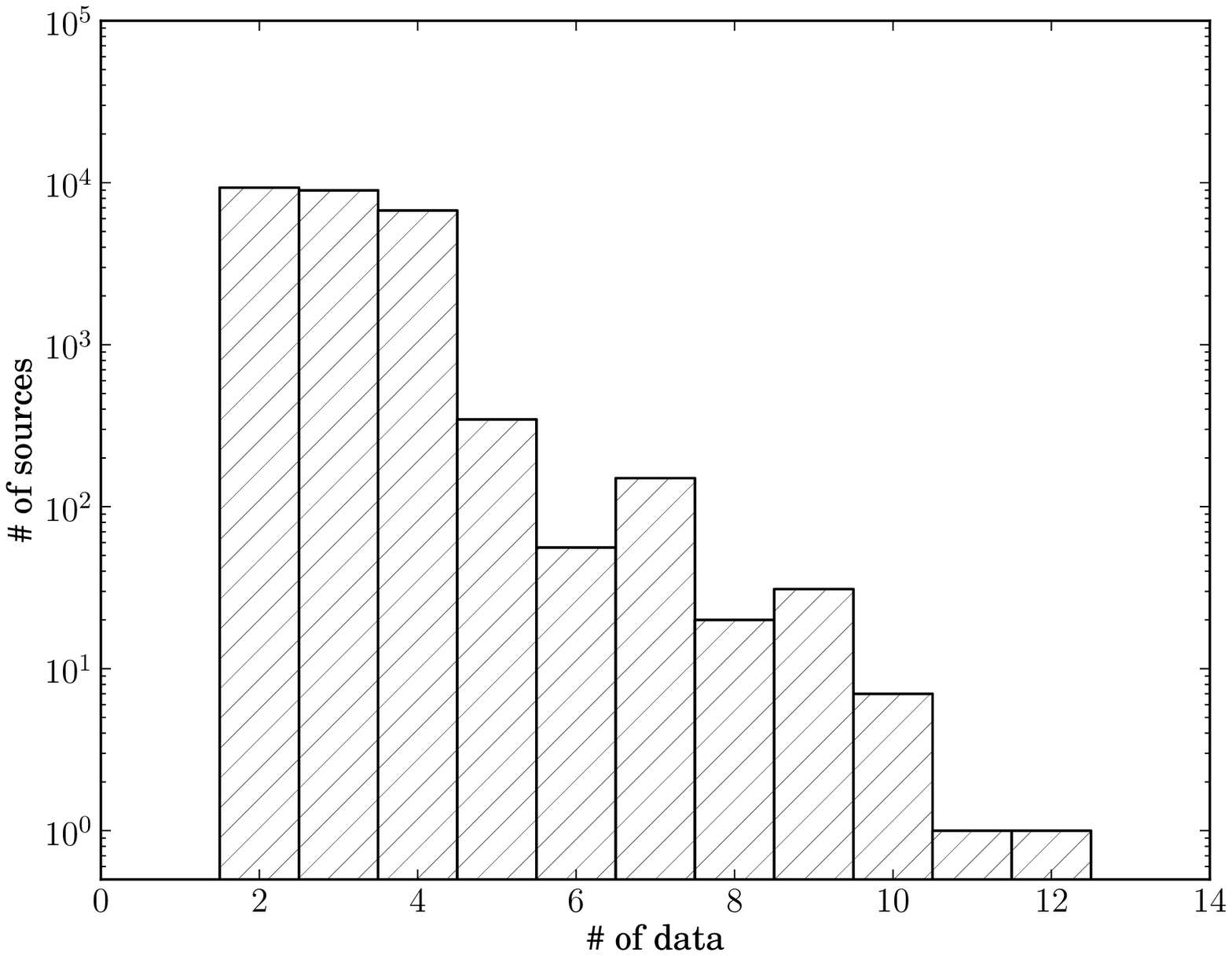}
\caption{Histograms showing the number of sources that have a given number of cross-matched measurements or data points in the SED. The histograms are shown for both polarized (top) and total intensity (bottom) SEDs. Note that the $y$-axes are shown as a log-scale. Sources with just one datum (i.e.\ no cross-matches), are not shown.}
\label{polHisto}
\end{figure}

As an example, we consider the case of the AT20G. Based on our matching criteria, the offset sky positions result in four matches, compared to 1,272 matches when no offset is added. We therefore estimate the FDR of the AT20G (when combined with the NVSS RM catalog) to be $\approx0.3$\%. In Table~\ref{tab:astrometry}, we list the measured astrometric radii that we used for the cross-matching of each catalog, together with the number of matches for each catalog, and the estimated FDRs. Histograms of the number of cross-matched measurements for each source are shown in Fig.~\ref{polHisto}. The histograms show that most polarized sources have $\le20$~measurements, while a few sources have up to 56~measurements. Similarly, most total intensity sources have $\le5$~measurements, while a few sources have up to 12~measurements. Note that some measurements are at similar or closely-spaced wavelengths.

\section{Spectral Energy Distributions: Fitting and Automated Classification}
\label{sourcefitting}

\subsection{The Model Selection Algorithm}
\begin{figure*}
\centering
\includegraphics[clip=true, trim=0.7cm 0.5cm 0cm 0.2cm, width=18.3cm]{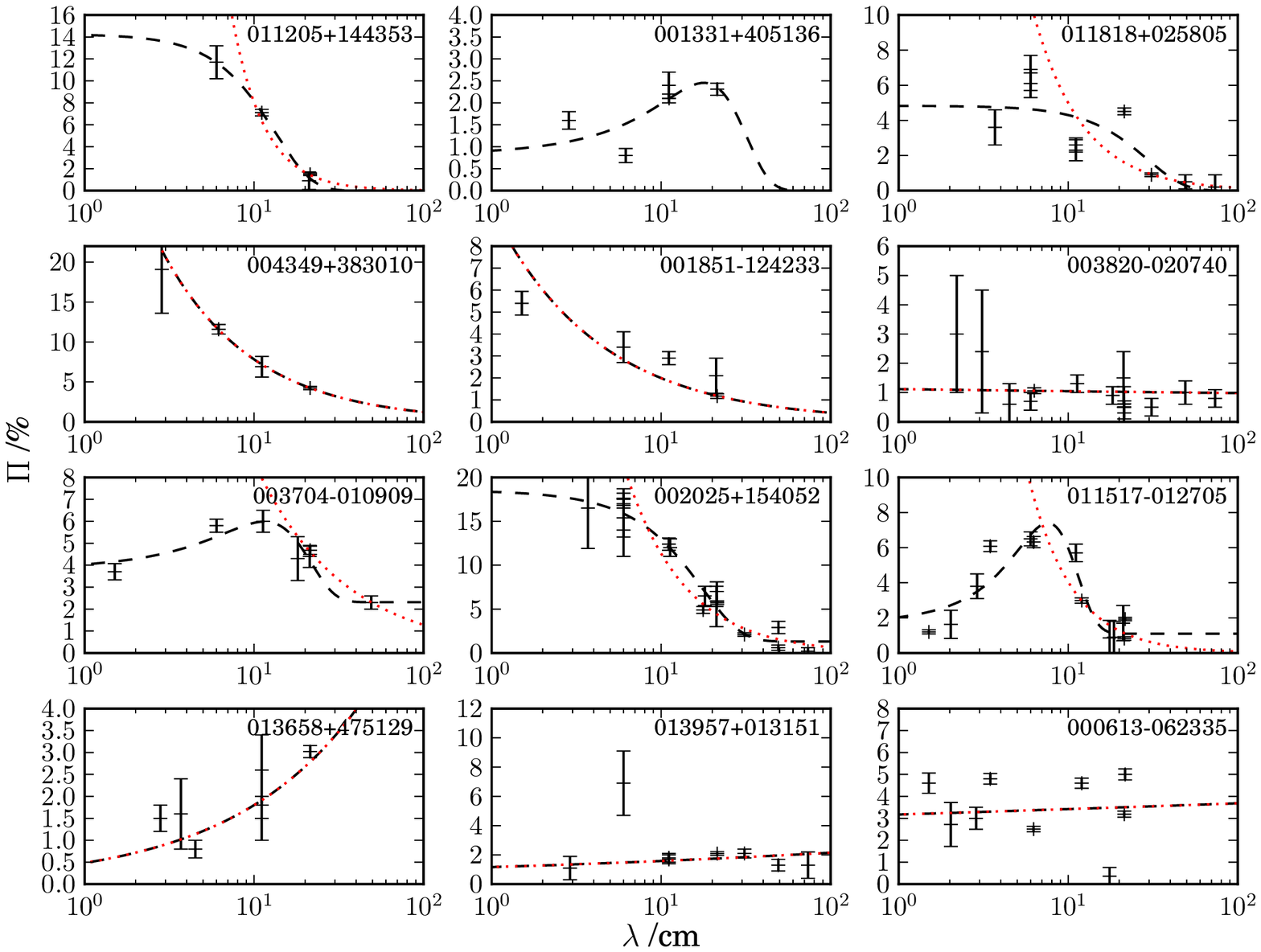}
\caption{Examples of the cross-matched SED data for nine sources from the catalog. The NVSS source name is shown in the upper right. Note that the $x$-axis is plotted as a log-scale. Two fitted polarized models are shown: (i) the model selected by the automated classification algorithm which is shown as a black dashed line, and (ii) the fitted polarization spectral index, $\beta$, which is shown as a red dotted line. The polarized spectral index is only calculated when certain conditions are met, as described in the main text. A selection of SEDs are shown for each specified model, including Gaussians (top row), depolarizing power laws (second row), Gaussians with a constant term (third row), and repolarizing power laws (bottom row). After a number of statistical tests, these sources are identified by the data quality flag as `accept' (left column), `caution' (middle column), and `poor' (right column).}
\label{SEDs}
\end{figure*}

\begin{figure*}
\centering
\includegraphics[clip=true, trim=0.7cm 0.5cm 0cm 0.2cm, width=18.3cm]{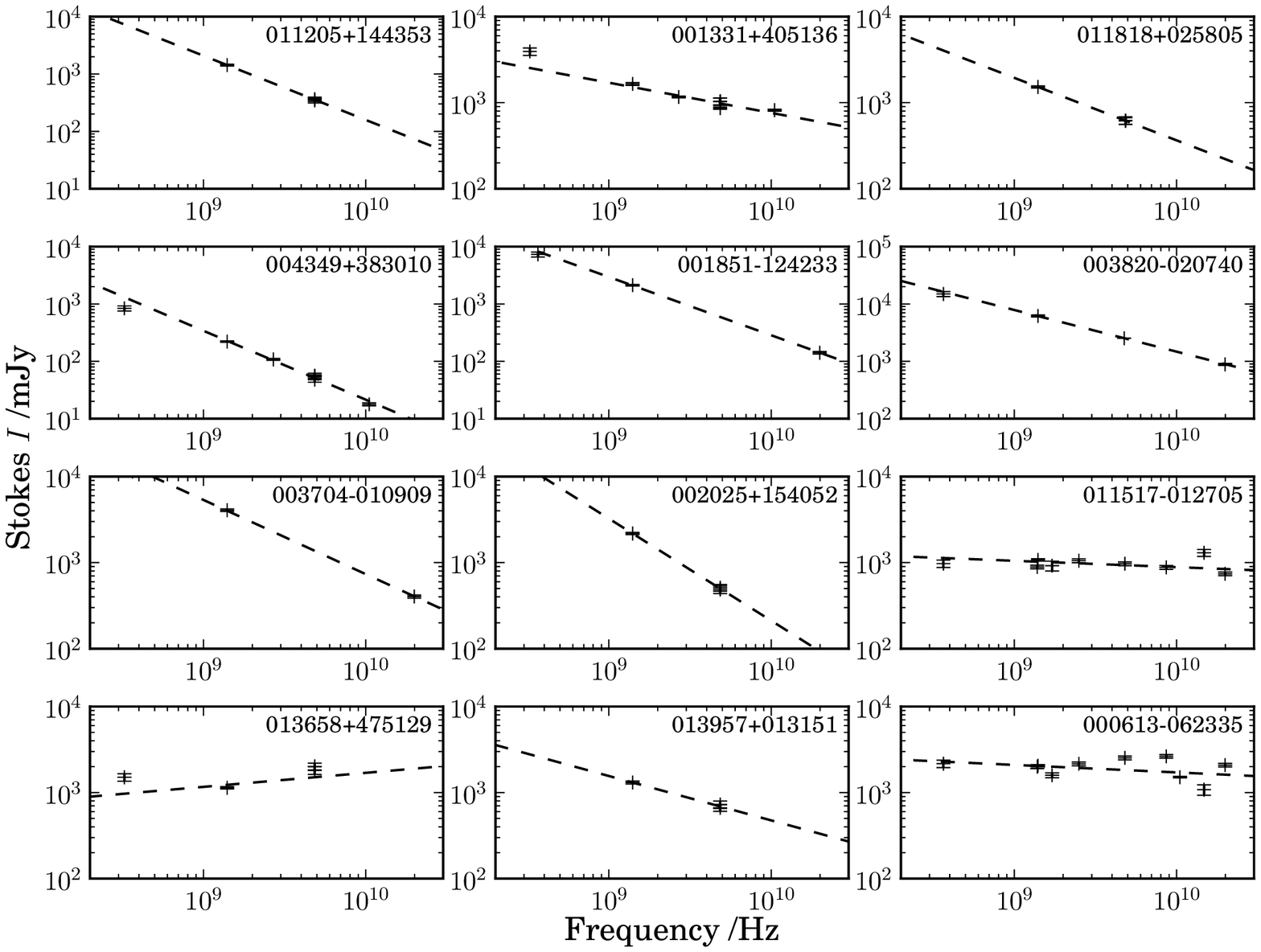}
\caption{Examples of fitted total intensity models from the catalog with the original SED data also shown. The NVSS source name is shown in the upper right. Both the $y$- and $x$-axes are plotted as a log-scale. The sources displayed are identical to those shown in Fig.~\ref{SEDs}. The model SED selected by the automated classification algorithm is shown as a dashed line -- in all of the displayed cases the total intensity spectrum is classified as a regular power law.}
\label{SEDs-I}
\end{figure*}

The cross-matching in Section~\ref{crossmatching} provides 951 SEDs, each of which yields measurements of the polarized fraction as a function of observing wavelength (as shown in Fig.~\ref{SEDs}), and of the total intensity as a function of frequency (as shown in Fig.~\ref{SEDs-I}). The SEDs show a significant number of different behaviors, with sources showing a monotonic decline, distinct maxima, or even oscillatory behavior with varying wavelength. We now wish to fit a depolarization model to each SED and here focus on the fitting and model selection procedure -- the physics that motivates each model is detailed in Appendix~\ref{appendix-b}. Across the wavelength range of 0.4~GHz to 100~GHz, we only use sources that have three or more data -- often providing minimal constraints to our fits. Fitting to such empirical data is one of the most ubiquitous and notorious problems in astronomy \citep{1990ApJ...364..104I,1992ApJ...397...55F,1996ApJ...470..706A,2007ApJ...665.1489K}. 

Model fitting can assist in further reducing the FDR (the FDRs were measured in Section \ref{section:FDR}). Consider the case where the data exhibit `regular depolarization', which we define to be an unambiguous and monotonic decline in fractional polarization as a function of increasing wavelength, e.g.\ as shown by the sources in the top-left and middle-left of Fig.~\ref{SEDs}. We make the assumption that the distribution of $\Pi$ values\footnote{$\Pi=P/I=\sqrt{Q^2+U^2}/I$ -- we here neglect any correction for the effect of Rician bias -- see Section~\ref{ricianbias} for further details.} can be very approximately modeled as being drawn from a Rayleigh distribution -- such an assumption may be reasonable based on previous studies which show a distribution that is positive-definite, with a defined peak, and an asymmetric tail \citep[e.g.][]{2010ApJ...714.1689G,2013arXiv1309.2527M}, although such analyses require careful consideration of numerous systematics \citep[e.g.][]{halesmnras13}. In the simplest case, all values are drawn from distributions with a similar mean, $\sigma\sqrt{\frac{\pi}{2}}$, and mode, $\sigma$, for the population at each frequency. The probability of drawing three values from these distributions that appear to monotonically decline as a function of wavelength was estimated using Monte Carlo techniques to be 0.16. Consequently, our estimated typical FDR of $\approx5$\% (see Table \ref{tab:astrometry}) can be modified to $\approx0.8$\% (i.e.\ $5$\%$\times0.16$) for cases where an SED with three measurements is found to be a regular depolarizer. The FDR continues to decrease as a function of the number of measurements in the SED: with four data our estimated FDR can be modified to $\approx0.21$\%, while with five data our FDR becomes $\approx0.04$\%. Note that our catalog contains up to 56 measurements in a single SED (see Table~\ref{tab:surveyparameters}). These FDRs are best case scenarios, depending on our assumption that a Rayleigh distribution can be used to model the polarized fractions. At the very minimum, we estimate that at least 578 of the 951 sources in our catalog are regular depolarizers, and can be considered to have a reduced FDR. For many more sources this becomes complicated to assess due to the uncertainty of each datum. It should be noted that such an argument does not exclude that we may be probing different parts of the source at each observing frequency, especially as the recorded $\Pi$ in each input catalog is biased towards the most highly polarized region of a given source. Consequently, any measurement of different source components will affect our estimate of a reduced FDR -- we therefore maintain our worst-case assumption of a FDR of $\approx5$\% (see Section~\ref{systematics} for a discussion of systematic effects in the catalog).

We are therefore faced with the issue of both obtaining the parameters that define the fitted SED, while also attempting to determine the physical model that best describes each source. For this purpose of model selection, we therefore need to pick a number of models a-priori that may describe the physics of each source. The functional forms of the chosen models are described in detail in Section~\ref{fittingpolarizedSED}, alongside our justification for their use. The physics that motivates each of these models is explained in Appendix~\ref{appendix-b}. Distinguishing which of these is the \emph{correct} model has been well-described in the case of univariate data \citep{2007arXiv0706.1062C}, while in the case of bivariate data such as that presented here, linear regression is the most appropriate tool \citep[as described by e.g.][]{warton2006}. As each SED consists of independent data with differing $1\sigma$ uncertainties, we use a combination of weighted least squares (WLS) and the non-linear Weighted Levenberg--Marquadt algorithm (LMA) to perform the fitting of each model. We use WLS for cases where the model to be fit can be linearized, and use the LMA for non-linearized cases. The reported errors in the fits for the LMA are fundamentally approximate. Furthermore, the LMA only provides the parameters of a local rather than a global minima. For a significantly large sample of fitted SEDs, we argue that we will still be dominated by random, rather than systematic errors.

For the large data volumes expected with the next generation of telescopes, automated and robust classification algorithms will be essential. Such automated classification is complicated not just by an assessment of the quality of fit, but also by different physical models having varying numbers of degrees of freedom. In order to perform model selection in an automated fashion, we therefore require a methodology to distinguish between better fits, while simultaneously penalizing increasing model complexity. To do this, we assessed the Bayesian Information Criterion or BIC \citep{schwarz1978}.\footnote{The BIC is minimized by the model with the highest posterior probability, and can be considered as a rough approximation to the logarithm of the Bayes factor \citep[as detailed by][]{kassraftery1995,spiegelhalter2002}. In the case of model selection for models with equal numbers of parameters, the BIC simply reduces to maximum likelihood selection. The BIC is also asymptotically consistent as a selection criterion, i.e.\ given a sufficient family of models that includes the `correct' model, as the sample size increases, the probability that the BIC selects the correct model tends to one. Such models are not required to be nested.} The BIC evaluates
\begin{equation}
\textrm{BIC} \approx -2\ln L + k\ln n \,,
\label{BIC}
\end{equation}
where $\ln L$ is the log-likelihood of the data given the model, and $k\ln n$ is a parsimony term with $k$ giving the number of free parameters in the model, and $n$ giving the sample size.\footnote{We make the simplifying assumption that model errors are independent and distributed according to a normal distribution. Under the assumption of normality, an additional trivial constant term appears in the BIC, which is purely a function of the data. Consequently, it is not possible to compare the fit between different sources -- it is only possible to assess the quality of different model fits at describing a single SED.} For each physical scenario, we select as the `true model' that which has the lowest value of the BIC.\footnote{It is not possible to define a confidence interval or to evaluate the evidence against an alternative model, as such techniques are overly subjective \citep[see][for further detail]{raftery1995}.}

The model we select serves as a parameterization of the SED. Despite the challenges presented for SEDs with a small number of measurements, we can determine the model of greatest likelihood given the data. It is possible that the `true' SED has deviations from the model that our algorithm selects, and broadband polarization observations will be necessary to understand such rapid fluctuations as a function of $\lambda$. Deviations from our assumed models are beyond the scope of this paper, but can in principle be investigated using the statistical tests we used to assess the quality of an SED model fit.

\subsection{Statistical Tests}

Beyond a method for model selection, we also require a measurement of the quality of the model selection process. In the catalog we provide both the reduced-$\chi^2$ and the Kolmogorov--Smirnov statistic for assessment of the quality of the model that is selected by our classification algorithm. We also provide the associated $p$-value for each test. These statistical tests are important for revealing cases where no model describes the data well.

Note that reduced-$\chi^2$ is not appropriate for many SEDs, as $\chi^2$ statistics are derived in the limit of a sample size that tends towards infinity. However, as least squares methods are derived from likelihood theory -- and assume normally distributed residuals \citep[e.g.][]{sprent1969} -- the Kolmogorov--Smirnov statistic (KS-test) can be used as an alternative method to assess the normality of the residuals between our data and selected model.

In our particular case, we require comparison of the residuals to a normal distribution of unknown mean and variance -- the KS-test can be modified to deal with such a case \citep[as described by][]{lilliefors1967}. Furthermore, the KS-test assumes that the data and model are independent, while in this case the data have been used to derive the model. This does not affect the value of the KS-statistic, but causes all associated $p$-values to be in error. We therefore perform `bootstrap resampling'. We evaluate the KS-test \citep{lilliefors1967} for an SED, and then find the correct $p$-value by combining our data and model datasets, and randomly sampling from them (with replacement) to form two new samples of size $N$ (where $N$ is the number of SED datapoints). We then compute the KS-test statistic for these new self-generated samples. This process was repeated 10,000 times. The calculated $p$-value is the proportion of test statistics that are as extreme, or more so, than our original KS-test statistic. We find that the values correspond well with a $p$-value derived simply using tables and formulae \citep{dallal-wilkinson1986}. We do not indicate $p$-values $<0.01$ or $>0.2$; such values in the catalog indicate a lower or upper limit respectively. Unlike for $\chi^2$, a high sample size is not fundamental to the derivation of the KS-test. Nevertheless, the KS-test still has sensitivity to sample size, and in SEDs with few data the residuals will have to deviate substantially from normality in order for a low $p$-value to be output. 

The $p$-values indicate the probability of getting a result as extreme as the one obtained, if the null hypothesis is true (that the residuals are normally distributed or that the data are described by the assumed model, for the KS and $\chi^2$ tests respectively). Note that the $p$-value only specifies the probability with which one would reject the null-hypothesis, if it \emph{were} correct -- it unfortunately provides no information on the probability that the null hypothesis \emph{is} correct, i.e.\ we have calculated $P( \ge D | H_{0})$ and not $P(H_{0}|D)$. For this frequentist approach, in cases where the $p$-value is lower than some threshold (for example, $p<0.02$), we take this to indicate a low probability of the two samples being as different as they are (or more so), if drawn from the same distribution \citep[e.g.][]{fisher1922}.

Since each test has its limitations and so as to avoid overreliance on a single measure, we combine the output of all tests to produce a data quality flag. The flag value is evaluated based on a subjective combination of the $\chi^2$ of the SED model, the $p$-value of the $\chi^2$, and the $p$-value of the KS-test. The flag has a value of 1, 2, or 3 and is provided as an indicative measure only; these correspond to `accept', `caution', and `poor', respectively.

\subsection{Fitting the Polarized SED}
\label{fittingpolarizedSED}

The extensive number of depolarization models, as discussed at length in Appendix~\ref{appendix-b}, can be primarily summarised by the `Burn' \citep{1966MNRAS.133...67B}, `Tribble' \citep{1991MNRAS.250..726T}, `Rossetti--Mantovani' \citep{2008A&A...487..865R,2009A&A...502...61M}, `Repolarizer' \citep{2002ApJ...568...99H,2009A&A...502...61M,2012AJ....144..105H}, and `Spectral Depolarizer' \citep{1974MNRAS.168..137C} forms. However, fitting these physical models to our data would arguably be misguided, as many of them assume an optically-thin emitting region. Furthermore, all of these models make either the critical assumption that we detect the same emitting region at each frequency, or that the polarized fraction is a meaningful quantity with the measured peak in polarized intensity on the sky emanating from the same emission region as the total intensity peak. As there is no way to verify this, and due to the large frequency range used, it is possible that any attempt at source classification will be corrupted. For example, a source that \emph{truly} depolarizes following \citet{1966MNRAS.133...67B} may appear as a peaked spectral depolarizer as different emission regions are probed at different frequencies, and at high frequencies the SED may be altered as the result of the combined effects of Faraday depolarization together with beam depolarization due to multiple components, and disordered magnetic fields \citep[e.g.][]{2013arXiv1309.2527M}. We therefore argue that there is no reason to believe that any attempt to derive $\sigma_{\text{RM}}$ (the RM dispersion of the Faraday screen within a single beam -- see Appendix~\ref{appendix-b}) would be, or ever is, a reasonable physical probe without high resolution observations. In addition, our main goal is to use the SED to derive the rest-frame source properties, allowing us some flexibility in our fitting procedures. Note that we are merely seeking a smoothly interpolating fit that is functionally similar to a physical model.

For the purposes of fitting models to the data, we therefore ignore the underlying physics involved, and choose three models that serve as mathematical edifices in order to fit to the data. For the quality and sampling of data typically found in our catalog, we find that these three simple, generalized functional forms can produce SEDs that broadly mimic the wavelength-dependence of polarization of the various physical models described in Appendix~\ref{appendix-b}. For our fits, we consider three possibilities: a Gaussian, a power law, and a Gaussian with a constant term, as shown in equations \ref{gaussian}, \ref{powerlaw}, and \ref{gauss+const} respectively. Given the limited number of data in each SED, these models serve as empirical analogues to the physical models in which we are ultimately interested, and are given by
\begin{equation}
\Pi = c_{1}\exp \left[ \frac{-(\lambda-c_{2})^2}{2c_{3}^2} \right] \,,
\label{gaussian}
\end{equation}
\begin{equation}
\Pi = 10^{c_{1}}\lambda^{c_{2}} \,,
\label{powerlaw}
\end{equation}
and
\begin{equation}
\Pi = c_{1}\exp \left[ \frac{-(\lambda-c_{2})^2}{2c_{3}^2} \right] + c_{4} \,,
\label{gauss+const}
\end{equation}
where $\lambda$ is in units of centimetres, and the $c_{i}$ are coefficients for which we solve during the fitting process. The units of $c_{i}$ are given in Table~\ref{tab:modelsunits}. With appropriately chosen coefficients, the Gaussian model of equation~\ref{gaussian} shows almost indistinguishable wavelength-dependence of fractional polarization to a `Burn' law \citep[see equation~\ref{burnbabyburn};][]{1966MNRAS.133...67B} or to a 'Spectral Depolarizer' \citep[see Appendix~\ref{spectraldepolrepol};][]{1974MNRAS.168..137C}. The power law model of equation~\ref{powerlaw} provides a fit for a `Tribble' law \citep[see equation~\ref{tribbling2};][]{1991MNRAS.250..726T}, and is flexible enough to fit a `repolarizer' \citep[see Appendix~\ref{spectraldepolrepol};][]{2002ApJ...568...99H,2009A&A...502...61M,2012AJ....144..105H}. Similarly to equation~\ref{gaussian}, with appropriately chosen coefficients the model for a Gaussian with a constant term of equation~\ref{gauss+const} behaves similarly to the `Rossetti--Mantovani' law \citep[see equation \ref{partialcoverage};][]{2008A&A...487..865R,2009A&A...502...61M}. These models are all discussed in detail in Appendix~\ref{appendix-b}. We emphasize that for two of the functional forms for which we fit to the data, namely equations~\ref{gaussian} and \ref{gauss+const}, the fit parameters $c_{1}$, $c_{2}$, $c_{3}$, and $c_{4}$ should not be trivially equated with the physical coefficients included in the equations in the Appendices. This is due to the fact that we use a dependence on $\lambda^2$, rather than on $\lambda^4$ as per the physical models of Appendix~\ref{appendix-b}. As both functions look similar, distinguishing between such models would require data of exceptional quality. We also again emphasize that our goal is to obtain the rest-frame source properties using a smoothly interpolating predictive SED that covers a broad range of wavelengths, and that can be used for the purpose of a $k$-correction. In addition, we also highlight that that there is no reason to believe that any attempt to derive $\sigma_{\text{RM}}$ would be, or ever is, a reasonable physical probe without high resolution observations. Should a user of the catalog wish to obtain a quantity other than a source classification or $k$-correction from the SED, they are able to refit the SED using the raw data available in the catalog.

In all cases, fitting was only performed on sources for which data were available at three or more wavelengths and when the minimum wavelength separation between the three measurements was greater than 5~cm. Due to the differing number of degrees of freedom for the various models, attempts to solve equations \ref{gaussian} and \ref{powerlaw} were made only when there were three or more data, and attempts to solve equation \ref{gauss+const} were made only when there were five or more data. Various constraints were placed on the fitting process to ensure that only values of $c_{i}$ corresponding to physical solutions could be obtained. This is particularly important when fitting equation~\ref{gaussian}, which has three degrees of freedom and can be fit to three data. Nevertheless, this is necessary in order to attempt to classify peaked sources, i.e.\ spectral depolarizers.

The addition of term $c_{2}$ to equations~\ref{gaussian} and \ref{gauss+const} allows an opportunity to identify spectral depolarizers, or other `peaked' sources. In cases where this offset was found to be low ($c_{2}\le7.5$~cm), the data were refit without the inclusion of an offset. Repolarizing sources can be identified via the flexibility of equation~\ref{powerlaw}, which provides a polarization spectral index, $\beta = c_{2} > 0$ in such cases, where $\beta$ is defined such that $\Pi \propto \lambda^{\beta}$. However, such sources should be treated with caution and may represent an undersampled peaked source -- particularly if the SED lacks multiple measurements at lower frequencies. Example fitted model SEDs from the catalog are shown in Fig.~\ref{SEDs}. The SEDs in the catalog show a number of different forms, most of which are well described by our chosen models, but with a few unusual and possibly oscillatory cases also present.

The model selected by the BIC is indicated in the catalog for each SED, alongside the coefficients for this model, and the corresponding statistical tests. However, the best fit to the SED with a polarization spectral index, $\beta$, is always returned in the catalog unless the source was found to be peaked -- in which case, only measurements in the tail region (i.e.\ those with a wavelength larger than the peak value) were used to calculate $\beta$. A best fit power law is calculated whenever there are $\ge2$~polarization measurements. For some sources, the power law does not provide an accurate representation of the full SED; the model fits provide more detailed information.

\subsection{Fitting the Total Intensity SED}
\begin{figure}[htb]
\centering
\includegraphics[clip=true, trim=0cm 0cm 0cm 0cm, width=8.5cm]{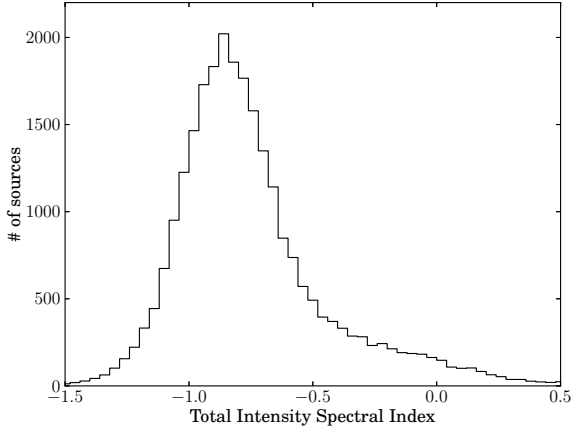}
\caption{A histogram of the number of sources in our sample with a given total intensity spectral index, $\alpha$. The sample is dominated by steep-spectrum sources.}
\label{I_histo}
\end{figure}

We choose two mathematical models to fit to our total intensity data. These are the power law, and the curved power law respectively, as given by
\begin{equation}
I = d_{1}\nu ^{d_{2}} \,,
\label{powerlawI}
\end{equation}
and
\begin{equation}
I = d_{1}\left(\frac{\nu}{\nu_{\textrm{ref}}}\right)^{\left[d_{2}+d_{3}\ln\left(\frac{\nu}{\nu_{\textrm{ref}}}\right)\right]} \,,
\label{powerlawIcurved}
\end{equation}
where $\nu_{\textrm{ref}}$ is a reference frequency that we define as 1.4~GHz, $I$ is the total intensity flux density in units of millijansky, $\nu$ is the observing frequency in hertz, and the $d_{i}$ are some coefficients for which we solve. The units of $d_{i}$ are given in Table~\ref{tab:modelsunits}. In all cases, we only attempt to estimate curvature in spectra with three or more data at different wavelengths and when the minimum distance between the three measurements is greater than 5~cm of wavelength. For sources with just two data points, we calculate a two-point spectral index based on equation \ref{powerlawI} and make no attempt to estimate the presence of curvature. Example total intensity SEDs from the catalog are shown in Fig.~\ref{SEDs-I}. A histogram of the number of sources with a given spectral index is shown in Fig.~\ref{I_histo}. The histogram shows a peaked distribution with a maximum at $\alpha\approx-0.8$. There is an extended tail in the $\alpha$-distribution towards flatter-spectral indices, caused by the flat-spectrum population. Nevertheless, our sample is clearly dominated by steep-spectrum objects.

Calculating the spectral index also allows us to estimate the non-thermal rest-frame luminosity of the source at 1.4~GHz. We do this using $L_{\nu}=4 \pi d_{L}(z)^2 S (1+z)^{-(\alpha+1)}$ \citep{1992ARA&A..30..575C,2009MNRAS.397.1101G}, where $d_{L}(z)$ is the luminosity distance. The calculation assumes that the dominant emission mechanism is the synchrotron process, and that this process follows a power law such that $S_{\nu}\propto\nu^{+\alpha}$. We also assume a $\Lambda$CDM model and a flat cosmology, using the Planck cosmological parameters such that $\Omega_\Lambda=0.685$ and $H_0=67.3$~km~s$^{-1}$~Mpc$^{-1}$ \citep{2013arXiv1303.5076P}. The FR-I/FR-II luminosity divide is typically taken to be $L_{1.4} \approx 10^{24.5}$~W~Hz$^{-1}$.

\subsection{Fitting the Rotation Measure}
\begin{figure*}[t]
\centering
\includegraphics[clip=true, trim=0cm 0cm 1.5cm 0.5cm, width=8.9cm]{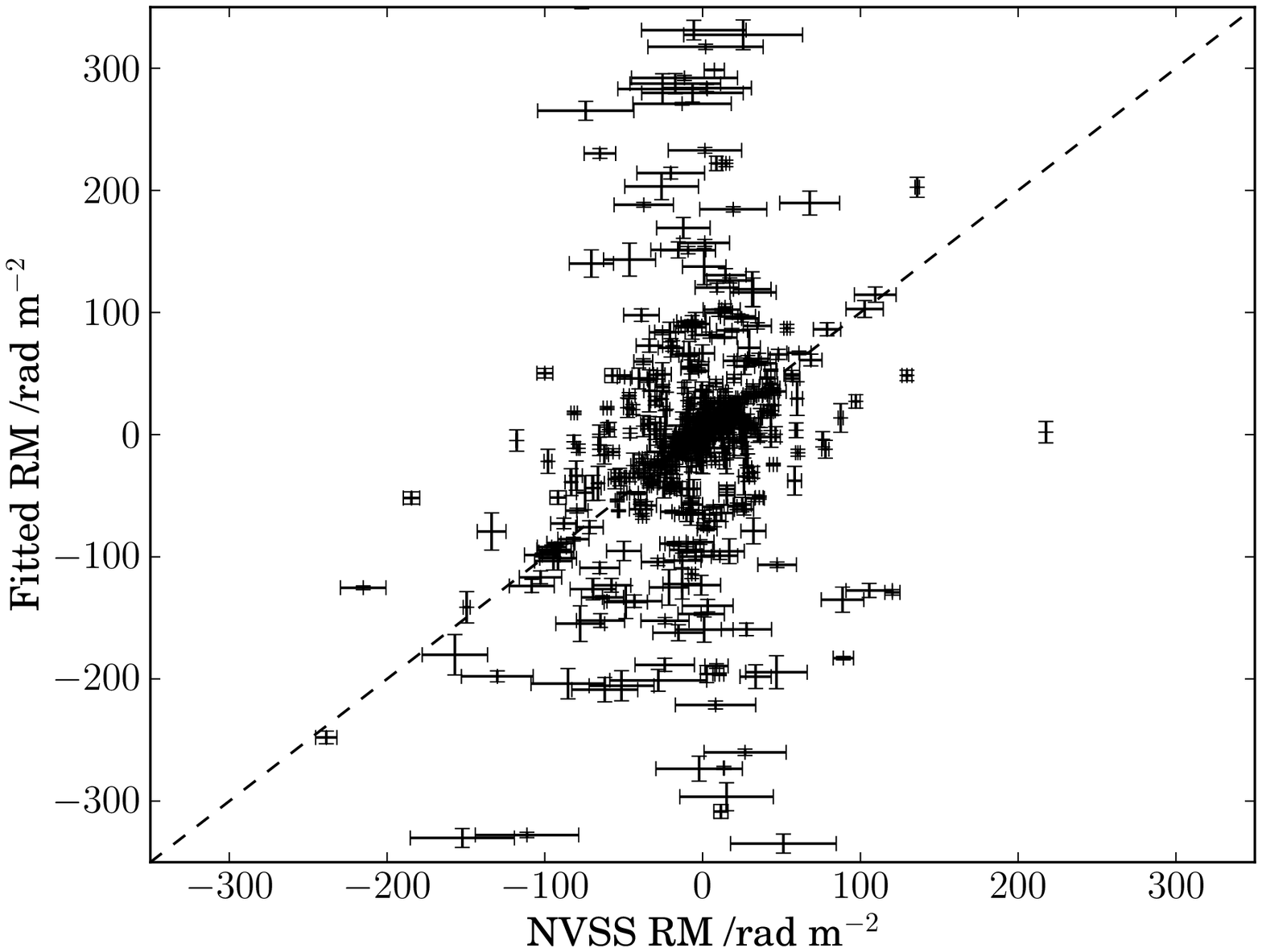}
\includegraphics[clip=true, trim=0cm 0cm 1.5cm 0.5cm, width=8.9cm]{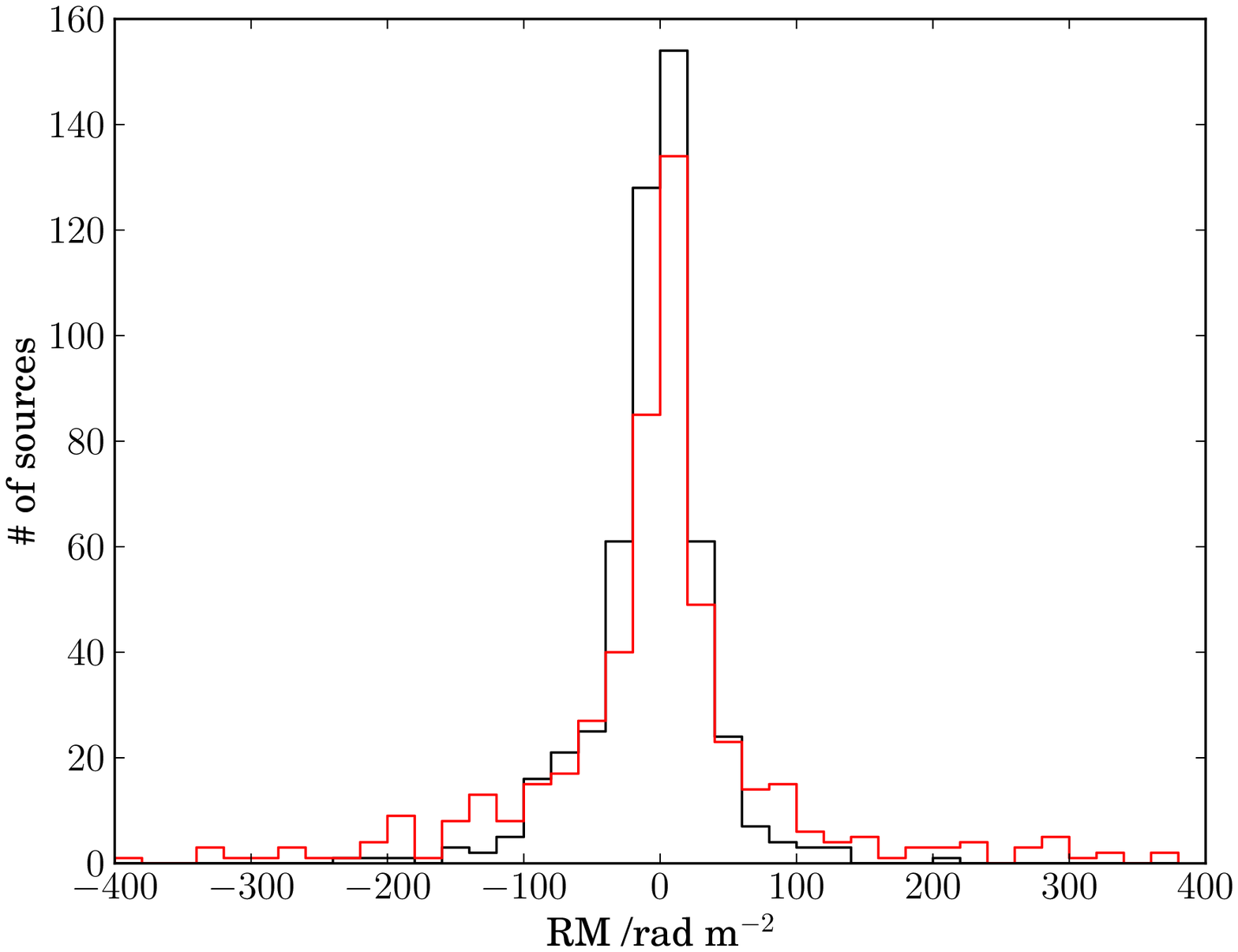}
\caption{Left: The fitted rotation measures as compared to the NVSS rotation measures of \citet{2009ApJ...702.1230T} for sources above a Galactic latitude of $|b|>20^{\circ}$. Note that sources with an |RM|$\ge$350~rad~m$^{-2}$ are outside of the plotted range. The dashed line shows a ratio of unity. Right: A histogram showing the distribution of fitted rotation measures (red line) as compared to the NVSS rotation measures of \citet{2009ApJ...702.1230T} (black line) for all fitted sources above a Galactic latitude of $|b|>20^{\circ}$.}
\label{RMmatch}
\end{figure*}
Under some circumstances, it can be advantageous to calculate the peak Faraday depth using the technique of RM Synthesis \citep[see][for extensive detail on the technique]{2005A&A...441.1217B}. This technique has three main advantages in comparison to fitting a straight line to a plot of the EVPA versus $\lambda^2$: (i) no `$n \pi$' ambiguities, (ii) recognition of several Faraday depth components in Faraday-space within the source (which can also cause the depolarization to be non-monotonic), and (iii) recognition of internal Faraday rotation, which can lead to deviations of the EVPA from a simple $\lambda^2$ law (see equation \ref{rotationmeasureequation}) and can cause broad structures in Faraday-space \citep[e.g.][]{2011MNRAS.414.2540F,2012A&A...543A.113B}.

However, as our catalog covers such a broad range in $\lambda$, and as there are large gaps in our $\lambda^2$-coverage, it is problematic to use RM Synthesis to calculate the RM \citep[e.g.][]{2005A&A...441.1217B,2011AJ....141..191F,2012MNRAS.421.3300O}. Instead, we fit a straight line to a plot of the EVPA versus $\lambda^2$, where the EVPA is defined between $0^\circ$ and $+180^\circ$ from North through East. Such a method assumes that the RM is constant as a function of $\lambda$. While deviations from a $\lambda^2$-law (see equation~\ref{rotationmeasureequation}) are beyond the scope of this paper, such effects can in principle be explored using the raw data available in the catalog. The fitting process is complicated by the `$n\pi$ ambiguity', with each datum having undergone wrapping by some integer multiple of $180^\circ$. We therefore need to maintain goodness of fit while carefully unwrapping these ambiguities. Various approaches to this have been attempted previously \citep[e.g.][]{1981ApJS...45...97S,2005MNRAS.360.1305R,2008MNRAS.386.1881N,2008A&A...487..865R}.

We undertake a `brute force' approach, and fit to the data for every possible combination of positive and negative wrapping per datum, with any number of wraps between zero and five being allowed. This is computationally expensive, particularly for sources with large numbers of measurements. For example, a source with $n$ data across the squared-wavelength coverage has $6^n$ wrap combinations, each of which (with the exception of zero wrap, which has no sign ambiguity) has a further $2^n$ sign combinations -- providing a total number of $11^n$ fitting attempts for a single source. However, only a fraction of these possibilities are physical; for a source that demonstrates a linear relationship between EVPA and $\lambda^2$ the sign of a wrap at higher frequency can not change towards low frequencies. Furthermore, if there is a wrap at high frequency then all lower frequency measurements must also be unwrapped by at least an equal integer amount. This property was exploited to reduce the number of brute force attempts; for a source with seven data this lowers the number of trial fits from 35,831,808 to just 10,836.

Such a technique is biased towards providing high RMs, as with a sufficient number of wraps the goodness of fit can be improved to arbitrary levels. To counteract this, we utilize a modified version of the BIC discussed in Section~\ref{sourcefitting}, taking $k$ to be the square of the maximum number of wraps. Such a method penalizes high numbers of wraps substantially, unless they provide a substantial increase in fitting quality.

We only allow for a maximum of five possible wraps in our data, allowing for a maximum |RM|$\approx$370~rad~m$^{-2}$ (depending on the frequency coverage). Such high RMs are expected to be rare on the sky, and exist mostly near the Galactic plane \citep[e.g.][]{2009ApJ...702.1230T}. Furthermore, we use the NVSS RMs as our reference catalog -- which suffers from significant bandwidth depolarization for sources with |RM|$\gtrsim340$~rad~m$^{-2}$ \citep{2009ApJ...702.1230T}. In all cases, we only fit RMs to sources with three or more data at different wavelengths and when the minimum wavelength separation between the three measurements is greater than 5~cm. For each fit, we also attempt to calculate the intrinsic EVPA, i.e.\ the EVPA at infinite frequency. Due to the significant number of wraps required at low observing frequencies, all measurements at wavelengths greater than $35$~cm were excluded from the fitting.

All sources have EVPA measurements at 1.4~GHz from the NVSS catalog (see Section~\ref{crossmatching}), with the majority of other measurements being at higher frequencies. Due to the distribution of these measurements in frequency-space, our broadband RM fits are biased towards providing the Faraday structure at high frequency. It is difficult to define the exact frequency at which RMs are measured, and we take them to be a high-frequency RM at $\approx5$~GHz. These broadband RMs are compared against the NVSS RMs in Fig.~\ref{RMmatch}. The distributions of both the broadband and NVSS RMs are also shown for sources that have both values. While there are clearly outliers, many sources are also in reasonable agreement and cluster about the line of unity. The outliers appear associated with low NVSS RMs, which have a high corresponding broadband RM. Similar effects have been seen in other similar samples \citep{2009ApJ...702.1230T,2012ApJ...761..144B}. The differences could occur due to errors in the unwrapping of our data, uncertainty in the calculation of the RM in the NVSS data, time-variability of the RM towards a radio source, or `Faraday complexity' (i.e.\ a non-linear relationship between the EVPA and $\lambda^2$).

\section{The Catalog}
\label{catalog}
Data for all sources and the associated fitted and calculated properties are provided as machine-readable tables. These tables are published in their entirety in the electronic edition of The Astrophysical Journal Supplement Series. A portion is shown here for guidance regarding its form and content. There are two versions of the catalog available, the `Full-catalog' and the `SED-catalog' respectively. The Full-catalog contains 37,543 rows, providing an expanded version of the \citet{2009ApJ...702.1230T} data and contains all of the derived total intensity and polarized fraction spectral indices and other incorporated cross-matched and derived quantities. Note that the catalog contains no upper limits on polarized or total intensity measurements. The SED-catalog contains only the 951 sources for which a model has been fitted to the polarized SED. The column headings are identical in both catalog versions, and the SED-catalog is just a subset of rows from our primary data product, the Full-catalog. The column headings of both catalogs are described in Appendix~\ref{appendix-a}, including calculated quantities and ancilliary data provided during the cross-matching process. We emphasize the various statistical tests that are provided within the catalog, which are included so that a user may assess the quality of a particular model fit. Selected columns for the first 45 sources in the SED-catalog are presented in Tables~\ref{tab:thedata}~to~\ref{tab:thedata5}. All errors are the $1\sigma$ uncertainties and have been calculated using standard error propagation.
\begin{turnpage}
\begin{deluxetable*}{c c c c c c c c c c c}
\tablefontsize{\tiny}
\setlength{\tabcolsep}{0.015in} 
\tablewidth{0pt}
\tablecaption{Selected Columns from the First 45 Entries in the SED-catalog. Units are shown where appropriate. The corresponding column number in Appendix~\ref{appendix-a} (and the catalog itself) is shown in brackets in the Table header. This table is available in its entirety in the online journal.}
\tablehead{ \colhead{\#\tablenotemark{a}} & \colhead{Source\_Name} & \colhead{RA\tablenotemark{b}} & \colhead{Dec.} & \colhead{G$_{\text{lon}}$} & \colhead{G$_{\text{lat}}$} & \colhead{\#$_{\Pi}$\tablenotemark{c}} & \colhead{\#$_{I}$\tablenotemark{d}} & \colhead{$\lambda$} & \colhead{$\Pi$} & \colhead{$\Delta\Pi$} \\ 
\colhead{} & \colhead{} & \colhead{(J2000)} & \colhead{(J2000)} & \colhead{} & \colhead{} & \colhead{} & \colhead{} & \colhead{} & \colhead{} & \colhead{} \\
\colhead{} & \colhead{} & \colhead{} & \colhead{} & \colhead{/$^\circ$} & \colhead{/$^\circ$} & \colhead{} & \colhead{} & \colhead{ /cm } & \colhead{ /\% } & \colhead{ /\% } \\
\colhead{(1)} & \colhead{(2)} & \colhead{(3)} & \colhead{(4)} & \colhead{(7)} & \colhead{(8)} & \colhead{(9)} & \colhead{(10)} & \colhead{(11--79)} & \colhead{(80--148)} & \colhead{(149--217)}}
\startdata
104  &  000322-172711  &  00h03m22.00s  &  -17d27m11.40  &  71.5253  &  -75.2777  &  4  &  3  &  [ 21.41,  ---,  ---, 1.51, $\hdots$]  &  [ 1.47,  ---,  ---, 3.70, $\hdots$]  &  [ 0.11,  ---,  ---, 0.37, $\hdots$] \\
157  &  000457+124818  &  00h04m57.12s  &  +12d48m18.90  &  105.686  &  -48.5038  &  5  &  2  &  [ 21.41,  ---,  ---,  ---, $\hdots$]  &  [ 19.89,  ---,  ---,  ---, $\hdots$]  &  [ 0.84,  ---,  ---,  ---, $\hdots$] \\
186  &  000559+160946  &  00h05m59.41s  &  +16d09m46.70  &  107.3192  &  -45.3266  &  7  &  4  &  [ 21.41,  ---,  ---,  ---, $\hdots$]  &  [ 2.10,  ---,  ---,  ---, $\hdots$]  &  [ 0.13,  ---,  ---,  ---, $\hdots$] \\
195  &  000613-062335  &  00h06m13.87s  &  -06d23m35.20  &  93.5053  &  -66.6465  &  9  &  10  &  [ 21.41,  ---,  ---, 1.51, $\hdots$]  &  [ 3.19,  ---,  ---, 4.60, $\hdots$]  &  [ 0.14,  ---,  ---, 0.46, $\hdots$] \\
200  &  000622-000425  &  00h06m22.60s  &  -00d04m25.10  &  99.2806  &  -60.8593  &  14  &  3  &  [ 21.41,  ---,  ---, 1.51, $\hdots$]  &  [ 0.71,  ---,  ---, 2.50, $\hdots$]  &  [ 0.10,  ---,  ---, 0.25, $\hdots$] \\
246  &  000826-255912  &  00h08m26.21s  &  -25d59m12.30  &  37.2729  &  -80.3196  &  4  &  5  &  [ 21.41, 6.25, 3.47, 1.51, $\hdots$]  &  [ 1.47, 2.90, 4.20, 6.0, $\hdots$]  &  [ 0.14, 0.29, 0.42, 0.60, $\hdots$] \\
274  &  000935-321636  &  00h09m35.76s  &  -32d16m36.90  &  0.8704  &  -79.5645  &  3  &  5  &  [ 21.41, 6.25, 3.47,  ---, $\hdots$]  &  [ 4.36, 4.40, 3.10,  ---, $\hdots$]  &  [ 0.23, 0.44, 0.31,  ---, $\hdots$] \\
366  &  001259-395425  &  00h12m59.95s  &  -39d54m25.70  &  -27.5425  &  -74.9393  &  4  &  4  &  [ 21.41, 6.25, 3.47, 1.51, $\hdots$]  &  [ 2.74, 2.20, 3.30, 4.10, $\hdots$]  &  [ 0.16, 0.22, 0.33, 0.41, $\hdots$] \\
383  &  001331+405136  &  00h13m31.09s  &  +40d51m36.00  &  115.2409  &  -21.4448  &  5  &  7  &  [ 21.41,  ---,  ---,  ---, $\hdots$]  &  [ 2.31,  ---,  ---,  ---, $\hdots$]  &  [ 0.14,  ---,  ---,  ---, $\hdots$] \\
391  &  001354+204522  &  00h13m54.84s  &  +20d45m22.90  &  111.2333  &  -41.2615  &  3  &  3  &  [ 21.41,  ---,  ---,  ---, $\hdots$]  &  [ 5.41,  ---,  ---,  ---, $\hdots$]  &  [ 0.21,  ---,  ---,  ---, $\hdots$] \\
449  &  001559+390027  &  00h15m59.98s  &  +39d00m27.20  &  115.442  &  -23.3483  &  4  &  7  &  [ 21.41,  ---,  ---,  ---, $\hdots$]  &  [ 4.59,  ---,  ---,  ---, $\hdots$]  &  [ 0.21,  ---,  ---,  ---, $\hdots$] \\
460  &  001631+791651  &  00h16m31.61s  &  +79d16m51.70  &  121.245  &  16.5238  &  9  &  2  &  [ 21.41,  ---,  ---,  ---, $\hdots$]  &  [ 6.51,  ---,  ---,  ---, $\hdots$]  &  [ 0.25,  ---,  ---,  ---, $\hdots$] \\
519  &  001851-124233  &  00h18m51.38s  &  -12d42m33.50  &  93.4551  &  -73.6872  &  5  &  3  &  [ 21.41,  ---,  ---, 1.51, $\hdots$]  &  [ 1.17,  ---,  ---, 5.40, $\hdots$]  &  [ 0.11,  ---,  ---, 0.54, $\hdots$] \\
571  &  002025+154052  &  00h20m25.32s  &  +15d40m52.70  &  112.0476  &  -46.5339  &  16  &  3  &  [ 21.41,  ---,  ---,  ---, $\hdots$]  &  [ 5.72,  ---,  ---,  ---, $\hdots$]  &  [ 0.20,  ---,  ---,  ---, $\hdots$] \\
696  &  002430-292848  &  00h24m30.12s  &  -29d28m48.90  &  9.7191  &  -83.6231  &  6  &  3  &  [ 21.41,  ---,  ---,  ---, $\hdots$]  &  [ 2.03,  ---,  ---,  ---, $\hdots$]  &  [ 0.12,  ---,  ---,  ---, $\hdots$] \\
725  &  002526+391935  &  00h25m26.15s  &  +39d19m35.70  &  117.4617  &  -23.2689  &  4  &  7  &  [ 21.41,  ---,  ---,  ---, $\hdots$]  &  [ 6.02,  ---,  ---,  ---, $\hdots$]  &  [ 0.21,  ---,  ---,  ---, $\hdots$] \\
730  &  002549-260212  &  00h25m49.17s  &  -26d02m12.80  &  42.2744  &  -84.1701  &  12  &  3  &  [ 21.41,  ---,  ---,  ---, $\hdots$]  &  [ 0.79,  ---,  ---,  ---, $\hdots$]  &  [ 0.10,  ---,  ---,  ---, $\hdots$] \\
813  &  002831-002540  &  00h28m31.35s  &  -00d25m40.80  &  110.3451  &  -62.7384  &  3  &  1  &  [ 21.41,  ---,  ---,  ---, $\hdots$]  &  [ 8.06,  ---,  ---,  ---, $\hdots$]  &  [ 0.38,  ---,  ---,  ---, $\hdots$] \\
896  &  003133-010121  &  00h31m33.63s  &  -01d01m21.10  &  111.7545  &  -63.4621  &  3  &  1  &  [ 21.41,  ---,  ---,  ---, $\hdots$]  &  [ 14.19,  ---,  ---,  ---, $\hdots$]  &  [ 0.49,  ---,  ---,  ---, $\hdots$] \\
924  &  003235+394215  &  00h32m35.51s  &  +39d42m15.50  &  118.9947  &  -23.0234  &  4  &  7  &  [ 21.41,  ---,  ---,  ---, $\hdots$]  &  [ 1.78,  ---,  ---,  ---, $\hdots$]  &  [ 0.14,  ---,  ---,  ---, $\hdots$] \\
998  &  003508-200359  &  00h35m08.80s  &  -20d03m59.20  &  94.233  &  -82.0144  &  3  &  5  &  [ 21.41,  ---,  ---,  ---, $\hdots$]  &  [ 0.63,  ---,  ---,  ---, $\hdots$]  &  [ 0.10,  ---,  ---,  ---, $\hdots$] \\
1030  &  003606+183759  &  00h36m06.55s  &  +18d37m59.00  &  117.8735  &  -44.0881  &  7  &  3  &  [ 21.41,  ---,  ---,  ---, $\hdots$]  &  [ 1.94,  ---,  ---,  ---, $\hdots$]  &  [ 0.12,  ---,  ---,  ---, $\hdots$] \\
1031  &  003607+585548  &  00h36m07.89s  &  +58d55m48.90  &  120.9536  &  -3.8827  &  4  &  4  &  [ 21.41,  ---,  ---,  ---, $\hdots$]  &  [ 1.94,  ---,  ---,  ---, $\hdots$]  &  [ 0.12,  ---,  ---,  ---, $\hdots$] \\
1062  &  003704-010909  &  00h37m04.06s  &  -01d09m09.40  &  114.7753  &  -63.7966  &  7  &  3  &  [ 21.41,  ---,  ---, 1.51, $\hdots$]  &  [ 4.68,  ---,  ---, 3.70, $\hdots$]  &  [ 0.19,  ---,  ---, 0.37, $\hdots$] \\
1098  &  003758+240711  &  00h37m58.31s  &  +24d07m11.60  &  118.9968  &  -38.649  &  5  &  5  &  [ 21.41,  ---,  ---,  ---, $\hdots$]  &  [ 2.40,  ---,  ---,  ---, $\hdots$]  &  [ 0.13,  ---,  ---,  ---, $\hdots$] \\
1115  &  003820-020740  &  00h38m20.41s  &  -02d07m40.40  &  115.2268  &  -64.8035  &  14  &  4  &  [ 21.41,  ---,  ---,  ---, $\hdots$]  &  [ 0.61,  ---,  ---,  ---, $\hdots$]  &  [ 0.10,  ---,  ---,  ---, $\hdots$] \\
1144  &  003916-185405  &  00h39m16.95s  &  -18d54m05.30  &  103.5398  &  -81.312  &  3  &  4  &  [ 21.41, 6.25, 3.47,  ---, $\hdots$]  &  [ 2.57, 7.10, 6.40,  ---, $\hdots$]  &  [ 0.19, 0.71, 0.64,  ---, $\hdots$] \\
1190  &  004057-014631  &  00h40m57.62s  &  -01d46m31.60  &  116.8359  &  -64.5231  &  6  &  2  &  [ 21.41,  ---,  ---, 1.51, $\hdots$]  &  [ 3.48,  ---,  ---, 2.70, $\hdots$]  &  [ 0.15,  ---,  ---, 0.27, $\hdots$] \\
1268  &  004308+520334  &  00h43m08.79s  &  +52d03m34.40  &  121.6347  &  -10.7913  &  13  &  5  &  [ 21.41,  ---,  ---,  ---, $\hdots$]  &  [ 2.07,  ---,  ---,  ---, $\hdots$]  &  [ 0.12,  ---,  ---,  ---, $\hdots$] \\
1282  &  004349+383010  &  00h43m49.42s  &  +38d30m10.10  &  121.2969  &  -24.3448  &  4  &  7  &  [ 21.41,  ---,  ---,  ---, $\hdots$]  &  [ 4.23,  ---,  ---,  ---, $\hdots$]  &  [ 0.21,  ---,  ---,  ---, $\hdots$] \\
1289  &  004358+393756  &  00h43m58.38s  &  +39d37m56.60  &  121.3681  &  -23.2167  &  4  &  7  &  [ 21.41,  ---,  ---,  ---, $\hdots$]  &  [ 8.79,  ---,  ---,  ---, $\hdots$]  &  [ 0.35,  ---,  ---,  ---, $\hdots$] \\
1315  &  004441-353032  &  00h44m41.47s  &  -35d30m32.80  &  -47.7431  &  -81.4966  &  8  &  5  &  [ 21.41, 6.25, 3.47, 1.51, $\hdots$]  &  [ 1.46, 6.30, 7.80, 7.70, $\hdots$]  &  [ 0.11, 0.63, 0.78, 0.77, $\hdots$] \\
1421  &  004843+404458  &  00h48m43.83s  &  +40d44m58.50  &  122.3784  &  -22.1193  &  4  &  7  &  [ 21.41,  ---,  ---,  ---, $\hdots$]  &  [ 13.83,  ---,  ---,  ---, $\hdots$]  &  [ 0.52,  ---,  ---,  ---, $\hdots$] \\
1476  &  005041-092906  &  00h50m41.27s  &  -09d29m06.20  &  122.3217  &  -72.356  &  9  &  3  &  [ 21.41,  ---,  ---, 1.51, $\hdots$]  &  [ 3.16,  ---,  ---, 1.80, $\hdots$]  &  [ 0.14,  ---,  ---, 0.18, $\hdots$] \\
1561  &  005408-033355  &  00h54m08.40s  &  -03d33m55.90  &  124.6181  &  -66.4285  &  4  &  4  &  [ 21.41,  ---,  ---,  ---, $\hdots$]  &  [ 0.52,  ---,  ---,  ---, $\hdots$]  &  [ 0.10,  ---,  ---,  ---, $\hdots$] \\
1566  &  005427+404211  &  00h54m27.31s  &  +40d42m11.80  &  123.5494  &  -22.1649  &  4  &  7  &  [ 21.41,  ---,  ---,  ---, $\hdots$]  &  [ 1.80,  ---,  ---,  ---, $\hdots$]  &  [ 0.12,  ---,  ---,  ---, $\hdots$] \\
1606  &  005550+262436  &  00h55m50.38s  &  +26d24m36.00  &  124.1572  &  -36.4513  &  6  &  3  &  [ 21.41,  ---,  ---,  ---, $\hdots$]  &  [ 0.82,  ---,  ---,  ---, $\hdots$]  &  [ 0.11,  ---,  ---,  ---, $\hdots$] \\
1664  &  005802-323418  &  00h58m02.26s  &  -32d34m18.30  &  -71.3925  &  -84.3718  &  4  &  4  &  [ 21.41, 6.25, 3.47, 1.51, $\hdots$]  &  [ 4.12, 8.90, 10.10, 11.40, $\hdots$]  &  [ 0.25, 0.89, 1.01, 1.14, $\hdots$] \\
1698  &  005905+000651  &  00h59m05.53s  &  +00d06m51.50  &  127.1063  &  -62.6954  &  14  &  5  &  [ 21.41,  ---,  ---,  ---, $\hdots$]  &  [ 3.91,  ---,  ---,  ---, $\hdots$]  &  [ 0.15,  ---,  ---,  ---, $\hdots$] \\
1738  &  010005+394935  &  01h00m05.28s  &  +39d49m35.70  &  124.7362  &  -23.0149  &  4  &  6  &  [ 21.41,  ---,  ---,  ---, $\hdots$]  &  [ 16.67,  ---,  ---,  ---, $\hdots$]  &  [ 0.65,  ---,  ---,  ---, $\hdots$] \\
1817  &  010246+462438  &  01h02m46.64s  &  +46d24m38.10  &  124.9692  &  -16.4163  &  4  &  6  &  [ 21.41,  ---,  ---,  ---, $\hdots$]  &  [ 4.56,  ---,  ---,  ---, $\hdots$]  &  [ 0.18,  ---,  ---,  ---, $\hdots$] \\
1820  &  010250+255216  &  01h02m50.22s  &  +25d52m16.60  &  126.14  &  -36.9295  &  4  &  4  &  [ 21.41,  ---,  ---,  ---, $\hdots$]  &  [ 0.72,  ---,  ---,  ---, $\hdots$]  &  [ 0.11,  ---,  ---,  ---, $\hdots$] \\
1852  &  010422-123515  &  01h04m22.49s  &  -12d35m15.60  &  135.3358  &  -75.147  &  5  &  3  &  [ 21.41,  ---,  ---,  ---, $\hdots$]  &  [ 5.88,  ---,  ---,  ---, $\hdots$]  &  [ 0.23,  ---,  ---,  ---, $\hdots$] \\
1870  &  010509-381057  &  01h05m09.33s  &  -38d10m57.20  &  -70.8025  &  -78.5774  &  3  &  1  &  [ 21.41,  ---,  ---,  ---, $\hdots$]  &  [ 8.13,  ---,  ---,  ---, $\hdots$]  &  [ 0.29,  ---,  ---,  ---, $\hdots$] \\
1900  &  010622-015538  &  01h06m22.98s  &  -01d55m38.30  &  131.6477  &  -64.5459  &  4  &  3  &  [ 21.41,  ---,  ---, 1.51, $\hdots$]  &  [ 7.62,  ---,  ---, 4.30, $\hdots$]  &  [ 0.25,  ---,  ---, 0.43, $\hdots$] 
\enddata
\tablenotetext{a}{\small{Corresponding row number in \citet{2009ApJ...702.1230T}.}}
\tablenotetext{b}{\small{Decimal coordinates are also provided within the machine-readable catalog.}}
\tablenotetext{c}{\small{Number of polarized fraction measurements.}}
\tablenotetext{d}{\small{Number of total intensity measurements.}}
\label{tab:thedata}
\end{deluxetable*}
\end{turnpage}

\begin{turnpage}
\begin{deluxetable*}{c c c c c c c c}
\tablefontsize{\tiny}
\setlength{\tabcolsep}{0.015in} 
\tablewidth{0pt}
\tablecaption{Selected Columns from the First 45 Entries in the SED-catalog (continued from Table~\ref{tab:thedata})}
\tablehead{ \colhead{\#} & \colhead{Pol\_Angle} & \colhead{Pol\_Angle\_err} & \colhead{I\_Lambda} & \colhead{I\_Flux} & \colhead{I\_Flux\_err} & \colhead{Depol\_} & \colhead{Depol\_} \\
\colhead{} & \colhead{} & \colhead{} & \colhead{} & \colhead{} & \colhead{} & \colhead{Type\tablenotemark{e}} & \colhead{Physics\tablenotemark{f}}\\
\colhead{} & \colhead{/$^\circ$} & \colhead{/$^\circ$} & \colhead{/cm} & \colhead{/mJy} & \colhead{/mJy} & \colhead{} & \colhead{}\\
\colhead{(1)} & \colhead{(218--286)} & \colhead{(287--355)} & \colhead{(356--376)} & \colhead{(377--397)} & \colhead{(398--418)} & \colhead{(419)} & \colhead{(420)} }
\startdata
104  &  [ 126.6,  ---,  ---, 18.0, $\hdots$]  &  [ 0.20,  ---,  ---, 2.57, $\hdots$]  &  [ 21.41,  ---,  ---, 1.51, $\hdots$]  &  [ 2414.80,  ---,  ---, 386.0, $\hdots$]  &  [ 70.0,  ---,  ---, 40.0, $\hdots$]  &  power  &  tribble   \\
157  &  [ 158.3,  ---,  ---,  ---, $\hdots$]  &  [ 0.10,  ---,  ---,  ---, $\hdots$]  &  [ 21.41,  ---,  ---,  ---, $\hdots$]  &  [ 1069.50,  ---,  ---,  ---, $\hdots$]  &  [ 40.0,  ---,  ---,  ---, $\hdots$]  &  power  &  increase  \\
186  &  [ 131.1,  ---,  ---,  ---, $\hdots$]  &  [ 0.60,  ---,  ---,  ---, $\hdots$]  &  [ 21.41,  ---,  ---,  ---, $\hdots$]  &  [ 804.40,  ---,  ---,  ---, $\hdots$]  &  [ 27.0,  ---,  ---,  ---, $\hdots$]  &  gauss  &  burn      \\
195  &  [ 36.4,  ---,  ---, 32.0, $\hdots$]  &  [ 0.20,  ---,  ---, 0.99, $\hdots$]  &  [ 21.41,  ---,  ---, 1.51, $\hdots$]  &  [ 2050.70,  ---,  ---, 2084.0, $\hdots$]  &  [ 60.0,  ---,  ---, 100.0, $\hdots$]  &  power  &  increase  \\
200  &  [ 157.9,  ---,  ---, 127.0, $\hdots$]  &  [ 0.30,  ---,  ---, 14.45, $\hdots$]  &  [ 21.41,  ---,  ---, 1.51, $\hdots$]  &  [ 3897.60,  ---,  ---, 425.0, $\hdots$]  &  [ 120.0,  ---,  ---, 17.0, $\hdots$]  &  gau+t  &  mantovani \\
246  &  [ 107.6, 50.0, 138.0, 154.0, $\hdots$]  &  [ 1.90, 10.0, 11.45, 7.43, $\hdots$]  &  [ 21.41, 6.25, 3.47, 1.51, $\hdots$]  &  [ 468.20, 360.0, 259.0, 120.0, $\hdots$]  &  [ 14.0, 18.0, 14.0, 8.0, $\hdots$]  &  power  &  tribble   \\
274  &  [ 95.0, 64.0, 62.0,  ---, $\hdots$]  &  [ 0.50, 4.57, 6.89,  ---, $\hdots$]  &  [ 21.41, 6.25, 3.47, 1.51, $\hdots$]  &  [ 388.30, 313.0, 275.0, 221.0, $\hdots$]  &  [ 12.0, 16.0, 14.0, 10.0, $\hdots$]  &  gauss  &  peaked    \\
366  &  [ 113.0, 89.0, 100.0, 91.0, $\hdots$]  &  [ 0.70, 1.98, 2.42, 2.70, $\hdots$]  &  [ 21.41, 6.25, 3.47, 1.51, $\hdots$]  &  [ 493.60, 2015.0, 2009.0, 1609.0, $\hdots$]  &  [ 17.0, 100.0, 100.0, 70.0, $\hdots$]  &  power  &  tribble   \\
383  &  [ 86.3,  ---,  ---,  ---, $\hdots$]  &  [ 0.30,  ---,  ---,  ---, $\hdots$]  &  [ 21.41,  ---,  ---,  ---, $\hdots$]  &  [ 1650.90,  ---,  ---,  ---, $\hdots$]  &  [ 60.0,  ---,  ---,  ---, $\hdots$]  &  gauss  &  peaked    \\
391  &  [ 73.0,  ---,  ---,  ---, $\hdots$]  &  [ 0.30,  ---,  ---,  ---, $\hdots$]  &  [ 21.41,  ---,  ---,  ---, $\hdots$]  &  [ 624.80,  ---,  ---,  ---, $\hdots$]  &  [ 21.0,  ---,  ---,  ---, $\hdots$]  &  gauss  &  peaked    \\
449  &  [ 168.3,  ---,  ---,  ---, $\hdots$]  &  [ 0.90,  ---,  ---,  ---, $\hdots$]  &  [ 21.41,  ---,  ---,  ---, $\hdots$]  &  [ 263.60,  ---,  ---,  ---, $\hdots$]  &  [ 8.0,  ---,  ---,  ---, $\hdots$]  &  power  &  tribble   \\
460  &  [ 62.2,  ---,  ---,  ---, $\hdots$]  &  [ 0.40,  ---,  ---,  ---, $\hdots$]  &  [ 21.41,  ---,  ---,  ---, $\hdots$]  &  [ 3650.30,  ---,  ---,  ---, $\hdots$]  &  [ 120.0,  ---,  ---,  ---, $\hdots$]  &  power  &  tribble   \\
519  &  [ 155.2,  ---,  ---, 83.0, $\hdots$]  &  [ 0.40,  ---,  ---, 20.75, $\hdots$]  &  [ 21.41,  ---,  ---, 1.51, $\hdots$]  &  [ 2122.30,  ---,  ---, 141.0, $\hdots$]  &  [ 60.0,  ---,  ---, 7.0, $\hdots$]  &  power  &  tribble   \\
571  &  [ 161.1,  ---,  ---,  ---, $\hdots$]  &  [ 0.10,  ---,  ---,  ---, $\hdots$]  &  [ 21.41,  ---,  ---,  ---, $\hdots$]  &  [ 2180.0,  ---,  ---,  ---, $\hdots$]  &  [ 70.0,  ---,  ---,  ---, $\hdots$]  &  gau+t  &  mantovani \\
696  &  [ 151.9,  ---,  ---,  ---, $\hdots$]  &  [ 0.20,  ---,  ---,  ---, $\hdots$]  &  [ 21.41,  ---,  ---, 1.51, $\hdots$]  &  [ 2922.80,  ---,  ---, 126.0, $\hdots$]  &  [ 100.0,  ---,  ---, 8.0, $\hdots$]  &  power  &  tribble   \\
725  &  [ 13.8,  ---,  ---,  ---, $\hdots$]  &  [ 0.20,  ---,  ---,  ---, $\hdots$]  &  [ 21.41,  ---,  ---,  ---, $\hdots$]  &  [ 707.90,  ---,  ---,  ---, $\hdots$]  &  [ 21.0,  ---,  ---,  ---, $\hdots$]  &  power  &  increase  \\
730  &  [ 140.4,  ---,  ---,  ---, $\hdots$]  &  [ 0.90,  ---,  ---,  ---, $\hdots$]  &  [ 21.41,  ---,  ---, 1.51, $\hdots$]  &  [ 8753.30,  ---,  ---, 984.0, $\hdots$]  &  [ 260.0,  ---,  ---, 60.0, $\hdots$]  &  power  &  tribble   \\
813  &  [ 107.7,  ---,  ---,  ---, $\hdots$]  &  [ 0.30,  ---,  ---,  ---, $\hdots$]  &  [ 21.41,  ---,  ---,  ---, $\hdots$]  &  [ 645.10,  ---,  ---,  ---, $\hdots$]  &  [ 22.0,  ---,  ---,  ---, $\hdots$]  &  power  &  tribble   \\
896  &  [ 24.8,  ---,  ---,  ---, $\hdots$]  &  [ 0.10,  ---,  ---,  ---, $\hdots$]  &  [ 21.41,  ---,  ---,  ---, $\hdots$]  &  [ 463.40,  ---,  ---,  ---, $\hdots$]  &  [ 16.0,  ---,  ---,  ---, $\hdots$]  &  power  &  increase  \\
924  &  [ 95.9,  ---,  ---,  ---, $\hdots$]  &  [ 1.50,  ---,  ---,  ---, $\hdots$]  &  [ 21.41,  ---,  ---,  ---, $\hdots$]  &  [ 387.20,  ---,  ---,  ---, $\hdots$]  &  [ 14.0,  ---,  ---,  ---, $\hdots$]  &  power  &  tribble   \\
998  &  [ 164.7,  ---,  ---,  ---, $\hdots$]  &  [ 3.80,  ---,  ---,  ---, $\hdots$]  &  [ 21.41, 6.25, 3.47, 1.51, $\hdots$]  &  [ 1957.10, 560.0, 282.0, 87.0, $\hdots$]  &  [ 60.0, 28.0, 15.0, 6.0, $\hdots$]  &  power  &  tribble   \\
1030  &  [ 10.7,  ---,  ---,  ---, $\hdots$]  &  [ 0.50,  ---,  ---,  ---, $\hdots$]  &  [ 21.41,  ---,  ---,  ---, $\hdots$]  &  [ 1984.10,  ---,  ---,  ---, $\hdots$]  &  [ 70.0,  ---,  ---,  ---, $\hdots$]  &  power  &  tribble   \\
1031  &  [ 76.3,  ---,  ---,  ---, $\hdots$]  &  [ 0.20,  ---,  ---,  ---, $\hdots$]  &  [ 21.41,  ---,  ---,  ---, $\hdots$]  &  [ 2541.90,  ---,  ---,  ---, $\hdots$]  &  [ 90.0,  ---,  ---,  ---, $\hdots$]  &  power  &  increase  \\
1062  &  [ 20.7,  ---,  ---, 45.0, $\hdots$]  &  [ 0.10,  ---,  ---, 6.92, $\hdots$]  &  [ 21.41,  ---,  ---, 1.51, $\hdots$]  &  [ 4067.10,  ---,  ---, 404.0, $\hdots$]  &  [ 140.0,  ---,  ---, 17.0, $\hdots$]  &  gau+t  &  mantovani \\
1098  &  [ 55.2,  ---,  ---,  ---, $\hdots$]  &  [ 0.40,  ---,  ---,  ---, $\hdots$]  &  [ 21.41,  ---,  ---,  ---, $\hdots$]  &  [ 996.10,  ---,  ---,  ---, $\hdots$]  &  [ 30.0,  ---,  ---,  ---, $\hdots$]  &  power  &  tribble   \\
1115  &  [ 6.7,  ---,  ---,  ---, $\hdots$]  &  [ 0.80,  ---,  ---,  ---, $\hdots$]  &  [ 21.41,  ---,  ---, 1.51, $\hdots$]  &  [ 6187.80,  ---,  ---, 885.0, $\hdots$]  &  [ 240.0,  ---,  ---, 40.0, $\hdots$]  &  power  &  tribble   \\
1144  &  [ 148.2, 117.0, 111.0,  ---, $\hdots$]  &  [ 1.30, 4.85, 6.90,  ---, $\hdots$]  &  [ 21.41, 6.25, 3.47, 1.51, $\hdots$]  &  [ 236.60, 182.0, 151.0, 105.0, $\hdots$]  &  [ 7.0, 9.0, 8.0, 7.0, $\hdots$]  &  gauss  &  peaked    \\
1190  &  [ 8.8,  ---,  ---, 44.0, $\hdots$]  &  [ 0.50,  ---,  ---, 3.38, $\hdots$]  &  [ 21.41,  ---,  ---, 1.51, $\hdots$]  &  [ 573.20,  ---,  ---, 467.0, $\hdots$]  &  [ 17.0,  ---,  ---, 22.0, $\hdots$]  &  power  &  increase  \\
1268  &  [ 116.3,  ---,  ---,  ---, $\hdots$]  &  [ 0.10,  ---,  ---,  ---, $\hdots$]  &  [ 21.41,  ---,  ---,  ---, $\hdots$]  &  [ 11534.0,  ---,  ---,  ---, $\hdots$]  &  [ 400.0,  ---,  ---,  ---, $\hdots$]  &  power  &  tribble   \\
1282  &  [ 161.8,  ---,  ---,  ---, $\hdots$]  &  [ 1.0,  ---,  ---,  ---, $\hdots$]  &  [ 21.41,  ---,  ---,  ---, $\hdots$]  &  [ 220.80,  ---,  ---,  ---, $\hdots$]  &  [ 7.0,  ---,  ---,  ---, $\hdots$]  &  power  &  tribble   \\
1289  &  [ 177.8,  ---,  ---,  ---, $\hdots$]  &  [ 0.70,  ---,  ---,  ---, $\hdots$]  &  [ 21.41,  ---,  ---,  ---, $\hdots$]  &  [ 146.90,  ---,  ---,  ---, $\hdots$]  &  [ 4.0,  ---,  ---,  ---, $\hdots$]  &  power  &  tribble   \\
1315  &  [ 9.7, 74.0, 73.0, 70.0, $\hdots$]  &  [ 0.30, 1.19, 1.74, 7.78, $\hdots$]  &  [ 21.41, 6.25, 3.47, 1.51, $\hdots$]  &  [ 2560.10, 991.0, 532.0, 240.0, $\hdots$]  &  [ 80.0, 50.0, 27.0, 12.0, $\hdots$]  &  gau+t  &  mantovani \\
1421  &  [ 5.6,  ---,  ---,  ---, $\hdots$]  &  [ 0.40,  ---,  ---,  ---, $\hdots$]  &  [ 21.41,  ---,  ---,  ---, $\hdots$]  &  [ 223.0,  ---,  ---,  ---, $\hdots$]  &  [ 8.0,  ---,  ---,  ---, $\hdots$]  &  power  &  increase  \\
1476  &  [ 74.6,  ---,  ---, 26.0, $\hdots$]  &  [ 0.40,  ---,  ---, 8.67, $\hdots$]  &  [ 21.41,  ---,  ---, 1.51, $\hdots$]  &  [ 814.0,  ---,  ---, 511.0, $\hdots$]  &  [ 24.0,  ---,  ---, 24.0, $\hdots$]  &  power  &  increase  \\
1561  &  [ 32.5,  ---,  ---,  ---, $\hdots$]  &  [ 3.20,  ---,  ---,  ---, $\hdots$]  &  [ 21.41,  ---,  ---, 1.51, $\hdots$]  &  [ 2210.80,  ---,  ---, 119.0, $\hdots$]  &  [ 70.0,  ---,  ---, 6.0, $\hdots$]  &  power  &  tribble   \\
1566  &  [ 92.9,  ---,  ---,  ---, $\hdots$]  &  [ 0.50,  ---,  ---,  ---, $\hdots$]  &  [ 21.41,  ---,  ---,  ---, $\hdots$]  &  [ 888.10,  ---,  ---,  ---, $\hdots$]  &  [ 30.0,  ---,  ---,  ---, $\hdots$]  &  power  &  tribble   \\
1606  &  [ 72.9,  ---,  ---,  ---, $\hdots$]  &  [ 1.20,  ---,  ---,  ---, $\hdots$]  &  [ 21.41,  ---,  ---,  ---, $\hdots$]  &  [ 1361.70,  ---,  ---,  ---, $\hdots$]  &  [ 50.0,  ---,  ---,  ---, $\hdots$]  &  power  &  tribble   \\
1664  &  [ 157.4, 145.0, 143.0, 140.0, $\hdots$]  &  [ 1.20, 1.46, 1.19, 3.24, $\hdots$]  &  [ 21.41, 6.25, 3.47, 1.51, $\hdots$]  &  [ 185.90, 267.0, 309.0, 327.0, $\hdots$]  &  [ 6.0, 14.0, 16.0, 14.0, $\hdots$]  &  power  &  tribble   \\
1698  &  [ 72.7,  ---,  ---,  ---, $\hdots$]  &  [ 0.10,  ---,  ---,  ---, $\hdots$]  &  [ 21.41,  ---,  ---,  ---, $\hdots$]  &  [ 2508.80,  ---,  ---,  ---, $\hdots$]  &  [ 80.0,  ---,  ---,  ---, $\hdots$]  &  gau+t  &  mantovani \\
1738  &  [ 68.4,  ---,  ---,  ---, $\hdots$]  &  [ 0.80,  ---,  ---,  ---, $\hdots$]  &  [ 21.41,  ---,  ---,  ---, $\hdots$]  &  [ 71.50,  ---,  ---,  ---, $\hdots$]  &  [ 2.20,  ---,  ---,  ---, $\hdots$]  &  power  &  tribble   \\
1817  &  [ 37.9,  ---,  ---,  ---, $\hdots$]  &  [ 0.20,  ---,  ---,  ---, $\hdots$]  &  [ 21.41,  ---,  ---,  ---, $\hdots$]  &  [ 1322.0,  ---,  ---,  ---, $\hdots$]  &  [ 40.0,  ---,  ---,  ---, $\hdots$]  &  power  &  tribble   \\
1820  &  [ 160.3,  ---,  ---,  ---, $\hdots$]  &  [ 0.70,  ---,  ---,  ---, $\hdots$]  &  [ 21.41,  ---,  ---,  ---, $\hdots$]  &  [ 1093.30,  ---,  ---,  ---, $\hdots$]  &  [ 30.0,  ---,  ---,  ---, $\hdots$]  &  gauss  &  burn      \\
1852  &  [ 154.8,  ---,  ---,  ---, $\hdots$]  &  [ 0.10,  ---,  ---,  ---, $\hdots$]  &  [ 21.41,  ---,  ---, 1.51, $\hdots$]  &  [ 2125.10,  ---,  ---, 152.0, $\hdots$]  &  [ 80.0,  ---,  ---, 7.0, $\hdots$]  &  power  &  tribble   \\
1870  &  [ 62.8,  ---,  ---,  ---, $\hdots$]  &  [ 0.10,  ---,  ---,  ---, $\hdots$]  &  [ 21.41,  ---,  ---,  ---, $\hdots$]  &  [ 1238.70,  ---,  ---,  ---, $\hdots$]  &  [ 40.0,  ---,  ---,  ---, $\hdots$]  &  power  &  increase  \\
1900  &  [ 158.7,  ---,  ---, 179.0, $\hdots$]  &  [ 0.20,  ---,  ---, 0.18, $\hdots$]  &  [ 21.41,  ---,  ---, 1.51, $\hdots$]  &  [ 602.60,  ---,  ---, 261.0, $\hdots$]  &  [ 18.0,  ---,  ---, 13.0, $\hdots$]  &  gauss  &  peaked
\tablenotetext{e,f}{\small{See Table~\ref{tab:models} for details on the relationship between Depol\_Type, Depol\_Physics, and the coefficients $c_{i}$.}}
\enddata
\label{tab:thedata2}
\end{deluxetable*}
\end{turnpage}

\begin{turnpage}
\begin{deluxetable*}{c c c c c c c c c c c c c c c c c c c c c}
\tablefontsize{\tiny}
\setlength{\tabcolsep}{0.015in} 
\tablewidth{0pt}
\tablecaption{Selected Columns from the First 45 Entries in the SED-catalog (continued from Table~\ref{tab:thedata})}
\tablehead{ \colhead{$\#$} & \colhead{$c_{1}$} & \colhead{$\Delta c_{1}$} & \colhead{$c_{2}$} & \colhead{$\Delta c_{2}$} & \colhead{$c_{3}$} & \colhead{$\Delta c_{3}$} & \colhead{$c_{4}$} & \colhead{$\Delta c_{4}$} & \colhead{KS$_{\text{SED}}$} & \colhead{$p$(KS$_{\text{SED}}$)} & \colhead{$\chi_{\text{SED}}^2$} & \colhead{$p$($\chi_{\text{SED}}^2$)} & \colhead{Data$\_$} & \colhead{$\beta_{c_{1}}$} & \colhead{$\beta$} & \colhead{$\Delta\beta$} & \colhead{$\chi_{\beta}^2$} & \colhead{$p(\chi_{\beta}^2)$} & \colhead{KS$_{\beta}$} & \colhead{$p$(KS$_{\beta}$)}  \\
\colhead{} & \colhead{} & \colhead{} & \colhead{} & \colhead{} & \colhead{} & \colhead{} & \colhead{} & \colhead{} & \colhead{} & \colhead{} & \colhead{} & \colhead{} & \colhead{Flag} & \colhead{} & \colhead{} & \colhead{} & \colhead{} & \colhead{} & \colhead{} & \colhead{}
\\
\colhead{} & \colhead{} & \colhead{} & \colhead{} & \colhead{} & \colhead{} & \colhead{} & \colhead{} & \colhead{} & \colhead{} & \colhead{} & \colhead{} & \colhead{} & \colhead{} & \colhead{} & \colhead{} & \colhead{} & \colhead{} & \colhead{} & \colhead{} & \colhead{}
\\
\colhead{(1)} & \colhead{(421)} & \colhead{(422)} & \colhead{(423)} & \colhead{(424)} & \colhead{(425)} & \colhead{(426)} & \colhead{(427)} & \colhead{(428)} & \colhead{(429)} & \colhead{(430)} & \colhead{(431)} & \colhead{(432)} & \colhead{(433)} & \colhead{(434)} & \colhead{(435)} & \colhead{(436)} & \colhead{(437)} & \colhead{(438)} & \colhead{(439)} & \colhead{(440)} }
\startdata
104  &  0.61  &  0.12  &  -0.33  &  0.07  &  ---  &  ---  &  ---  &  ---  &  0.191  &  0.200  &  0.597  &  0.550  &  1  &  0.609  &  -0.33  &  0.07  &  0.597  &  0.550  &  0.191  &  0.200 \\
157  &  0.6  &  0.7  &  0.4  &  0.6  &  ---  &  ---  &  ---  &  ---  &  0.218  &  0.200  &  25.513  &  0.000  &  3  &  0.601  &  0.4  &  0.6  &  25.513  &  0.000  &  0.218  &  0.200 \\
186  &  2.5  &  0.3  &  0.0  &  0.0  &  41.6  &  11.0  &  ---  &  ---  &  0.206  &  0.200  &  2.330  &  0.040  &  1  &  2.153  &  -1.4  &  0.5  &  7.067  &  0.000  &  0.291  &  0.074 \\
195  &  0.5  &  0.17  &  0.03  &  0.16  &  ---  &  ---  &  ---  &  ---  &  0.211  &  0.200  &  31.766  &  0.000  &  3  &  0.501  &  0.03  &  0.16  &  31.766  &  0.000  &  0.211  &  0.200 \\
200  &  3.3  &  0.8  &  8.7  &  0.9  &  3.5  &  1.4  &  0.8  &  0.25  &  0.222  &  0.060  &  3.385  &  0.000  &  2  &  0.000  &  -0.0  &  0.0005  &  25.171  &  0.000  &  0.294  &  0.010 \\
246  &  0.9  &  0.18  &  -0.55  &  0.11  &  ---  &  ---  &  ---  &  ---  &  0.306  &  0.200  &  0.263  &  0.769  &  1  &  0.899  &  -0.55  &  0.11  &  0.263  &  0.769  &  0.306  &  0.200 \\
274  &  6.6  &  0.7  &  13.8  &  1.4  &  8.4  &  0.8  &  ---  &  ---  &  0.385  &  0.108  &  0.000  &  1.000  &  1  &  ---  &  ---  &  ---  &  ---  &  ---  &  0.329  &  0.200 \\
366  &  0.5  &  0.1  &  -0.076  &  0.015  &  ---  &  ---  &  ---  &  ---  &  0.223  &  0.200  &  7.865  &  0.000  &  2  &  0.511  &  -0.076  &  0.015  &  7.865  &  0.000  &  0.223  &  0.200 \\
383  &  2.5  &  0.5  &  17.9  &  6.0  &  12.0  &  8.0  &  ---  &  ---  &  0.260  &  0.200  &  9.138  &  0.000  &  2  &  ---  &  ---  &  ---  &  ---  &  ---  &  0.211  &  0.200 \\
391  &  11.1  &  1.1  &  14.0  &  1.4  &  6.2  &  0.6  &  ---  &  ---  &  0.237  &  0.200  &  0.000  &  1.000  &  1  &  ---  &  ---  &  ---  &  ---  &  ---  &  0.336  &  0.200 \\
449  &  0.77  &  0.15  &  -0.072  &  0.014  &  ---  &  ---  &  ---  &  ---  &  0.228  &  0.200  &  7.772  &  0.000  &  2  &  0.770  &  -0.072  &  0.014  &  7.772  &  0.000  &  0.228  &  0.200 \\
460  &  0.7  &  0.5  &  -0.5  &  0.4  &  ---  &  ---  &  ---  &  ---  &  0.269  &  0.059  &  77.742  &  0.000  &  3  &  0.717  &  -0.5  &  0.4  &  77.742  &  0.000  &  0.269  &  0.059 \\
519  &  1.0  &  0.4  &  -0.69  &  0.29  &  ---  &  ---  &  ---  &  ---  &  0.256  &  0.200  &  8.855  &  0.000  &  2  &  0.985  &  -0.69  &  0.29  &  8.855  &  0.000  &  0.256  &  0.200 \\
571  &  16.3  &  5.0  &  0.0  &  0.0  &  12.3  &  9.0  &  1.4  &  5.0  &  0.134  &  0.200  &  5.585  &  0.000  &  2  &  2.300  &  -1.25  &  0.15  &  12.858  &  0.000  &  0.265  &  0.010 \\
696  &  0.8  &  1.2  &  -0.4  &  0.9  &  ---  &  ---  &  ---  &  ---  &  0.132  &  0.200  &  4.316  &  0.002  &  1  &  0.812  &  -0.4  &  0.9  &  4.316  &  0.002  &  0.132  &  0.200 \\
725  &  0.31  &  0.06  &  0.31  &  0.06  &  ---  &  ---  &  ---  &  ---  &  0.252  &  0.200  &  18.989  &  0.000  &  3  &  0.310  &  0.31  &  0.06  &  18.989  &  0.000  &  0.252  &  0.200 \\
730  &  0.24  &  0.06  &  -0.26  &  0.06  &  ---  &  ---  &  ---  &  ---  &  0.168  &  0.200  &  132.132  &  0.000  &  3  &  0.245  &  -0.26  &  0.06  &  132.132  &  0.000  &  0.168  &  0.200 \\
813  &  1.0  &  0.2  &  -0.089  &  0.018  &  ---  &  ---  &  ---  &  ---  &  0.278  &  0.200  &  1.499  &  0.221  &  1  &  1.018  &  -0.089  &  0.018  &  1.499  &  0.221  &  0.278  &  0.200 \\
896  &  0.17  &  0.03  &  0.73  &  0.15  &  ---  &  ---  &  ---  &  ---  &  0.196  &  0.200  &  7.418  &  0.006  &  1  &  0.165  &  0.73  &  0.15  &  7.418  &  0.006  &  0.196  &  0.200 \\
924  &  1.8  &  0.4  &  -1.15  &  0.23  &  ---  &  ---  &  ---  &  ---  &  0.377  &  0.048  &  12.966  &  0.000  &  2  &  1.771  &  -1.15  &  0.23  &  12.966  &  0.000  &  0.377  &  0.048 \\
998  &  0.38  &  0.08  &  -0.44  &  0.09  &  ---  &  ---  &  ---  &  ---  &  0.384  &  0.109  &  4.539  &  0.033  &  1  &  0.381  &  -0.44  &  0.09  &  4.539  &  0.033  &  0.384  &  0.109 \\
1030  &  1.1  &  0.3  &  -0.59  &  0.26  &  ---  &  ---  &  ---  &  ---  &  0.166  &  0.200  &  2.635  &  0.022  &  1  &  1.055  &  -0.59  &  0.26  &  2.635  &  0.022  &  0.166  &  0.200 \\
1031  &  -0.21  &  0.04  &  0.38  &  0.08  &  ---  &  ---  &  ---  &  ---  &  0.304  &  0.200  &  2.120  &  0.120  &  1  &  -0.206  &  0.38  &  0.08  &  2.120  &  0.120  &  0.304  &  0.200 \\
1062  &  3.7  &  0.8  &  11.5  &  1.2  &  8.5  &  2.1  &  2.3  &  0.7  &  0.202  &  0.200  &  2.427  &  0.033  &  1  &  1.771  &  -0.83  &  0.17  &  42903.144  &  0.000  &  0.357  &  0.010 \\
1098  &  1.9  &  0.3  &  -1.15  &  0.25  &  ---  &  ---  &  ---  &  ---  &  0.440  &  0.010  &  12.231  &  0.000  &  3  &  1.907  &  -1.15  &  0.25  &  12.231  &  0.000  &  0.440  &  0.010 \\
1115  &  0.047  &  0.014  &  -0.028  &  0.008  &  ---  &  ---  &  ---  &  ---  &  0.238  &  0.030  &  3.121  &  0.000  &  3  &  0.047  &  -0.028  &  0.008  &  3.121  &  0.000  &  0.238  &  0.030 \\
1144  &  7.2  &  0.7  &  8.1  &  0.8  &  9.3  &  0.9  &  ---  &  ---  &  0.253  &  0.200  &  0.000  &  1.000  &  1  &  ---  &  ---  &  ---  &  ---  &  ---  &  0.221  &  0.200 \\
1190  &  0.1  &  0.4  &  0.2  &  0.4  &  ---  &  ---  &  ---  &  ---  &  0.161  &  0.200  &  21.713  &  0.000  &  3  &  0.117  &  0.2  &  0.4  &  21.713  &  0.000  &  0.161  &  0.200 \\
1268  &  0.3  &  0.2  &  -0.22  &  0.14  &  ---  &  ---  &  ---  &  ---  &  0.224  &  0.074  &  23.464  &  0.000  &  3  &  0.323  &  -0.22  &  0.14  &  23.464  &  0.000  &  0.224  &  0.074 \\
1282  &  1.7  &  0.3  &  -0.81  &  0.16  &  ---  &  ---  &  ---  &  ---  &  0.375  &  0.051  &  0.124  &  0.883  &  1  &  1.700  &  -0.81  &  0.16  &  0.124  &  0.883  &  0.375  &  0.051 \\
1289  &  1.2  &  0.24  &  -0.19  &  0.04  &  ---  &  ---  &  ---  &  ---  &  0.255  &  0.200  &  3.116  &  0.044  &  1  &  1.200  &  -0.19  &  0.04  &  3.116  &  0.044  &  0.255  &  0.200 \\
1315  &  7.3  &  0.6  &  0.0  &  0.0  &  6.1  &  0.4  &  1.4  &  0.7  &  0.251  &  0.156  &  3.984  &  0.001  &  2  &  0.977  &  -0.6  &  0.3  &  15.823  &  0.000  &  0.274  &  0.079 \\
1421  &  1.12  &  0.22  &  0.0103  &  0.0021  &  ---  &  ---  &  ---  &  ---  &  0.419  &  0.014  &  1.593  &  0.203  &  1  &  1.121  &  0.0103  &  0.0021  &  1.593  &  0.203  &  0.419  &  0.014 \\
1476  &  0.3  &  0.15  &  0.17  &  0.15  &  ---  &  ---  &  ---  &  ---  &  0.282  &  0.037  &  27.938  &  0.000  &  3  &  0.305  &  0.17  &  0.15  &  27.938  &  0.000  &  0.282  &  0.037 \\
1561  &  0.46  &  0.09  &  -0.54  &  0.11  &  ---  &  ---  &  ---  &  ---  &  0.331  &  0.159  &  0.133  &  0.875  &  1  &  0.456  &  -0.54  &  0.11  &  0.133  &  0.875  &  0.331  &  0.159 \\
1566  &  1.5  &  0.3  &  -0.96  &  0.19  &  ---  &  ---  &  ---  &  ---  &  0.422  &  0.012  &  11.255  &  0.000  &  3  &  1.537  &  -0.96  &  0.19  &  11.255  &  0.000  &  0.422  &  0.012 \\
1606  &  0.18  &  0.18  &  -0.17  &  0.17  &  ---  &  ---  &  ---  &  ---  &  0.469  &  0.010  &  0.790  &  0.531  &  1  &  0.180  &  -0.17  &  0.17  &  0.790  &  0.531  &  0.469  &  0.010 \\
1664  &  1.21  &  0.24  &  -0.44  &  0.09  &  ---  &  ---  &  ---  &  ---  &  0.222  &  0.200  &  3.843  &  0.021  &  1  &  1.211  &  -0.44  &  0.09  &  3.843  &  0.021  &  0.222  &  0.200 \\
1698  &  3.9  &  1.0  &  9.0  &  2.6  &  10.1  &  4.0  &  1.8  &  1.0  &  0.193  &  0.175  &  8.846  &  0.000  &  2  &  1.924  &  -1.1  &  0.24  &  111270.044  &  0.000  &  0.323  &  0.010 \\
1738  &  1.7  &  0.3  &  -0.32  &  0.06  &  ---  &  ---  &  ---  &  ---  &  0.380  &  0.044  &  11.261  &  0.000  &  3  &  1.654  &  -0.32  &  0.06  &  11.261  &  0.000  &  0.380  &  0.044 \\
1817  &  0.9  &  0.18  &  -0.17  &  0.03  &  ---  &  ---  &  ---  &  ---  &  0.410  &  0.018  &  1.624  &  0.197  &  1  &  0.897  &  -0.17  &  0.03  &  1.624  &  0.197  &  0.410  &  0.018 \\
1820  &  9.14  &  0.22  &  0.0  &  0.0  &  9.92  &  0.22  &  ---  &  ---  &  0.238  &  0.200  &  1.235  &  0.291  &  1  &  2.486  &  -1.9  &  0.4  &  2.032  &  0.131  &  0.314  &  0.200 \\
1852  &  0.96  &  0.23  &  -0.16  &  0.19  &  ---  &  ---  &  ---  &  ---  &  0.308  &  0.144  &  1.959  &  0.118  &  1  &  0.960  &  -0.16  &  0.19  &  1.959  &  0.118  &  0.308  &  0.144 \\
1870  &  0.67  &  0.13  &  0.18  &  0.04  &  ---  &  ---  &  ---  &  ---  &  0.261  &  0.200  &  0.212  &  0.645  &  1  &  0.666  &  0.18  &  0.04  &  0.212  &  0.645  &  0.261  &  0.200 \\
1900  &  11.82  &  0.29  &  13.4  &  0.2  &  8.35  &  0.25  &  ---  &  ---  &  0.235  &  0.200  &  0.303  &  0.738  &  1  &  ---  &  ---  &  ---  &  ---  &  ---  &  0.431  &  0.010
\enddata
\label{tab:thedata3}
\end{deluxetable*}
\end{turnpage}

\begin{turnpage}
\begin{deluxetable*}{c c c c c c c c c c c c c c c c c c c c c}
\tablefontsize{\tiny}
\setlength{\tabcolsep}{0.015in} 
\tablewidth{0pt}
\tablecaption{Selected Columns from the First 45 Entries in the SED-catalog (continued from Table~\ref{tab:thedata})}
\tablehead{ \colhead{\#} & \colhead{$z$} & \colhead{$\Delta z$} & \colhead{Obj.} & \colhead{I\_} & \colhead{$d_{1}$} & \colhead{$\alpha$} & \colhead{$d_{3}$} & \colhead{$\Delta d_{1}$} & \colhead{$\Delta \alpha$} & \colhead{$\Delta d_{3}$} & \colhead{$\chi_{\alpha}^2$} & \colhead{$\chi_{\alpha}^2$} & \colhead{$p$($\chi_{\alpha}^2$)} & \colhead{$p$($\chi_{\alpha}^2$)} & \colhead{RM} & \colhead{$\Delta$RM} & \colhead{EVPA} & \colhead{$\Delta$EVPA} & \colhead{$\chi_{\text{RM}}^2$} & \colhead{p($\chi_{\text{RM}}^2$)}  \\
\colhead{} & \colhead{} & \colhead{} & \colhead{} & \colhead{Class} & \colhead{} & \colhead{} & \colhead{} & \colhead{} & \colhead{} & \colhead{} & \colhead{(reg)} & \colhead{(cur)} & \colhead{(reg)} & \colhead{(cur)} & \colhead{(broad)} & \colhead{(broad)} & \colhead{(broad)} & \colhead{(broad)} & \colhead{(broad)} & \colhead{(broad)}
\\
\colhead{} & \colhead{} & \colhead{} & \colhead{} & \colhead{} & \colhead{} & \colhead{} & \colhead{} & \colhead{} & \colhead{} & \colhead{} & \colhead{} & \colhead{} & \colhead{} & \colhead{} & \colhead{/rad~m$^{-2}$} & \colhead{/rad~m$^{-2}$} & \colhead{/$^\circ$} & \colhead{/$^\circ$} & \colhead{} & \colhead{}
\\
\colhead{(1)} & \colhead{(441)} & \colhead{(442)} & \colhead{(443)} & \colhead{(444)} & \colhead{(445)} & \colhead{(446)} & \colhead{(447)} & \colhead{(448)} & \colhead{(449)} & \colhead{(450)} & \colhead{(451)} & \colhead{(452)} & \colhead{(453)} & \colhead{(454)} & \colhead{(455)} & \colhead{(456)} & \colhead{(457)} & \colhead{(458)} & \colhead{(459)} & \colhead{(460)} }
\startdata
104  &  1.465  &  0.003  &  QSO         &  R  &  9.8  &  -0.71  &  ---  &  2.0  &  0.14  &  ---  &  1.033  &  ---  &  0.309  &  ---  &  -26.9  &  3.0  &  17.2  &  2.0  &  1.057  &  0.348 \\
157  &  ---  &  ---  &              &  2  &  11.0  &  -0.88  &  ---  &  1.3  &  0.18  &  ---  &  ---  &  ---  &  ---  &  ---  &  -5.3  &  5.0  &  172.27  &  0.24  &  14.129  &  0.000 \\
186  &  0.4509  &  0.001  &  QSO         &  R  &  9.3  &  -0.7  &  ---  &  1.9  &  0.14  &  ---  &  0.211  &  85.798  &  0.810  &  0.000  &  -33.4  &  1.7  &  38.94  &  0.08  &  0.293  &  0.830 \\
195  &  0.3466  &  0.0  &              &  R  &  4.1  &  -0.09  &  ---  &  0.7  &  0.07  &  ---  &  6.138  &  6.972  &  0.000  &  0.000  &  -409.2  &  7.0  &  31.5  &  0.3  &  134.518  &  0.000 \\
200  &  1.0365  &  0.0002  &  QSO         &  R  &  11.2  &  -0.83  &  ---  &  2.2  &  0.17  &  ---  &  0.911  &  ---  &  0.340  &  ---  &  20.3  &  1.9  &  104.55  &  0.09  &  5.166  &  0.000 \\
246  &  ---  &  ---  &              &  R  &  6.3  &  -0.4  &  ---  &  1.7  &  0.17  &  ---  &  19.863  &  53.460  &  0.000  &  0.000  &  -430.2  &  50.0  &  157.8  &  18.0  &  1.080  &  0.340 \\
274  &  0.0254  &  0.0  &              &  R  &  4.8  &  -0.23  &  ---  &  0.7  &  0.07  &  ---  &  4.084  &  15.658  &  0.007  &  0.000  &  12.9  &  1.3  &  61.1  &  6.0  &  0.000  &  0.999 \\
366  &  ---  &  ---  &              &  R  &  -2.1  &  0.53  &  ---  &  0.4  &  0.11  &  ---  &  25.123  &  206.815  &  0.000  &  0.000  &  8.0  &  0.9  &  91.9  &  11.0  &  7.730  &  0.000 \\
383  &  0.255  &  0.002  &  G           &  R  &  6.4  &  -0.35  &  ---  &  0.8  &  0.08  &  ---  &  3.331  &  26.047  &  0.005  &  0.000  &  -55.2  &  4.0  &  51.19  &  0.18  &  10.098  &  0.000 \\
391  &  ---  &  ---  &              &  R  &  11.4  &  -0.94  &  ---  &  2.3  &  0.19  &  ---  &  0.740  &  ---  &  0.390  &  ---  &  -31.2  &  3.0  &  155.1  &  16.0  &  0.640  &  0.424 \\
449  &  1.721  &  0.007  &  QSO         &  R  &  11.0  &  -0.94  &  ---  &  0.3  &  0.04  &  ---  &  0.739  &  0.880  &  0.594  &  0.475  &  -123.9  &  14.0  &  133.8  &  15.0  &  2.599  &  0.074 \\
460  &  0.8404  &  0.0004  &  G           &  2  &  12.5  &  -0.98  &  ---  &  1.5  &  0.06  &  ---  &  ---  &  ---  &  ---  &  ---  &  112.7  &  2.7  &  126.17  &  0.12  &  11.003  &  0.000 \\
519  &  1.589  &  0.001  &  G           &  R  &  12.6  &  -1.0  &  ---  &  2.5  &  0.2  &  ---  &  1.469  &  ---  &  0.226  &  ---  &  8.7  &  5.0  &  132.36  &  0.21  &  4.295  &  0.005 \\
571  &  2.0225  &  0.0003  &  QSO         &  R  &  14.2  &  -1.19  &  ---  &  2.8  &  0.24  &  ---  &  0.119  &  ---  &  0.730  &  ---  &  -20.3  &  0.7  &  34.4  &  0.03  &  1.785  &  0.051 \\
696  &  0.4061  &  0.0  &  QSO         &  R  &  14.1  &  -1.16  &  ---  &  2.8  &  0.23  &  ---  &  20.450  &  ---  &  0.000  &  ---  &  18.9  &  4.0  &  102.31  &  0.18  &  4.246  &  0.005 \\
725  &  1.946  &  0.002  &  QSO         &  R  &  4.6  &  -0.19  &  ---  &  0.4  &  0.04  &  ---  &  1.592  &  1.506  &  0.158  &  0.197  &  -98.5  &  11.0  &  92.6  &  11.0  &  1.675  &  0.187 \\
730  &  0.3216  &  3e-05  &  G           &  R  &  11.2  &  -0.8  &  ---  &  2.2  &  0.16  &  ---  &  15.003  &  ---  &  0.000  &  ---  &  47.6  &  6.0  &  14.66  &  0.26  &  12.417  &  0.000 \\
813  &  ---  &  ---  &              &  n  &  ---  &  ---  &  ---  &  ---  &  ---  &  ---  &  ---  &  ---  &  ---  &  ---  &  287.3  &  29.0  &  72.8  &  7.0  &  7.347  &  0.007 \\
896  &  ---  &  ---  &              &  n  &  ---  &  ---  &  ---  &  ---  &  ---  &  ---  &  ---  &  ---  &  ---  &  ---  &  -8.5  &  0.9  &  47.3  &  5.0  &  0.740  &  0.390 \\
924  &  ---  &  ---  &              &  R  &  11.9  &  -1.02  &  ---  &  0.7  &  0.07  &  ---  &  4.222  &  11.187  &  0.001  &  0.000  &  -101.2  &  12.0  &  1.7  &  0.2  &  0.107  &  0.898 \\
998  &  0.518  &  0.0  &  QSO         &  R  &  13.7  &  -1.13  &  ---  &  0.9  &  0.09  &  ---  &  3.530  &  24.427  &  0.014  &  0.000  &  -51.5  &  5.0  &  119.9  &  12.0  &  0.076  &  0.782 \\
1030  &  1.469  &  0.001  &  QSO         &  R  &  12.9  &  -1.05  &  ---  &  2.6  &  0.21  &  ---  &  0.084  &  ---  &  0.772  &  ---  &  124.6  &  5.0  &  43.37  &  0.22  &  4.974  &  0.000 \\
1031  &  ---  &  ---  &              &  R  &  12.8  &  -1.0  &  ---  &  2.6  &  0.2  &  ---  &  0.373  &  1.463  &  0.689  &  0.226  &  31.7  &  4.0  &  173.1  &  20.0  &  1.505  &  0.222 \\
1062  &  0.0733  &  0.0002  &  G           &  R  &  11.5  &  -0.86  &  ---  &  2.3  &  0.17  &  ---  &  1.319  &  ---  &  0.251  &  ---  &  -10.3  &  1.2  &  47.72  &  0.05  &  1.353  &  0.248 \\
1098  &  2.2627  &  9.3e-05  &  QSO         &  R  &  8.9  &  -0.64  &  ---  &  0.4  &  0.05  &  ---  &  0.198  &  6.717  &  0.898  &  0.001  &  3.3  &  2.9  &  46.6  &  0.13  &  3.213  &  0.022 \\
1115  &  0.2196  &  0.0  &  G           &  R  &  10.4  &  -0.73  &  ---  &  2.1  &  0.15  &  ---  &  0.436  &  0.508  &  0.647  &  0.476  &  86.6  &  5.0  &  138.79  &  0.24  &  21.714  &  0.000 \\
1144  &  ---  &  ---  &              &  R  &  5.0  &  -0.28  &  ---  &  1.0  &  0.06  &  ---  &  0.700  &  29.100  &  0.497  &  0.000  &  13.5  &  1.4  &  112.6  &  11.0  &  0.215  &  0.643 \\
1190  &  1.178  &  0.0  &  QSO         &  2  &  3.5  &  -0.08  &  ---  &  0.4  &  0.05  &  ---  &  ---  &  ---  &  ---  &  ---  &  -15.0  &  2.7  &  48.16  &  0.12  &  8.165  &  0.000 \\
1268  &  0.174  &  0.001  &  G           &  R  &  10.9  &  -0.76  &  ---  &  1.2  &  0.12  &  ---  &  1.597  &  41.617  &  0.188  &  0.000  &  90.5  &  2.3  &  58.51  &  0.11  &  4.281  &  0.000 \\
1282  &  ---  &  ---  &              &  R  &  13.3  &  -1.2  &  ---  &  0.7  &  0.08  &  ---  &  7.080  &  11.788  &  0.000  &  0.000  &  -103.5  &  12.0  &  73.7  &  8.0  &  0.735  &  0.480 \\
1289  &  ---  &  ---  &              &  R  &  9.8  &  -0.83  &  ---  &  0.4  &  0.04  &  ---  &  0.804  &  2.670  &  0.547  &  0.030  &  -96.1  &  11.0  &  70.2  &  8.0  &  0.920  &  0.398 \\
1315  &  ---  &  ---  &              &  R  &  11.5  &  -0.88  &  ---  &  0.5  &  0.05  &  ---  &  1.057  &  8.916  &  0.366  &  0.000  &  47.1  &  1.4  &  65.86  &  0.06  &  9.713  &  0.000 \\
1421  &  ---  &  ---  &              &  R  &  10.8  &  -0.92  &  ---  &  0.9  &  0.09  &  ---  &  9.667  &  11.889  &  0.000  &  0.000  &  -85.5  &  10.0  &  50.3  &  6.0  &  0.297  &  0.743 \\
1476  &  1.5338  &  0.0019  &  BLLac       &  R  &  4.7  &  -0.19  &  ---  &  0.9  &  0.04  &  ---  &  3.626  &  ---  &  0.057  &  ---  &  283.4  &  1.3  &  50.13  &  0.06  &  6.806  &  0.000 \\
1561  &  0.2104  &  0.0  &  G           &  R  &  13.3  &  -1.09  &  ---  &  2.7  &  0.22  &  ---  &  5.759  &  11.652  &  0.003  &  0.001  &  151.3  &  17.0  &  175.1  &  20.0  &  0.516  &  0.597 \\
1566  &  ---  &  ---  &              &  C  &  2.94  &  -0.962  &  -0.0769  &  0.27  &  0.027  &  0.0022  &  2.628  &  0.838  &  0.022  &  0.501  &  -109.1  &  13.0  &  19.4  &  2.2  &  3.269  &  0.038 \\
1606  &  0.1952  &  0.0002  &  G           &  R  &  19.6  &  -1.8  &  ---  &  4.0  &  0.4  &  ---  &  0.042  &  ---  &  0.838  &  ---  &  -198.1  &  10.0  &  53.7  &  0.4  &  5.625  &  0.000 \\
1664  &  ---  &  ---  &              &  R  &  0.19  &  0.23  &  ---  &  0.04  &  0.05  &  ---  &  0.652  &  28.141  &  0.521  &  0.000  &  5.6  &  0.6  &  142.8  &  16.0  &  0.619  &  0.539 \\
1698  &  0.7186  &  4e-05  &  QSO         &  R  &  7.85  &  -0.49  &  ---  &  0.28  &  0.03  &  ---  &  0.094  &  0.141  &  0.964  &  0.868  &  -6.3  &  0.8  &  89.25  &  0.04  &  5.063  &  0.000 \\
1738  &  ---  &  ---  &              &  R  &  8.94  &  -0.77  &  ---  &  0.09  &  0.01  &  ---  &  0.085  &  0.100  &  0.987  &  0.960  &  -94.3  &  11.0  &  136.1  &  16.0  &  0.212  &  0.809 \\
1817  &  ---  &  ---  &              &  R  &  14.5  &  -1.24  &  ---  &  0.6  &  0.07  &  ---  &  0.247  &  8.609  &  0.912  &  0.000  &  -62.9  &  7.0  &  23.1  &  2.7  &  1.379  &  0.252 \\
1820  &  0.0784  &  0.0  &  G           &  R  &  12.6  &  -1.05  &  ---  &  2.5  &  0.21  &  ---  &  0.272  &  2.098  &  0.762  &  0.147  &  -33.6  &  4.0  &  68.5  &  8.0  &  0.135  &  0.874 \\
1852  &  0.387  &  0.001  &  G           &  R  &  12.2  &  -1.0  &  ---  &  2.4  &  0.2  &  ---  &  9.087  &  ---  &  0.003  &  ---  &  3.5  &  1.0  &  145.72  &  0.04  &  0.539  &  0.655 \\
1870  &  ---  &  ---  &              &  n  &  ---  &  ---  &  ---  &  ---  &  ---  &  ---  &  ---  &  ---  &  ---  &  ---  &  1.31  &  0.13  &  59.4  &  6.0  &  2.888  &  0.089 \\
1900  &  2.201  &  0.0  &  QSO         &  R  &  5.8  &  -0.33  &  ---  &  1.2  &  0.07  &  ---  &  2.437  &  ---  &  0.119  &  ---  &  -7.8  &  0.9  &  179.1  &  21.0  &  0.081  &  0.922
\enddata
\label{tab:thedata4}
\end{deluxetable*}
\end{turnpage}

\begin{turnpage}
\begin{deluxetable*}{c c c c c c c c c c c c c c c c c c c}
\tablefontsize{\tiny}
\setlength{\tabcolsep}{0.015in} 
\tablewidth{0pt}
\tablecaption{Selected Columns from the First 45 Entries in the SED-catalog (continued from Table~\ref{tab:thedata})}
\tablehead{ \colhead{\#} & \colhead{RM} & \colhead{$\Delta$RM} & \colhead{$I$} & \colhead{$\Delta I$} & \colhead{$P$} & \colhead{$\Delta P$} & \colhead{$\Pi$} & \colhead{$\Delta \Pi$} & \colhead{$\theta_{\text{maj}}$} & \colhead{$\theta_{\text{maj}}$} & \colhead{$\Delta\theta_{\text{maj}}$} & \colhead{$\theta_{\text{min}}$} & \colhead{$\theta_{\text{min}}$} & \colhead{$\Delta\theta_{\text{min}}$} & \colhead{PA} & \colhead{$\Delta$PA\tablenotemark{g}} & \colhead{$\log_{10} (L_{\nu})$ \tablenotemark{h}} & \colhead{$\log_{10} (\Delta L_{\nu})$}    \\
\colhead{} & \colhead{(NVSS)} & \colhead{(NVSS)} & \colhead{(NVSS)} & \colhead{(NVSS)} & \colhead{(NVSS)} & \colhead{(NVSS)} & \colhead{(NVSS)} & \colhead{(NVSS)} & \colhead{limit} & \colhead{} & \colhead{} & \colhead{limit} & \colhead{} & \colhead{} & \colhead{} & \colhead{} & \colhead{} & \colhead{}
\\
\colhead{} & \colhead{/rad~m$^{-2}$} & \colhead{/rad~m$^{-2}$} & \colhead{/mJy} & \colhead{/mJy} & \colhead{/mJy} & \colhead{/mJy} & \colhead{/\%} & \colhead{/\%} & \colhead{} & \colhead{/arcsec} & \colhead{/arcsec} & \colhead{} & \colhead{/arcsec} & \colhead{/arcsec} & \colhead{/$^\circ$} & \colhead{/$^\circ$} & \colhead{} & \colhead{}
\\
\colhead{(1)} & \colhead{(461)} & \colhead{(462)} & \colhead{(463)} & \colhead{(464)} & \colhead{(465)} & \colhead{(466)} & \colhead{(467)} & \colhead{(468)} & \colhead{(469)} & \colhead{(470)} & \colhead{(471)} & \colhead{(472)} & \colhead{(473)} & \colhead{(474)} & \colhead{(475)} & \colhead{(476)} & \colhead{(477)} & \colhead{(478)}  }
\startdata
104  &  -33.2  &  2.4  &  2414.8  &  72.4  &  32.64  &  0.35  &  1.47  &  0.11  &  <  &  15.6  &  ---  &  <  &  13.9  &  ---  &  ---  &  ---  &  28.4  &  6.0 \\
157  &  -11.6  &  6.0  &  1069.5  &  35.1  &  103.2  &  2.73  &  19.9  &  0.8  &     &  86.7  &  1.0  &     &  20.4  &  2.2  &  -81.1  &  0.0  &  ---  &  --- \\
186  &  -28.5  &  5.0  &  804.4  &  27.0  &  14.25  &  0.25  &  2.1  &  0.13  &     &  30.8  &  1.6  &  <  &  17.5  &  ---  &  -63.6  &  0.2  &  26.8  &  5.0 \\
195  &  12.0  &  1.2  &  2050.7  &  61.5  &  61.84  &  0.32  &  3.19  &  0.14  &  <  &  19.6  &  ---  &  <  &  16.6  &  ---  &  ---  &  ---  &  26.8  &  27.0 \\
200  &  22.5  &  3.0  &  3897.6  &  116.9  &  26.17  &  0.35  &  0.7  &  0.1  &  <  &  17.1  &  ---  &  <  &  15.5  &  ---  &  ---  &  ---  &  28.3  &  6.0 \\
246  &  13.6  &  12.0  &  468.2  &  14.1  &  6.46  &  0.37  &  1.47  &  0.14  &  <  &  16.9  &  ---  &  <  &  13.9  &  ---  &  ---  &  ---  &  ---  &  --- \\
274  &  -2.9  &  8.0  &  388.3  &  11.7  &  17.65  &  0.67  &  4.36  &  0.23  &  <  &  18.8  &  ---  &  <  &  15.3  &  ---  &  ---  &  ---  &  23.8  &  23.0 \\
366  &  10.2  &  6.0  &  493.6  &  17.4  &  13.16  &  0.34  &  2.74  &  0.16  &     &  14.4  &  3.0  &  <  &  17.0  &  ---  &  -33.7  &  1.6  &  ---  &  --- \\
383  &  -39.3  &  2.0  &  1650.9  &  64.2  &  35.06  &  0.36  &  2.31  &  0.14  &     &  18.4  &  2.4  &     &  15.3  &  2.8  &  -76.1  &  0.6  &  26.5  &  26.0 \\
391  &  -36.9  &  2.3  &  624.8  &  20.5  &  27.16  &  0.3  &  5.41  &  0.21  &     &  38.1  &  1.4  &  <  &  16.3  &  ---  &  77.0  &  0.1  &  ---  &  --- \\
449  &  -108.3  &  5.0  &  263.6  &  7.9  &  11.3  &  0.31  &  4.59  &  0.21  &  <  &  16.7  &  ---  &  <  &  14.2  &  ---  &  ---  &  ---  &  27.7  &  26.0 \\
460  &  -22.1  &  1.9  &  3650.3  &  122.1  &  180.33  &  1.74  &  6.51  &  0.25  &     &  31.7  &  1.6  &  <  &  16.8  &  ---  &  25.3  &  0.0  &  28.1  &  27.0 \\
519  &  -34.6  &  4.0  &  2122.3  &  63.7  &  24.6  &  0.33  &  1.17  &  0.11  &  <  &  17.3  &  ---  &  <  &  15.3  &  ---  &  ---  &  ---  &  28.6  &  6.0 \\
571  &  -16.8  &  1.1  &  2180.0  &  65.4  &  130.67  &  0.64  &  5.7  &  0.2  &  <  &  19.6  &  ---  &  <  &  16.1  &  ---  &  ---  &  ---  &  28.9  &  6.0 \\
696  &  40.9  &  1.4  &  2922.8  &  95.0  &  50.91  &  0.31  &  2.03  &  0.12  &     &  42.0  &  1.3  &  <  &  16.7  &  ---  &  -11.4  &  0.0  &  27.3  &  5.0 \\
725  &  -101.6  &  1.5  &  707.9  &  21.2  &  40.7  &  0.27  &  6.02  &  0.21  &  <  &  15.9  &  ---  &  <  &  15.7  &  ---  &  ---  &  ---  &  27.9  &  27.0 \\
730  &  55.5  &  1.8  &  8753.3  &  262.6  &  67.24  &  0.7  &  0.8  &  0.1  &  <  &  17.8  &  ---  &  <  &  16.7  &  ---  &  ---  &  ---  &  27.5  &  5.0 \\
813  &  -17.3  &  8.0  &  645.1  &  21.9  &  31.21  &  0.97  &  8.1  &  0.4  &     &  54.6  &  1.2  &     &  27.7  &  1.7  &  76.3  &  0.2  &  ---  &  --- \\
896  &  -10.4  &  1.0  &  463.4  &  15.6  &  60.49  &  0.29  &  14.2  &  0.5  &     &  30.3  &  1.6  &  <  &  17.3  &  ---  &  -32.6  &  0.3  &  ---  &  --- \\
924  &  -91.6  &  10.0  &  387.2  &  13.7  &  6.47  &  0.29  &  1.78  &  0.14  &     &  15.0  &  2.9  &  <  &  18.1  &  ---  &  46.5  &  2.4  &  ---  &  --- \\
998  &  -91.6  &  6.0  &  1957.1  &  58.7  &  11.79  &  0.31  &  0.6  &  0.1  &  <  &  18.6  &  ---  &  <  &  16.3  &  ---  &  ---  &  ---  &  27.4  &  26.0 \\
1030  &  1.9  &  2.1  &  1984.1  &  68.0  &  33.1  &  0.29  &  1.94  &  0.12  &     &  24.2  &  1.9  &  <  &  18.8  &  ---  &  -16.7  &  0.1  &  28.5  &  6.0 \\
1031  &  6.5  &  1.8  &  2541.9  &  85.0  &  41.26  &  0.35  &  1.94  &  0.12  &     &  31.9  &  1.6  &  <  &  15.7  &  ---  &  51.6  &  0.0  &  ---  &  --- \\
1062  &  -13.8  &  0.5  &  4067.1  &  138.5  &  167.96  &  0.35  &  4.68  &  0.19  &     &  26.1  &  1.8  &  <  &  16.2  &  ---  &  -27.4  &  0.0  &  25.8  &  5.0 \\
1098  &  33.4  &  2.5  &  996.1  &  29.9  &  22.74  &  0.26  &  2.4  &  0.13  &  <  &  19.8  &  ---  &  <  &  17.3  &  ---  &  ---  &  ---  &  28.4  &  27.0 \\
1115  &  -4.3  &  2.8  &  6187.8  &  239.1  &  33.58  &  0.41  &  0.6  &  0.1  &     &  19.9  &  2.2  &     &  16.1  &  2.7  &  -70.1  &  0.1  &  27.0  &  5.0 \\
1144  &  1.4  &  13.0  &  236.6  &  7.1  &  5.86  &  0.32  &  2.57  &  0.19  &  <  &  16.5  &  ---  &  <  &  15.3  &  ---  &  ---  &  ---  &  ---  &  --- \\
1190  &  4.8  &  4.0  &  573.2  &  17.2  &  19.61  &  0.31  &  3.48  &  0.15  &  <  &  19.0  &  ---  &  <  &  18.0  &  ---  &  ---  &  ---  &  27.4  &  27.0 \\
1268  &  -78.3  &  0.6  &  11534.0  &  373.7  &  191.62  &  0.55  &  2.07  &  0.12  &     &  43.3  &  1.3  &  <  &  16.6  &  ---  &  -71.0  &  0.0  &  27.0  &  26.0 \\
1282  &  -94.6  &  7.0  &  220.8  &  6.6  &  8.51  &  0.28  &  4.23  &  0.21  &  <  &  17.1  &  ---  &  <  &  15.8  &  ---  &  ---  &  ---  &  ---  &  --- \\
1289  &  -92.7  &  5.0  &  146.9  &  4.4  &  12.4  &  0.29  &  8.8  &  0.3  &  <  &  19.8  &  ---  &  <  &  16.6  &  ---  &  ---  &  ---  &  ---  &  --- \\
1315  &  18.9  &  4.0  &  2560.1  &  76.8  &  31.79  &  0.51  &  1.46  &  0.11  &  <  &  15.7  &  ---  &  <  &  14.1  &  ---  &  ---  &  ---  &  ---  &  --- \\
1421  &  -81.9  &  1.6  &  223.0  &  8.0  &  23.46  &  0.18  &  13.8  &  0.5  &     &  33.4  &  1.6  &     &  29.6  &  1.7  &  57.1  &  3.3  &  ---  &  --- \\
1476  &  -19.0  &  2.7  &  814.0  &  24.4  &  24.79  &  0.3  &  3.16  &  0.14  &  <  &  18.3  &  ---  &  <  &  16.4  &  ---  &  ---  &  ---  &  27.8  &  6.0 \\
1561  &  -15.4  &  7.0  &  2210.8  &  66.3  &  11.03  &  0.33  &  0.5  &  0.1  &  <  &  18.8  &  ---  &  <  &  15.2  &  ---  &  ---  &  ---  &  26.5  &  5.0 \\
1566  &  -65.2  &  5.0  &  888.1  &  31.3  &  15.13  &  0.3  &  1.8  &  0.12  &     &  15.1  &  2.9  &  <  &  17.7  &  ---  &  -36.5  &  0.8  &  ---  &  --- \\
1606  &  33.6  &  10.0  &  1361.7  &  45.4  &  9.51  &  0.46  &  0.82  &  0.11  &     &  33.1  &  1.5  &  <  &  16.5  &  ---  &  -31.9  &  0.1  &  26.3  &  5.0 \\
1664  &  23.8  &  10.0  &  185.9  &  5.6  &  7.1  &  0.34  &  4.12  &  0.25  &  <  &  16.8  &  ---  &  <  &  16.2  &  ---  &  ---  &  ---  &  ---  &  --- \\
1698  &  -10.1  &  0.8  &  2508.8  &  75.3  &  86.79  &  0.34  &  3.91  &  0.15  &  <  &  17.0  &  ---  &  <  &  16.0  &  ---  &  ---  &  ---  &  27.7  &  27.0 \\
1738  &  -94.2  &  6.0  &  71.5  &  2.2  &  11.48  &  0.27  &  16.7  &  0.7  &  <  &  18.9  &  ---  &  <  &  17.3  &  ---  &  ---  &  ---  &  ---  &  --- \\
1817  &  -76.5  &  1.3  &  1322.0  &  42.7  &  45.98  &  0.29  &  4.56  &  0.18  &     &  44.3  &  1.3  &  <  &  15.4  &  ---  &  -57.7  &  0.1  &  ---  &  --- \\
1820  &  -80.1  &  12.0  &  1093.3  &  32.8  &  7.54  &  0.38  &  0.72  &  0.11  &  <  &  18.5  &  ---  &  <  &  17.4  &  ---  &  ---  &  ---  &  25.3  &  5.0 \\
1852  &  6.9  &  0.6  &  2125.1  &  75.0  &  118.61  &  0.33  &  5.88  &  0.23  &     &  14.5  &  3.0  &  <  &  17.4  &  ---  &  29.8  &  0.4  &  27.1  &  5.0 \\
1870  &  0.6  &  0.9  &  1238.7  &  41.1  &  85.22  &  0.37  &  8.13  &  0.29  &     &  34.1  &  1.5  &  <  &  15.8  &  ---  &  -15.2  &  0.1  &  ---  &  --- \\
1900  &  -9.5  &  1.5  &  602.6  &  18.1  &  44.59  &  0.3  &  7.62  &  0.25  &  <  &  16.5  &  ---  &  <  &  16.1  &  ---  &  ---  &  ---  &  28.0  &  6.0
\enddata
\tablenotetext{g}{\small{The NVSS position angle error is sometimes listed as zero. This is the value provided in the original NVSS survey.}}
\tablenotetext{h}{\small{Calculated assuming the source is dominated by optically-thin synchrotron emission.}}
\label{tab:thedata5}
\end{deluxetable*}
\end{turnpage}

\section{Catalog Systematics and Limitations}
\label{systematics}
We here seek to present a collection of systematic effects that may guide the interpretation of data within the catalog. In order to ensure that the catalog is providing an accurate measure of astrophysical effects, we discuss these systematics, limitations, and other possible issues that could affect the quality of the catalog and of which a user of the catalog needs to be aware. Note that each source in the catalog is considered unresolved at all wavelengths, and is still unresolved at the 45~arcsec resolution of the NVSS. The angular size of each source in the NVSS is included in the catalog.

\subsection{Beam Depolarization}
Many of the sources included via the cross-matching process are likely affected by beam depolarization. This occurs as all observations have an intrinsic angular resolution that is determined by the size of the resolving beam. Consider an extended source that is not uniformly polarized: if the radio telescope has a resolution that is coarser (i.e.\ an observing beam size that is larger) than the angular scale over which the source polarization is coherent, then a given synthesized beam contains regions with different EVPAs. The synthesized beam therefore averages out the polarization of the source, and the measured polarization will be less than the true source polarization (see Appendix~\ref{externaldispersion}). This `beam depolarization' cannot be corrected for, even using a polarized SED with data at multiple wavelengths, which cannot separate the astrophysical and beam-induced contributions. The only way to obtain the true source polarization would be to observe with a telescope that has better angular resolution relative to the original observations.

\subsection{Multiple Source Components}
Complications can arise due to the spectral index and the multiple components of a radio source. An unresolved source may contain many sub-structures, so that within the resolving beam is a flat-spectrum compact nucleus and two steep-spectrum lobes/jets. These components become fainter or brighter with observing frequency, such that it is not trivial to confirm that we are observing the same dominant source component at all frequencies. Again, observations with increased angular resolution would be useful in separating out these various source contributions. Furthermore, for future studies, RM Synthesis can in principle distinguish different emitting regions via their separation in the Faraday spectrum, even if the components are not angularly resolved \citep[e.g.][]{2012A&A...543A.113B}. The polarized fraction can also help in separating these components, as $\Pi$ is typically higher at frequencies where steep-spectrum jets/lobes dominate \citep{2002A&A...396..463M}. Similarly, measurements of total intensity spectra may also help in resolving this issue. Not only is there an expected difference in the spectral index from different emitting regions, but sources that contain multiple components within an individual resolution element may also show spectral curvature -- although a number of other mechanisms can also cause such curvature \citep[e.g.][]{1970R&QE...13..162S}.

\subsection{Time-Variability}
The data in the catalog are obtained over many different epochs, with the time between observations spanning up to $\approx50$~years\footnote{\citet{1980A&AS...39..379T} compiled all of the published polarization measurements in the literature prior to December 1978.}. Both flux- and angle- variability in the sample may be significant. Nevertheless, it is feasible that time-variable sources will experience similar amounts of positive and negative perturbations in total intensity and/or polarized fraction, i.e.\ increases and decreases will occur in equal measure. In addition, it is also feasible that the time-variations themselves will be simultaneously similar for both the total and polarized intensity components -- leading to an unchanged polarized fraction, i.e.\ an increase in total intensity is directly proportional to the increase in polarized intensity. Together with the large number of compiled measurements, this will serve such that any time-variability will just appear as increased variance in the estimates of depolarization and spectral index. Furthermore, the catalog has also been cross-matched with optical identifications for each source \citep{2012arXiv1209.1438H}. It should be possible to distinguish between source types with differing variability, and a user of the catalog is able to study sources by classification e.g.\ galaxies and blazars. For the purposes of statistical studies, any derived results are again only affected in terms of increased sample variance. To assist the user in distinguishing potentially time-variable sources, we provide a data quality flag that is calculated based on statistical tests during fitting of the SED (see Section~\ref{sourcefitting}).

\subsection{Rician bias}
\label{ricianbias}
Consider a source that has no intrinsic polarization, so that $P = Q = U = 0$. As the measurements will be derived from images of $Q$ and $U$ that contain Gaussian noise, it is clear that a measurement of $P$ from an image will always yield $P > 0$. This over-estimation of polarized intensity due to $P$ being positive-definite is known as `Rician bias' \citep[e.g.][]{1985A&A...142..100S}. For higher s/n ratios, the amount of this bias becomes increasingly small. Nevertheless, at lower s/n ratios an estimator of the bias is essential to retrieve the true polarized intensity.

The effects of Rician bias, and the bias estimator have often changed in their use over the period in which the measurements have been collected. In many cases, the exact method used has not been stated in the literature. As all of our sources are detected at $\ge8\sigma$ at 1.4~GHz \citep{2009ApJ...702.1230T}, and these sources tend to increase in polarized fraction at higher frequency, we make the assumption that each source is of sufficient s/n that the Rician statistics are approximately Gaussian -- under such circumstances the bias in measuring a positive-definite quantity can be considered to be negligible. Additional systematics in our catalog may also result from the different ways in which source properties have been measured and parameterized, i.e.\ the source-finding technique used. We assume that such effects behave as random errors on each datum, and have negligible effects on an SED.

\subsection{Single Dish Measurements}
Modern single-dish measurements are typically done in `on-off' or `dual-beam' mode, so that any smooth Galactic background can in principle be subtracted. Nevertheless, the processing and observational details have not been stated in the literature for the single dish measurements that are included in the catalog -- some of which were taken $\approx50$~years ago. These measurements may not have removed a contribution from surrounding diffuse background components. Such a systematic would increasingly affect our measurements in proximity to the Galactic plane. Any subtle correlations in the data at low Galactic latitudes should be treated with caution.

\subsection{Errors in Polarized Fractions}
Many individual catalogs also ignore the total intensity contribution to the error in polarized fraction. Following standard error propagation, the fractional uncertainties in the polarized and total intensity measurements should be summed in quadrature as given by
\begin{equation}
\left(\frac{ \sigma_{\Pi} }{\Pi}\right)^2 = \left(\frac{ \sigma_{P} }{P}\right)^2 + \left(\frac{ \sigma_{I} }{I}\right)^2 \,,
\end{equation}
where $\Pi$ is the polarized fraction, $P$ is the polarized intensity (equal to $\sqrt{Q^2+U^2}$, assuming that Rician bias is negligible), $I$ is the total intensity, and $\sigma_{i}$ is the one sigma uncertainty in these quantities. For sources that are weak in total intensity, or equivalently have a large fractional error in $\sigma_{I}/I$, neglecting this total intensity contribution can significantly underestimate the error in the polarized fraction. For our catalog, this additional contribution has been recalculated when enough information has been available to do so. 

In practice, antenna feeds never have a perfect response to polarization. These imperfections can be modeled as leakage from Stokes $I$ into polarization, and is known as `instrumental polarization' or `polarization leakage'. Error contributions from this instrumental polarization are by convention also typically ignored when calculating the errors in the polarized fraction. However, if leakages are calculated accurately to within $0.3$\% then we would expect an additional error in each measurement. Such an error can be trivially handled as an additive component to the uncertainty in each datum. The SEDs that we fit to the data in Section~\ref{sourcefitting} are potentially affected by the stated precision of each datum, which we use for weighting our data. 

\subsection{Effect of Outliers}
The effects of outliers in our data are also of interest. There are two types of outliers: (i) those that are inaccurate, and (ii) those that are both inaccurate and imprecise. Case (i) relates to occasional false cross-matches, or poor measurements in the original surveys that occur due to calibration or source fitting issues. These cause systematic errors in our derived source parameters. Our low FDR, combined with the large number of measurements that are believed to be accurate, ensures that we still get reasonable values. This was tested on a simulated distribution of EVPA measurements as a function of $\lambda^2$, and we find that a single outlier affects the calculated RM by less than 10~rad~m$^{-2}$ for sources with five measurements. Case (ii) is more simply dealt with, as it occurs due to large measurement errors that in all cases are appropriately weighted during our fitting procedure. The effects of outliers therefore tend to be reduced by the fitting and weighting procedure, but still have the property of increasing the variance in any estimated parameters. The catalog therefore works well for statistical measurements of polarized source properties -- our FDR suggests that the number of inaccurate matches represents $<5$\% of the total sample. Nevertheless, we recommend that appropriate caution is used if attempting to derive physical properties from the polarization SED of a single source.

\section{Results}
\label{results}
We have produced a new, broadband radio polarization catalog. Our catalog uses the NVSS RM catalog \citep{2009ApJ...702.1230T} as a reference, with our `Full-catalog' incorporating 37,543 radio sources, of which 25,649 have measurements of both the Rotation Measure (RM) and total intensity spectral index. The catalog is expected to consist of mostly classical radio galaxies and quasars. Normal galaxies are only expected to constitute a significant fraction of the 1.4~GHz polarized source counts at $\approx$1~mJy in total intensity, and at 1 to $10$~\textmu Jy in polarized intensity \citep[e.g.][]{shanepaper,2014arXiv1402.3637R} -- approximately two orders of magnitude below the NVSS sensitivity limit. In this catalog, we have provided power-law depolarization measurements for 1,171 radio sources with $\ge2$ polarization measurements at independent frequencies, and a full, modeled SED in both polarized and total intensity for 951 radio sources with $\ge3$ multiwavelength polarization data. These 951 sources with modeled SEDs are available as a secondary `SED-catalog', constituting a subset of our primary data product. The catalog includes spectroscopic redshifts for 620 of these polarized SEDs. Based on the model selection criteria and various statistical tests described in the main text, we consider 533 of these SEDs to be excellent -- an improvement of an order of magnitude over any other published sample of polarized SEDs. The breakdown of SEDs in our catalog by different depolarization models suggests that 42.2\% of sources are best described by a Tribble-law, 20.6\% by repolarization, 19.4\% by spectral depolarization, 9.0\% by a Rossetti--Mantovani law, and 8.8\% by a Burn-law. See Section~\ref{sourcefitting} and Table~\ref{tab:models} for further detail on how these models are derived. Note that these add up to 100\%, as we have not included a `no best fit' category. Even if the data quality flag is taken into account, and `poor' fitting models are excluded from our analysis, the percentages are changed such that 42.5\% of sources are best described by a Tribble-law, 16.0\% by repolarization, 21.9\% by spectral depolarization, 11.0\% by a Rossetti--Mantovani law, and 8.6\% by a Burn-law. Note that the fraction of repolarizers is substantially affected -- this is unsurprising given that we use a power law to fit the repolarizers, and that this is based on the flexibility of the power law, rather than an exact physical model of what could be expected from different emission regions. Note that such depolarization models are only applicable if the same emitting region is probed at all frequencies -- as we shall discuss in this section, a key physical inference from our catalog is that this is likely not the case.

\begin{figure*}
\centering
\includegraphics[clip=true, trim=1.0cm 1.0cm 0cm 1.0cm, width=16cm]{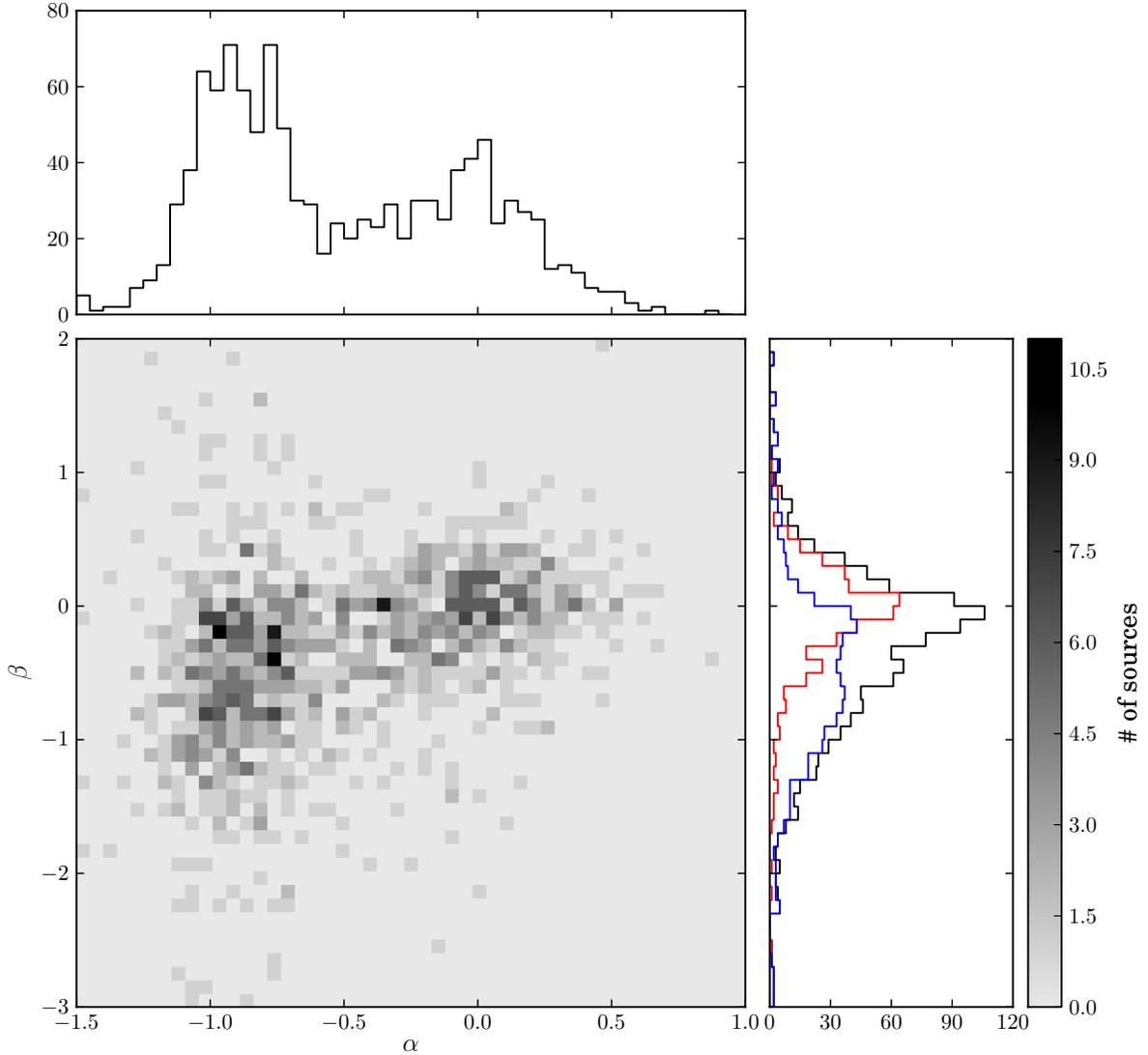}
\caption{A density plot of sources in our catalog that have both a measured total intensity spectral index, $\alpha$, and an estimated polarized fraction spectral index, $\beta$ (note that we have assumed the SED can be reasonably parameterized using this estimated polarization spectral index, see e.g.\ Fig.~\ref{SEDs}). Histograms of $\alpha$ and $\beta$, are shown to the top and right respectively. The distribution of depolarization spectral indices are shown in the right panel for flat-spectrum ($\alpha\ge-0.4$, red line), steep-spectrum ($\alpha\le-0.6$, blue line), and all (black line) sources. The $\alpha$ histogram (top panel) differs from that shown in Fig.~\ref{I_histo}; in the latter all spectral indices are displayed irrespective of whether the source has corresponding depolarization ($\beta$) information.}
\label{I_versus_P_alphas}
\end{figure*}

A density plot showing the number of sources with given total intensity and polarized spectral indices is shown in Fig.~\ref{I_versus_P_alphas}. There is a clear divide in the depolarization of radio sources that occurs at $\alpha\approx-0.5$, with steep-spectrum sources showing larger depolarization ($\beta<0$), and flat-spectrum sources tending to retain an essentially constant polarized fraction as a function of $\lambda$ ($\beta\approx0$). We split the sample into two subsets of $\beta$ values, consisting of either flat-spectrum ($\alpha\ge-0.4$) or steep-spectrum ($\alpha\le-0.6$) sources. Sources with $-0.6<\alpha<-0.4$ are not included in order to minimize cross-contamination between our samples, although our conclusions are not affected by changing this $\alpha$-threshold. The two-sample Kolmogorov--Smirnov test on $\beta$ provides a $p$-value of $1.6\times10^{-30}$, which indicates a very low probability of the two samples being this different, or more so, if drawn from the same distribution. This is consistent with a population of both core- and jet-dominated sources, in which steep-spectrum sources measure the optically thin lobes and jets ejected by an AGN, while flat-spectrum sources measure the optically thick region surrounding the central black hole of a radio galaxy. As the total intensity spectral index is an intrinsic source property, this implies that depolarization must mainly occur within the local source environment, rather than being due to intervening Faraday screens. We argue that such an effect is unlikely to be due to confounding variables, with independent samples selected at 20~GHz (i.e.\ core-dominated sources) also finding no statistically significant evidence of a relationship between the fraction of polarization and frequency \citep{2013arXiv1309.2527M}. One may argue that this could be caused by beam depolarization, with the flat-spectrum central engine being more compact, and tending to have smaller RM dispersion than the steep-spectrum sources. In such a case, the compact sources would show \emph{some} depolarization; instead, the distribution of $\beta$ for flat-spectrum sources is approximately symmetric about a mean value of zero (see the red histogram in Fig.~\ref{I_versus_P_alphas}). In fact, the distribution appears very slightly skewed with a few more sources showing $\beta>0$ rather than $\beta<0$. For pure beam depolarization, $\beta$ must \emph{always} be $<0$. Consequently, while there are possible alternative interpretations of our results, any such hypotheses are restricted in that they must be able to explain this approximately symmetric distribution of the flat-spectrum sources around $\beta\approx0$. As the most consistent explanation for the repolarizing ($\beta>0$) sources is that this is a consequence of different emitting regions being probed at different frequencies, the symmetrical distribution appears to indicate that different emitting regions are also giving rise to the depolarizing ($\beta<0$) sources.

Previous studies have suggested that intervening galaxies with a clumpy interstellar medium are important for explaining depolarization of background sources \citep[e.g.][]{2008Natur.454..302B,2012ApJ...761..144B}. As the likelihood of intersecting an intervenor increases for sources at high $z$, one could hypothesise from Fig.~\ref{I_versus_P_alphas} that steep-spectrum sources are more distant. However, such a scenario cannot explain why the flat-spectrum sources have an approximately symmetrical distribution that is centred about $\beta\approx0$. Similarly, it also does not explain the presence of repolarizing sources. As a further point in favour of our interpretation, due to relativistic beaming, one expects the flat-spectrum (core-dominated) sources to be the more distant for a flux-limited selected sample such as the NVSS. Furthermore, there is evidence from the catalog that allows one to completely rule out the currently proposed models of `partial coverage' (Farnes et al. in preparation).

Our data also show that repolarization can occur in some sources, although importantly this is predominantly associated with flat-spectrum (i.e.\ presumably core-dominated) sources. Our classification algorithm finds that 203 sources (21\%) of our sample show repolarization. Fig.~\ref{I_versus_P_alphas} shows that the lobe-dominated sources are mostly located with $\beta\le0$, however $\approx50$\% of the population of core-dominated sources have $\beta>0$. Together with the flat-spectrum sources tending to retain a relatively constant polarized fraction as a function of $\lambda$, we interpret this as a consequence of the core's optical thickness, with synchrotron self-absorption leading to the observation of different emitting regions within the source. This results in the observation that a source is repolarizing (and even depolarizing), when we are instead likely detecting regions of increasingly ordered magnetic field that are at greater distances from the central engine. Optically thick emission, although rarely considered, can result in more complicated behavior in a polarized SED -- including oscillations and even constant polarized fractions at increasing wavelengths \citep[e.g.][]{1967ApJ...150..647P}. An example of a possible oscillating source, for which our algorithm fits a flat value of $\beta$, is shown in the SED for NVSS~J003820--020740 in Fig.~\ref{SEDs}. Our data therefore support the hypothesis that flat-spectrum radio polarization measurements probe polarization structure in the optically thick quasar core, with the sources tending to provide a constant polarized fraction over a large range in wavelength merely due to the sparse sampling in $\lambda$-space which undersamples the oscillatory behaviour. As repolarization is also present for some lobe-dominated sources, it appears to indicate that a combination of spectral index and depolarization effects can always cause one to probe different polarized emitting regions at different frequencies. Nevertheless, there are only a relatively small number of repolarizing steep-spectrum objects, unlike the symmetric flat-spectrum distribution about $\beta=0$. The steep-spectrum sample \emph{cannot} therefore be similar to the flat-spectrum sample but with some additional depolarizing component. We argue that the most consistent physical description is of a divide based on the local source environment, with the large number of repolarizers showing that we do not sample similar emitting regions at different frequencies.

As a substantial number of sources in our sample are repolarizing, this indicates that the run of polarized fraction with observing wavelength is substantially affected by different emitting regions within a source. From an observational perspective this is problematic, as attempts to derive any particular depolarization law from a run of polarized fraction with $\lambda$ must attempt to separate different emitting regions. This is normally not considered, as the depolarization is typically considered to be most strongly affected by screens of material along the line of sight \citep[e.g.][]{1966MNRAS.133...67B,1991MNRAS.250..726T,2008Natur.454..302B,2008A&A...487..865R,2009A&A...502...61M,2012arXiv1209.1438H}. Nevertheless, the results from our new and larger sample suggest that Faraday depolarization in extragalactic sources is largely an `internal' effect that occurs within the local source environment. \footnote{Note the potentially confusing nomenclature -- `internal' does not necessarily refer to `internal Faraday depolarization', but rather to being within the vicinity of the source.} This has implications for the physical interpretation of polarized SEDs, and complicates the interpretation of broadband polarization observations for the purposes of measuring SEDs and RMs. In such a scenario, a balance must be obtained between broadband measurements (which provide a larger lever in $\lambda^2$-space), and narrow-band observations (which are more likely to probe the same source component across a frequency range of interest). This has implications for future broadband polarimetric surveys. High-resolution follow-up observations will be required to further understand the effects of different emitting regions on polarized SEDs.

\begin{figure*}
\centering
\includegraphics[clip=false, trim=1.0cm 0.0cm 0cm 1.0cm, width=16.5cm]{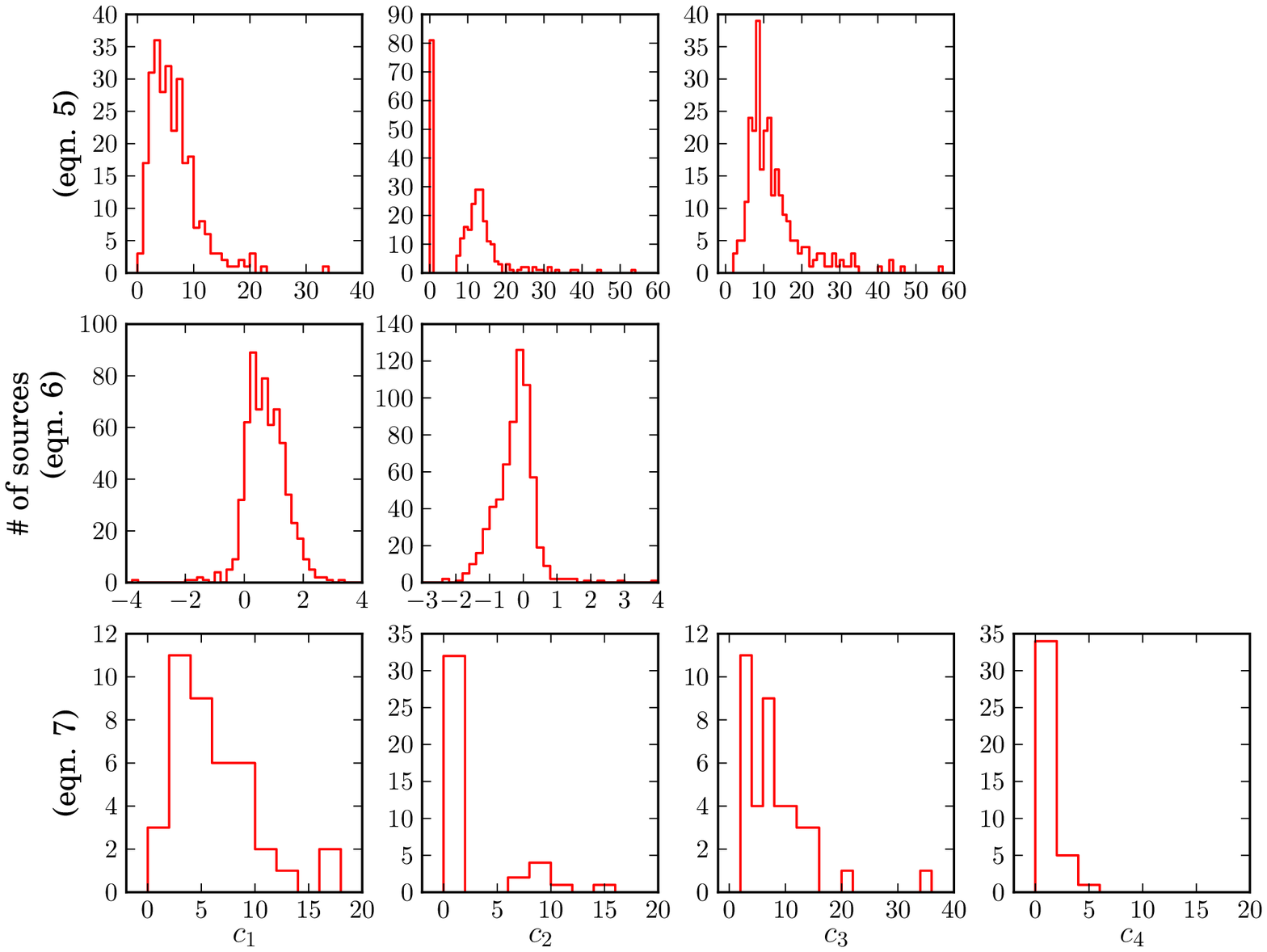}
\caption{A plot of the distribution of parameters for the various models selected by the BIC. Histograms of the number of sources are shown for $c_{1}$ (first column), $c_{2}$ (second column), $c_{3}$ (third column), and $c_{4}$ (fourth column). The physical model that is selected as corresponding to the coefficient values can be determined using Table~\ref{tab:models}. The units of the $c_{i}$ coefficients are given in Table~\ref{tab:modelsunits}. The histograms are separated according to the model selected by the BIC, with those modeled by a Gaussian (top row), power law (middle row), and Gaussian with constant term (bottom row). The top, middle, and bottom rows therefore correspond to equations~\ref{gaussian}, \ref{powerlaw}, and \ref{gauss+const} respectively. Note that the number of parameters differs for each model.}
\label{c_plot}
\end{figure*}

For the steep-spectrum objects in our catalog, there is weak evidence of a bimodal distribution in $\beta$ (see Fig.~\ref{I_versus_P_alphas}), with two distinct peaks. If true, this would suggest two populations of depolarizer, possibly consistent with previous studies \citep[e.g.][]{1988Natur.331..147G,1988Natur.331..149L,1991MNRAS.250..198G,1991MNRAS.249..343L}. Stronger conclusions on the Garrington--Laing and Liu--Pooley effects will require extensive further study of these data. This will require a cleaner sample that is separated by source type (e.g.\ by using the optical identifications that are included in the catalog -- see Appendix~\ref{appendix-a}), and through high-resolution follow-up observations to identify the morphology of both the source and the polarized emitting region. Such studies, albeit with smaller sample size, have recently identified relations between RM, depolarization and the complexity of the Faraday spectrum for polarized sources \citep{2013A&A...559A..27G} -- concluding that the depolarization originated within the sources, for instance in their radio lobes, or in intervening galaxies on the line of sight. As a point of further interest, plots of the distribution of parameters in our catalog for the various models selected by the BIC are shown in Fig.~\ref{c_plot}. This shows that the majority of sources have a maximum polarized fraction of $\approx5$\%, with some sources having polarized fractions up to $30$\%. While the majority of sources show regular depolarization, those sources that are peaked tend to have a maxima in polarized fraction at a wavelength of $10$ to $15$~cm. Sources that maintain a constant polarized fraction out to large wavelengths (a Rossetti--Mantovani law, i.e.\ that include a coefficient $c_{4}$), tend to remain polarized at only the $1$ to $3$\% level. The physical model that is selected as corresponding to the coefficient values can be determined using Table~\ref{tab:models}.

\section{Discussion and Conclusions}
\label{discussion}
We have presented a multiwavelength catalog of radio polarization that increases the number of well-defined polarized SEDs by over an order of magnitude. The resulting polarized SEDs have a frequency range of 0.4~GHz to 100~GHz, and our best sources are constrained by up to 56 independent polarization measurements. We have used a K-Dimensional tree for the cross-matching, reducing the computational expense by a factor of $(\log n)/n$, and have used an automated classification algorithm based on the Bayesian Information Criterion to distinguish between different models for the Faraday depolarization. The catalog also contains constraints on the total intensity spectral index and curvature, the broadband RM, spectroscopic redshift, angular size, and estimated non-thermal rest-frame luminosity at 1.4~GHz.

The catalog will allow a number of parameters to be explored in the effort to understand cosmic magnetic fields. In this paper, we have found that our sample is consistent with two populations of core- and jet-dominated sources based on the clustering in the plane of the polarized fraction versus total intensity spectral indices. This is consistent with the optically-thin jet/lobe-dominated sources undergoing significantly more depolarization relative to the optically-thick core-dominated sources. Such a connection implies that radio source depolarization predominantly occurs within the local source environment, rather than being due to intervening Faraday screens. Importantly, the catalog will be of particular use in $k$-correcting polarized SEDs into the source rest-frame. These $k$-corrections can be performed by using the statistical tests provided in the catalog to select good quality SEDs, and to use the depolarization model selected by the BIC. The $k$-corrected polarized fraction, i.e.\ in the source rest-frame, for equivalent emission to that at 1.4~GHz and $z=0$ is then the corresponding polarized fraction at $\lambda= 21.414/(1+z)$~cm. The $k$-corrected Faraday depth can be obtained using a multiplicative factor of $(1+z)^2$ to correct for the effects of cosmological expansion, however this assumes that all of the Faraday rotation is occurring at the source. Such a correction will break down if there is any Galactic or intervening contribution. The $(1+z)^2$ correction to the Faraday depth also assumes a linear relationship between the EVPA and $\lambda^2$ -- if the relationship is non-linear and the Faraday depth measured using a narrowband, then it will be necessary to $k$-correct the polarization angle SED in a similar way to $k$-correcting the polarized fraction, i.e.\ by sliding the SED and then reestimating the Faraday depth. This catalog is an enabling step for such $k$-corrected studies, as an increased sample of polarized SEDs will not be available until the advent of the SKA and other next generation facilities such as ASKAP -- which will yield the necessary broadband polarization data in combination with a redshift. The results of the $k$-corrections themselves constitute an extensive additional study, as such studies of magnetic field evolution are hindered by the ability to classify different source types (e.g.\ normal versus active galaxies, by viewing angle, core- versus lobe- dominated), by luminosity effects/Malmquist bias, and possibly by other evolutionary effects such as changes in the bulk Lorentz factor of radio jets. Furthermore, as $k$-correction of the RM into the rest-frame by a factor of $(1+z)^2$ assumes that the Faraday rotating medium is all local to the source -- reliable removal of Galactic contributions, together with measures of the magnetoionic content along typical lines of sight, are essential to probe the evolution of cosmic magnetism. Other parameters such as the total intensity spectral indices will assist in the process of source classification and correcting for luminosity effects, although higher resolution data are necessary to reveal how reliably this separates core- and lobe-dominated sources. In addition to $k$-corrections, there are also many subsidiary applications, including studies of where Faraday rotation occurs along the line of sight, the spectral index versus redshift \citep[e.g.][]{2006MNRAS.371..852K}, measurements of the Galactic magnetic field \citep[e.g.][]{2008ApJ...676...70K}, and investigations of intervening Mg~II absorption systems and their effects on SED type and depolarization \citep[e.g.][]{2008Natur.454..302B,2012ApJ...761..144B}.

Previous similar catalogs have not focused on polarization measurements \citep[e.g.][]{2008AJ....136..684K}. This therefore is the most comprehensive catalog of polarization measurements available to date. Our catalog is reliable, with more than 95\% of polarization measurements being correctly associated, and more than 93\% of total intensity measurements. The reliability is further improved by the source fitting procedures. Nevertheless, while we have selected our catalog as a 1.4~GHz flux-limited polarized sample, selection effects are still likely considerable and are hard to constrain. The sky coverage of the accumulated data is patchy, and typically taken via shallow observations on the brightest known sources. As the brightest sources have typically been observed repeatedly, we obtain many more measurements for the brightest sources, and consequently the SEDs of the fainter sources are less well-constrained.

Despite these inevitable effects, the listed polarized SEDs can reveal the physical form of depolarization, and allow for physical models of the intervening magnetic field structure to be ascertained. The predominant depolarizing mechanism is important for modelling polarized source counts at low radio frequencies, which will have a substantial impact on the number of polarized sources detectable by facilities such as LOFAR, MWA, and the SKA \citep{2004NewAR..48.1289B,2007mru..confE..69S,2011AAS...21714232H,halesmnras13}.

Alongside the multiple possible analyses of the catalog we have presented, additional future work will allow for expansion of these data. Numerous other polarization measurements are available in the literature. Furthermore, our catalog currently uses the NVSS RM catalog of \citet{2009ApJ...702.1230T} as the reference data; if a source is not listed as polarized in the \citet{2009ApJ...702.1230T} catalog at 1.4~GHz, then polarization measurements at other frequencies are not accumulated into our catalog. Furthermore, polarization measurements in the Southern sky (at declinations $<-40^{\circ}$) are currently not included in our catalog. Future work could therefore use a larger reference catalog. For example, the Sydney University Molonglo Sky Survey (SUMSS) \citep{1999AJ....117.1578B} explores a similar parameter space to the NVSS in the Southern sky (albeit without polarization information). A combination of the full total intensity NVSS measurements \citep{1998AJ....115.1693C}, together with the total intensity measurements from SUMSS, would therefore greatly expand our reference catalog. Using an NVSS+SUMSS reference catalog for cross-matching with additional polarization measurements accumulated from across the literature \citep[e.g.][]{2000A&A...363..141R,2010MNRAS.401.1388J,2011ApJ...732...45S,2013arXiv1309.2527M}, would lead to a much larger sample of polarized SEDs -- including those sources that undergo the most extreme depolarization. More recent polarization surveys, such as the S-band Polarization All Sky Survey (S-PASS) at 2.3~GHz, will provide $\approx5,000$ polarized sources in the Southern sky \citep{2011JApA...32..457C,2013Natur.493...66C} -- corresponding to a significant increase in sample size. In addition, possible upcoming polarization surveys will also be able to make substantial contributions to such an effort -- detecting up to  2.2$\times$10$^5$ sources in polarized intensity at 2 to 4~GHz \citep[e.g.][]{2014arXiv1401.1875M}. Such cross-matched catalogs, particularly when combined with large samples of redshifts, will continue to constitute useful resources for probing magnetic fields and other astrophysical phenomena.

\section{Acknowledgements}
We thank Paul Hancock for useful discussions on cross-matching, Phil Kronberg for helpful advice on fitting broadband rotation measures, Shane O'Sullivan for insightful conversations on radio source depolarization, and Xiaohui Sun for helpful comments on other existing polarization measurements. We are also grateful to the anonymous referee for helpful comments on the original version of the manuscript. J.S.F. \& B.M.G. acknowledge the support of the Australian Research Council through grants DP0986386 \& FL100100114. Parts of this research were conducted by the Australian Research Council Centre of Excellence for All-sky Astrophysics (CAASTRO), through project number CE110001020. The National Radio Astronomy Observatory is a facility of the National Science Foundation operated under cooperative agreement by Associated Universities, Inc. The Australia Telescope Compact Array is part of the Australia Telescope National Facility which is funded by the Commonwealth of Australia for operation as a National Facility managed by CSIRO. The Westerbork Synthesis Radio Telescope is operated by the ASTRON (Netherlands Institute for Radio Astronomy) with support from the Netherlands Foundation for Scientific Research (NWO). This research has made use of: the SIMBAD database, operated at Centre de Donn\'ees astronomiques de Strasbourg, France; the NASA/IPAC Extragalactic Database (NED) which is operated by the Jet Propulsion Laboratory, California Institute of Technology, under contract with NASA; NASA's Astrophysics Data System Abstract Service; the SDSS, for which funding has been provided by the Alfred P. Sloan Foundation, the Participating Institutions, the National Science Foundation, and the U.S. Department of Energy Office of Science; the 2dF/6dF QSO Redshift Surveys (2QZ/6QZ), which were compiled by the 2QZ/6QZ teams from observations made with the 2-degree Field and 6-degree Field on the Anglo-Australian Telescope. 


\appendix

\section{Depolarization Models}\label{appendix-b}
The polarization properties of sources at radio wavelengths are dominated by various forms of astrophysical depolarization that can be described by a number of models. For the purpose of model selection, we need to pick a number of models a-priori that may describe the physics of each source. This depolarization can be a reduction in the polarized fraction that occurs either internally\footnote{By `internal' depolarization we refer to a reduction in the polarized fraction that occurs directly within the radio source or within the source's immediate environment, such that the depolarization is related to the source itself in some manner.} to the radio source or in its foreground i.e.\ externally -- somewhere along the line of sight between us and the source environment. When the depolarization occurs externally, we consider this to be the result of a \emph{foreground depolarizing screen}. The effects of external screens must be removed if we are to understand the internal structure of radio sources, where thermal plasma is mixed with the magnetoionic medium responsible for the radio emission. This separation of internal and foreground effects is complicated as both have a similar wavelength dependence \citep[e.g.][]{1991MNRAS.250..726T}. Consequently, the relationship between polarized fraction, $\Pi$, and wavelength, $\lambda$, can be an insensitive test of the RM structure, unless information in Stokes $Q$/$U$ is considered simultaneously \citep[e.g.][]{2012MNRAS.421.3300O}. Nevertheless, statistical measures of SED properties can still be used to ascertain the predominant form of depolarization. We now explain the various forms of depolarization in detail. Unless otherwise stated, we focus on depolarization in optically thin emission \citep[e.g.][]{1966MNRAS.133...67B}. Optically thick emission could result in more complicated behavior in a polarized SED -- including oscillations and even constant polarized fractions at increasing wavelengths \citep[e.g.][]{1967ApJ...150..647P}.

\subsection{Disordered Magnetic Fields}
In the case of optically thin synchrotron radiation, random fluctuations from a disordered magnetic field within a source can lead to a reduced fractional polarization given by
\begin{equation}
\Pi = \left( \frac{3-3\alpha}{5-3\alpha} \right) \frac{H_{0}^2}{H_{0}^2+H_{r}^2} \,,
\label{internal}
\end{equation}
where $H_{0}$ and $H_{r}$ are the uniform and random field strengths respectively \citep[e.g.][]{1966MNRAS.133...67B}, and the total intensity spectral index, $\alpha$, is defined such that $S_{\nu}\propto\nu^{+\alpha}$. Extensions to equation~\ref{internal} have been proposed \citep{2003A&A...411...99B}. As the sources in our catalog are always unresolved, this effect is independent of wavelength -- we consider it no further in this paper.

\subsection{Internal Faraday depolarization}
Internal Faraday depolarization (also called `depth-' or `front--back-' depolarization) takes places when the emitting and Faraday rotating regions within a source are intermixed. This internal depolarization can be separated into two types: (i) differential Faraday rotation, and (ii) internal Faraday dispersion. For (i), in the presence of a regular magnetic field, the plane of polarization of emission from the far side of the region undergoes a different amount of Faraday rotation compared to emission from the near side. The process is known as `differential Faraday rotation' and the sum through the entire region (along the line of sight) results in depolarization \citep[e.g.][]{1998MNRAS.299..189S,2011MNRAS.418.2336A}. While this process leads to depolarization, it is characterized by `bobbing' of the polarized fraction as a function of increasing wavelength. This bobbing is difficult to identify in SEDs with sparse frequency sampling, and could likely only be measured using a wide and continuous observing bandwidth. For (ii), in the presence of a turbulent magnetic field, the plane of polarization experiences random fluctuations as it propagates through the region. This process is known as `internal Faraday dispersion' and this addition of random anisotropic magnetic fields results in further depolarization \citep[e.g.][]{1998MNRAS.299..189S,2012MNRAS.421.3300O}. However, observational evidence suggests that internal Faraday effects are likely not the dominant cause of depolarization in extragalactic radio sources \citep{1980AJ.....85..368C,1988A&A...194...79S,1988Natur.331..147G,1991MNRAS.250..726T} -- we therefore consider it no further in this paper.

\subsection{External Faraday depolarization}
\label{externaldispersion}
External Faraday depolarization occurs if the Faraday rotation in front of the emitting region varies significantly across the source over a solid angle smaller than the beam \citep[e.g.][]{1966MNRAS.133...67B,1998MNRAS.299..189S}. Such a Faraday screen consists of a magnetoionic region that is devoid of relativistic particles and that exists somewhere along the line of sight between the observer and the source \citep[e.g.][]{1998MNRAS.299..189S}. If this Faraday screen contains a constant regular field then the region causes Faraday rotation of the polarized emission from background sources, but does not cause any depolarization. Nevertheless, any deviation from a constant field within the synthesized beam will create a RM gradient and subsequently cause depolarization. Such anisotropy of the $B$ field is caused by either turbulent, or systematically varying regular fields. For turbulent magnetic fields or \emph{external Faraday dispersion}, depolarization occurs when there are many turbulent regions within the synthesized beam. For regular magnetic fields or \emph{beam depolarization}, depolarization will occur if there are any variations in the strength or orientation of the field within the synthesized beam \citep[e.g.][]{2012MNRAS.421.3300O}. The two situations are analogous and for the case of turbulent fields with the values of RM having a large spread over the beam, and for the case of a systematically varying regular field with a gradient in the external RM across the beam, the polarization is rotated through different angles in each direction at longer $\lambda$, resulting in a reduction in $\Pi$ \citep[see e.g.][for further detail]{1998MNRAS.299..189S}.

The effect of external Faraday depolarization was initially described by \citet{1966MNRAS.133...67B} with an equation of the form
\begin{equation}
\Pi(\lambda) = \Pi_{0}\exp{(-2\sigma_{\text{RM}}^2\lambda^4)} \,,
\label{burnbabyburn}
\end{equation}
where $\sigma_{\text{RM}}$ is the RM dispersion of the Faraday screen within a single beam. A smoothly interpolating simple approximation to a Burn law can be provided using a Gaussian model, see e.g.\ equation~\ref{gaussian} in Section~\ref{fittingpolarizedSED}. Nevertheless, observations show that significant polarized emission continues to be detectable at relatively large wavelengths \citep[e.g.][]{1991MNRAS.250..726T}, showing that the relationship between polarized fraction and wavelength cannot be as strong as a function of $\lambda^4$ for all sources. Evidence for deviations from a Burn law have also been found from diffuse polarized emission \citep{2013A&A...559A..27G}. The equations of \citet{1966MNRAS.133...67B} were extended by \citet{1991MNRAS.250..726T} to show that at long wavelengths ($\lambda > \lambda_{1/2}$, where $\lambda_{1/2}$ is the wavelength at which the polarized fraction is half its maximum value) the polarized fraction can decrease via a power law such that
\begin{equation}
\Pi(\lambda) = \Pi_{0}\left[ \frac{s_{0}/t}{2\sigma_{\text{RM}}\sqrt{2}} \right] \lambda^{-2} \,,
\label{tribbling}
\end{equation}
where $s_{0}/t$ is a measure of the resolution, with $t$ being related to the beam FWHM$=2t\sqrt{\ln 2}$ and $s_{0}$ being the size of the RM fluctuations. Importantly, \citet{1991MNRAS.250..726T} shows that depending on the precise form of the structure function of the RM fluctuations, the polarization can actually fall off such that $\Pi\propto\lambda^{-4/m}$ (and not only as $\Pi\propto\lambda^{-2}$), where $m$ has been shown observationally to approximately vary between $1$ and $4$ \citep{2004PhDT.........6G}. Furthermore, the Tribble models are all calculated in the source rest-frame, such that the regime $\lambda > \lambda_{1/2}$ is not well-defined without extensive prior measurements of a radio source. A more general solution is therefore given by
\begin{equation}
\Pi(\lambda) = A \lambda^{\beta} \,,
\label{tribbling2}
\end{equation}
where $A$ is some constant and $\beta$ is a polarization spectral index. Note that $\beta$ is defined in the opposite sense to the total intensity spectral index, $\alpha$, in that it is the exponent of observing frequency rather than wavelength. Such power laws have previously been used to define polarization spectral energy distributions \citep{1979Ap&SS..61..153E,2011MNRAS.413..132B}. A smoothly interpolating fit for a Tribble law can be provided using a power law model, see e.g.\ equation~\ref{powerlaw} in Section~\ref{fittingpolarizedSED}.

More recently, it has also been shown that the polarization can decrease as a function of $\lambda^4$ at smaller wavelengths -- as proposed by \citet{1966MNRAS.133...67B} -- but then remains unexpectedly constant out to longer wavelengths \citep[e.g.][]{2009A&A...502...61M}. There have been attempts to explain these SEDs as a result of `partial coverage', with the explanation that only a fraction of the source is covered by an inhomogeneous Faraday screen \citep[e.g.][]{2008A&A...487..865R,2009A&A...502...61M,2012ApJ...761..144B}. In an effort to derive $\sigma_{\text{RM}}$ for these sources, \citet{2008A&A...487..865R} make an empirical modification to equation \ref{burnbabyburn} so that
\begin{equation}
\Pi(\lambda) = \Pi_{0} \left[ f_{c}\exp{(-2\sigma_{\text{RM}}^2\lambda^4)} + (1-f_{c}) \right] \,,
\label{partialcoverage}
\end{equation}
where $f_{c}$ is interpreted as the covered (depolarizing) fraction of the source, with the uncovered fraction $(1-f_{c})$ retaining a constant $\Pi$ out to arbitrarily long wavelengths. This model was found to be more successful in reproducing the form of some polarized SEDs \citep[e.g.][]{2008A&A...487..865R,2009A&A...502...61M}. A smoothly interpolating simple approximation for a Rossetti--Mantovani law can be provided using a Gaussian with a constant term, see e.g.\ equation~\ref{gauss+const} in Section~\ref{fittingpolarizedSED}.

As a point of further interest, \citet{1966MNRAS.133...67B} also considered a partial coverage model. Under the assumption that depolarization could be originating from discrete clouds in the Galaxy, there would be $N$ average clouds along the line of sight. Following \citet{2008A&A...487..865R}, if $N\gg1$ the depolarization is similar to equation \ref{burnbabyburn}, while several lines of sight will not intersect any cloud if $N\ll1$. Consequently, wavelength-independent polarization may emerge through gaps in this `Faraday web'. Theoretically
\begin{equation}
\Pi(\lambda) = \Pi_{0}\exp{[-N(1-\exp{[2 F_{c}^2 \lambda^4]})]} \,,
\label{faradayweb}
\end{equation}
where $F_{c}$ is the RM of a single cloud. Such clouds may reasonably exist in the environment surrounding a radio source. Such a functional form also follows a similar decline to that of equation \ref{partialcoverage}, and high-quality data would be needed to distinguish between these similar models.

\subsection{Spectral Depolarization/Repolarization}
\label{spectraldepolrepol}
Differences in the total intensity spectral index of components in an unresolved source can result in different regions of a source being probed at different observing frequencies. The effect of a typical radio galaxy which has a flat-spectrum, weakly polarized core combined with steep-spectrum, more-highly polarized lobes/jets, can be observed in a significant number of sources \citep[e.g.][]{1974MNRAS.168..137C}. This `spectral depolarization' tends to give rise to a prominent peaked structure in a plot of $\Pi$ versus $\lambda$. In some cases, this may also give rise to behavior analogous to beating between two oscillatory components \citep{1984ApJ...283..540G,2011AJ....141..191F}. A smoothly interpolating simple approximation for a Spectral Depolarizer can be provided using a Gaussian model, see e.g.\ equation~\ref{gaussian} in Section~\ref{fittingpolarizedSED}.

Some sources also exhibit `inverse depolarization' or `repolarization', with the polarized fraction increasing at longer wavelengths \citep{2002ApJ...568...99H,2009A&A...502...61M,2012AJ....144..105H}. This is again likely the consequence of a multiple unresolved components within a compact source, and the probing of different emitting regions at different frequencies. Such repolarization can also be explained as increased ordering of the magnetic field in the component of the source nearest to the observer along the line of sight. At lower frequencies, the emitting region further into the source depolarizes -- effectively constituting a measurement of different source components at different observational frequencies. Increasing $\Pi$ could also arise in the unresolved, optically-thick core region where one is observing different parts of the jet at different wavelengths \citep[e.g.][]{1982IAUS...97..363K}. A smoothly interpolating simple approximation for a repolarizer can be provided using a power law model, see e.g.\ equation~\ref{powerlaw} in Section~\ref{fittingpolarizedSED}.

Other explanations have also been proposed, and repolarization could occur in jets that are well separated from the core in the optically thin part of the jet, and are also reasonably isolated from other strong jet features \citep{2012AJ....144..105H,2012ApJ...747L..24H}. Other suggested physical models propose that internal Faraday rotation acts to align the polarization from the far and near sides of a jet, leading to increased $\Pi$ at longer wavelengths. Helical and randomly tangled magnetic fields have also been suggested as being responsible \citep[e.g.][]{1998MNRAS.299..189S,2012ApJ...747L..24H}. 

\section{Catalog Quantities}\label{appendix-a}
Data for all sources and the associated fitted and calculated properties are provided as machine-readable tables. There are two versions of the catalog available, the `Full-catalog' and the `SED-catalog' respectively. The column headings are identical in both catalog versions, and the SED-catalog is just a subset of rows from our primary data product, the Full-catalog. The column headings of both catalogs are described in Table~\ref{tab:catkey}. Selected columns for the first 45 sources in the SED-catalog are presented in Tables~\ref{tab:thedata}~to~\ref{tab:thedata5} in Section~\ref{catalog}.
\tabletypesize{\tinyv}
\begin{deluxetable*}{ccclc}
\setlength{\tabcolsep}{0.02in} 
\tablewidth{0pt}
\tablecaption{Quantities in both of the catalogs, listed in order of the corresponding column numbers. All errors are the one sigma uncertainties.}
\tablehead{
\colhead{Column}  & \colhead{Format} & \colhead{Name} & \colhead{Description} & \colhead{Units} }
\startdata
\tabletypesize{\tinyv}
1 & I5  & Source\_Num & \parbox[t]{10.5cm}{Source number in the NVSS RM catalog of \citet{2009ApJ...702.1230T}, from 1 to 37543.}  & \nodata \\
2 & A13 & Source\_Name & \parbox[t]{10.5cm}{Source name in the original NVSS catalog of \citet{1998AJ....115.1693C}.}   & \nodata \\
3 & A12   & RA\_str            & Right ascension, from the NVSS (J2000).    & \nodata \\
4 & A12   & DEC\_str          & Declination, from the NVSS (J2000).            & \nodata \\
5 & F9.5   & RA                    & Right ascension, from the NVSS (J2000).    & $^{\circ}$ \\
6 & F8.5   & DEC                  & Declination, from the NVSS (J2000).            & $^{\circ}$ \\
7 & F8.4   & G\_lon              & Galactic Longitude, from the NVSS.           & $^{\circ}$ \\
8 & F8.4   & G\_lat               & Galactic Latitude, from the NVSS.               & $^{\circ}$ \\
9 & I2   & Num\_Data       & Number of polarized data.                         & \nodata \\
10 & I2 & Num\_Data\_I   & Number of total intensity data.                  & \nodata \\
11--79 & F6.3 & Lambda    & \parbox[t]{10.5cm}{Raw polarization wavelength data. Each element is listed in Table~\ref{tab:polelements}.}             & cm \\
80--148 & F5.2 & Pol\_Frac    & \parbox[t]{10.5cm}{Raw polarization fraction data. Each element is listed in Table~\ref{tab:polelements}.}             & \% \\
149--217 & F5.2 & Pol\_Frac\_err    & \parbox[t]{10.5cm}{Raw polarization fraction error data. Each element is listed in Table~\ref{tab:polelements}.}             & \% \\
218--286 & F6.2 & Pol\_Angle    & \parbox[t]{10.5cm}{Raw polarization angle data. Each element is listed in Table~\ref{tab:polelements}.}             & $^{\circ}$ \\
287--355 & F6.2 & Pol\_Angle\_err    & \parbox[t]{10.5cm}{Raw polarization angle error data. Each element is listed in Table~\ref{tab:polelements}.}             & $^{\circ}$ \\
356--376 & F6.3 & I\_Lambda    & \parbox[t]{10.5cm}{Raw total intensity wavelength data. Each element is listed in Table~\ref{tab:Ielements}.}             & cm \\
377--397 & F7.2 & I\_Flux    & \parbox[t]{10.5cm}{Raw total intensity flux density data. Each element is listed in Table~\ref{tab:Ielements}.}             & mJy\\
398--418 & F7.2 & I\_Flux\_err    & \parbox[t]{10.5cm}{Raw total intensity flux density error data. Each element is listed in Table~\ref{tab:Ielements}.}             & mJy \\
419 & A5 & Depol\_Type    & \parbox[t]{10.5cm}{The selected polarization SED model classification. The possible combinations are listed in Table~\ref{tab:models}.}             & \nodata \\
420 & A9 & Depol\_Physics    & \parbox[t]{10.5cm}{The selected physical model classification. This classification is determined using Depol\_Type, together with constraints on the coefficients $c_{i}$. The possible combinations are listed in Table~\ref{tab:models}.}             & \nodata \\
421 & F7.3 & c1    & \parbox[t]{10.5cm}{Coefficient $c_{1}$ for the selected polarization SED model.}             & See Table~\ref{tab:modelsunits} \\
422 & F7.3 & c2    & \parbox[t]{10.5cm}{Coefficient $c_{2}$ for the selected polarization SED model.}             & See Table~\ref{tab:modelsunits} \\
423 & F7.3 & c3    & \parbox[t]{10.5cm}{Coefficient $c_{3}$ for the selected polarization SED model.}             & See Table~\ref{tab:modelsunits} \\
424 & F7.3 & c4    & \parbox[t]{10.5cm}{Coefficient $c_{4}$ for the selected polarization SED model.}             & See Table~\ref{tab:modelsunits} \\
425 & F7.3 & c1\_err    & \parbox[t]{10.5cm}{Error in coefficient $c_{1}$ for the selected polarization SED model.}             & See Table~\ref{tab:modelsunits} \\
426 & F7.3 & c2\_err    & \parbox[t]{10.5cm}{Error in coefficient $c_{2}$ for the selected polarization SED model.}             & See Table~\ref{tab:modelsunits} \\
427 & F7.3 & c3\_err    & \parbox[t]{10.5cm}{Error in coefficient $c_{3}$ for the selected polarization SED model.}             & See Table~\ref{tab:modelsunits} \\
428 & F7.3 & c4\_err    & \parbox[t]{10.5cm}{Error in coefficient $c_{4}$ for the selected polarization SED model.}             & See Table~\ref{tab:modelsunits} \\
429 & F7.3 & Depol\_KS    & \parbox[t]{10.5cm}{Depolarization Kolmogorov--Smirnov statistic for the selected polarization SED model and the polarized fraction versus $\lambda$ data.}             & \nodata \\
430 & F7.3 & Depol\_KS\_Pval    & \parbox[t]{10.5cm}{Depolarization $p$-value for the selected polarization SED model and the polarized fraction versus $\lambda$ data from the Kolmogorov--Smirnov statistic.}             & \nodata \\
431 & F7.3 & Depol\_Chi2  & \parbox[t]{10.5cm}{Depolarization reduced $\chi^2$ for the selected polarization SED model and the polarized fraction versus $\lambda$ data.}             & \nodata \\
432 & F7.3 & Depol\_Chi2\_Pval    & \parbox[t]{10.5cm}{Depolarization $p$-value for the selected polarization SED model and the polarized fraction versus $\lambda$ data from the $\chi^2$ statistic.}             & \nodata \\
433 & I1 & Data\_Flag    & \parbox[t]{10.5cm}{A data quality flag that describes the model fit of an SED, calculated based on the combination of the other statistical tests. Has a value of 1, 2, or 3 and is provided as an indicative measure only; these correspond to `accept', `caution', and `poor' respectively.}             & \nodata \\
434 & F7.3 & Pol\_Beta\_c1    & \parbox[t]{10.5cm}{Polarization spectral index coefficient, $c_{1}$, such that $\Pi=10^{c_{1}}\lambda^\beta$.}             & \nodata \\
435 & F7.3 & Pol\_Beta    & \parbox[t]{10.5cm}{Polarization spectral index, $\beta$, such that $\Pi=10^{c_{1}}\lambda^\beta$.}             & \nodata \\
436 & F7.3 & Pol\_Beta\_err    & \parbox[t]{10.5cm}{Error in the polarization spectral index, $\beta$.}             & \nodata \\
437 & F7.3 & Pol\_Beta\_Chi2    & \parbox[t]{10.5cm}{The reduced $\chi^2$ for the polarization spectral index and the polarized fraction versus $\lambda$ data.}             & \nodata \\
438 & F7.3 & Pol\_Beta\_Chi2\_Pval    & \parbox[t]{10.5cm}{The $p$-value for the polarization spectral index and the polarized fraction versus $\lambda$ data from the $\chi^2$ statistic.}             & \nodata \\
439 & F7.3 & Pol\_Beta\_KS    & \parbox[t]{10.5cm}{The Kolmogorov--Smirnov statistic for the polarization spectral index and the polarized fraction versus $\lambda$ data.}             & \nodata \\
440 & F7.3 & Pol\_Beta\_KS\_Pval    & \parbox[t]{10.5cm}{The $p$-value for the polarization spectral index and the polarized fraction versus $\lambda$ data from the Kolmogorov--Smirnov statistic.}             & \nodata \\
441 & F7.3 & z    & \parbox[t]{10.5cm}{Redshift of the source from \citet{2012arXiv1209.1438H}.}             & \nodata \\
442 & F7.3 & z\_err    & \parbox[t]{10.5cm}{Error in the redshift of the source from \citet{2012arXiv1209.1438H}.}             & \nodata \\
443 & A8 & Selected\_Obj    & \parbox[t]{10.5cm}{Classification of the object type from \citet{2012arXiv1209.1438H}.}             & \nodata \\
444 & A1 & I\_Class   & \parbox[t]{10.5cm}{The selected total intensity spectral index classification, either `R' for a regular power law, `C' for a curved power law, or `2' for a two-point spectral index.}             & \nodata \\
445 & F7.3 & I\_alpha\_d1   & \parbox[t]{10.5cm}{Spectral index coefficient $d_{1}$ for the selected total intensity SED model.}             & See Table~\ref{tab:modelsunits} \\
446 & F7.3 & I\_alpha    & \parbox[t]{10.5cm}{Total intensity spectral index, $\alpha$, and equivalently the coefficient $d_{2}$ for the selected total intensity SED model.}             & See Table~\ref{tab:modelsunits} \\
447 & F7.3 & I\_alpha\_d3    & \parbox[t]{10.5cm}{Spectral index coefficient $d_{3}$ for the selected total intensity SED model.}             & See Table~\ref{tab:modelsunits} \\
448 & F7.3 & I\_alpha\_d1\_err    & \parbox[t]{10.5cm}{Error in the spectral index coefficient $d_{1}$ for the selected total intensity SED model.}             & See Table~\ref{tab:modelsunits} \\
449 & F7.3 & I\_alpha\_err    & \parbox[t]{10.5cm}{Error in the total intensity spectral index, $\alpha$, and equivalently the coefficient $d_{2}$ for the selected total intensity SED model.}             & See Table~\ref{tab:modelsunits} \\
450 & F7.3 & I\_alpha\_d3\_err    & \parbox[t]{10.5cm}{Error in the spectral index coefficient $d_{3}$ for the selected total intensity SED model.}             & See Table~\ref{tab:modelsunits} \\
451 & F7.3 & I\_Chi2    & \parbox[t]{10.5cm}{Spectral Index reduced $\chi^2$ for the regular power law model fit to the total intensity versus frequency data.}             & \nodata \\
452 & F7.3 & I\_Chi2\_Curved    & \parbox[t]{10.5cm}{Spectral Index reduced $\chi^2$ for the curved power law model fit to the total intensity versus frequency data.}             & \nodata \\
453 & F7.3 & I\_Pval    & \parbox[t]{10.5cm}{Spectral Index $p$-value for the regular power law model and the total intensity versus frequency data.}             & \nodata \\
454 & F7.3 & I\_Pval\_Curved    & \parbox[t]{10.5cm}{Spectral Index $p$-value for the curved power law model and the total intensity versus frequency data.}             & \nodata \\
455 & F7.2 &   RM\_Broad     & \parbox[t]{10.5cm}{Broadband RM calculated using the EVPA versus $\lambda^2$ data.} & rad~m$^{-2}$ \\
456 & F7.2 &  RM\_Broad\_err      & \parbox[t]{10.5cm}{Error in the broadband RM calculated using the EVPA versus $\lambda^2$ data.}             & rad~m$^{-2}$ \\
457 & F7.2 &  EVPA\_Broad      & \parbox[t]{10.5cm}{Intrinsic EVPA calculated using the EVPA versus $\lambda^2$ data.}             & $^{\circ}$ \\
458 & F7.2 &  EVPA\_Broad\_err      & \parbox[t]{10.5cm}{Error in the intrinsic EVPA calculated using the EVPA versus $\lambda^2$ data.}             & $^{\circ}$ \\
459 & F7.3 &    RM\_Broad\_Chi2    & \parbox[t]{10.5cm}{Broadband RM reduced $\chi^2$ for the selected broadband RM model and the EVPA versus $\lambda^2$ data.}             & \nodata \\
460 & F7.3 &    RM\_Broad\_Pval    & \parbox[t]{10.5cm}{Broadband RM $p$-value for the selected broadband RM model and the EVPA versus $\lambda^2$ data.}             & \nodata \\
461 & F6.1 &    NVSS\_RM    & \parbox[t]{10.5cm}{NVSS RM from \citet{2009ApJ...702.1230T}.}             & rad~m$^{-2}$ \\
462 & F6.1 &   NVSS\_RM\_err     & \parbox[t]{10.5cm}{Error in the NVSS RM from \citet{2009ApJ...702.1230T}.}             & rad~m$^{-2}$ \\
463 & F5.1 &    I    & \parbox[t]{10.5cm}{NVSS total intensity (Stokes $I$) from \citet{2009ApJ...702.1230T}.}             & mJy \\
464 & F5.1 &   I\_err     & \parbox[t]{10.5cm}{Error in the NVSS total intensity (Stokes $I$) from \citet{2009ApJ...702.1230T}.}             & mJy \\
465 & F5.1 &   P     & \parbox[t]{10.5cm}{NVSS polarized intensity, $P=(Q^2+U^2)$, from \citet{2009ApJ...702.1230T}.}             & mJy \\
466 & F5.1 &  P\_err      & \parbox[t]{10.5cm}{Error in the NVSS polarized intensity, $P=(Q^2+U^2)$, from \citet{2009ApJ...702.1230T}.}             & mJy \\
467 & F4.1 &  PI      & \parbox[t]{10.5cm}{NVSS fractional polarization, $\Pi=(Q^2+U^2)/I$, from \citet{2009ApJ...702.1230T}.}             & \% \\
468 & F4.1 &  PI\_err      & \parbox[t]{10.5cm}{Error in the NVSS fractional polarization, $\Pi=(Q^2+U^2)/I$, from \citet{2009ApJ...702.1230T}.}             & \% \\
469 & A1 &   Major\_axis\_limit     & \parbox[t]{10.5cm}{Upper limit on the major axis measurement from \citet{1998AJ....115.1693C}, either blank for a determined measurement, or `<' for an upper limit.}             & \nodata \\
470 & F4.1 &   Major\_axis     & \parbox[t]{10.5cm}{Major axis size from \citet{1998AJ....115.1693C}.}             & arcsec \\
471 & F4.1 &   Major\_axis\_err     & \parbox[t]{10.5cm}{Error in the major axis size from \citet{1998AJ....115.1693C}.}             & arcsec \\
472 & A1 &    Minor\_axis\_limit    & \parbox[t]{10.5cm}{Upper limit on the minor axis measurement from \citet{1998AJ....115.1693C}, either blank for a determined measurement, or `<' for an upper limit.}             & \nodata \\
473 & F4.1 &     Minor\_axis   & \parbox[t]{10.5cm}{Minor axis size from \citet{1998AJ....115.1693C}.}             & arcsec \\
474 & F4.1 &    Minor\_axis\_err    & \parbox[t]{10.5cm}{Error in the minor axis size from \citet{1998AJ....115.1693C}.}             & arcsec \\
475 & F5.1 &    NVSS\_PA    & \parbox[t]{10.5cm}{Position angle from \citet{1998AJ....115.1693C}.}             & $^{\circ}$ \\
476 & F5.1 &    NVSS\_PA\_err    & \parbox[t]{10.5cm}{Error in the position angle from \citet{1998AJ....115.1693C}.}             & $^{\circ}$ \\
477 & F5.2 &   L     & \parbox[t]{10.5cm}{The logarithm of the estimated non-thermal rest-frame luminosity of the source at 1.4~GHz.}             & $\log_{10}(L_{\nu}$ /W~Hz$^{-1})$ \\
478 & F5.2 &  L\_err      & \parbox[t]{10.5cm}{Error in the logarithm of the luminosity.}             & $\log_{10}(\Delta L_{\nu}$ /W~Hz$^{-1})$
\enddata
\label{tab:catkey}
\end{deluxetable*}

\tabletypesize{\small}
\begin{deluxetable*}{cccc}
\tablewidth{0pt}
\tablecaption{The data provided in each element of the polarization data arrays}
\tablehead{
\colhead{Element \#}  & \colhead{Abbrev.} & \colhead{Frequency}   & \colhead{Wavelength}  \\
\colhead{} & \colhead{}         & \colhead{/MHz} & \colhead{/cm} }
\startdata
1 & NVSS/NVSS RM      & 1400  & 21.41 \\
 2 & AT20G     & 4860  &  6.169 \\
 3 &   \nodata             & 8640   & 3.470  \\
 4 &   \nodata             & 20000  &   1.499 \\
 5 & T03         & 1400   & 21.41  \\
 6 &   \nodata             & 2496  & 12.01   \\
 7 &   \nodata             & 4800   & 6.246  \\
 8 &   \nodata             & 8640     & 3.470\\
 9 & Z99         & 4700  & 6.379 \\
 10 & B3-VLA    & 2695 & 11.12   \\
 11 & \nodata               & 4850   & 6.181 \\
 12 & \nodata               & 10500  & 2.855   \\
 13 & SN80/SN81/SN82      & 1580--14750 & 18.97-- 2.032\\
$\vdots$ &    $\vdots$   & $\vdots$  \\
 17 & \nodata      & 1580--14750 & 18.97--2.032 \\
18 & TI80      & 404--99930  & 74.21--0.300 \\
$\vdots$ & $\vdots$   & $\vdots$   \\
69 &  \nodata     & 404--99930  & 74.21--0.300
\enddata
\label{tab:polelements}
\end{deluxetable*}

\begin{deluxetable*}{cccc}
\tablewidth{0pt}
\tablecaption{The data provided in each element of the total intensity data arrays}
\tablehead{
\colhead{Element \#}  & \colhead{Abbrev.} & \colhead{Frequency} & \colhead{Wavelength}    \\
\colhead{} & \colhead{}         & \colhead{/MHz} & \colhead{/cm} }
\startdata
1 & NVSS/NVSS RM      & 1400  & 21.414  \\
 2 & AT20G     & 4860  & 6.169  \\
 3 &  \nodata              & 8640   & 3.470  \\
 4 &  \nodata              & 20000  & 1.499  \\
 5 & T03         & 1400 & 21.414    \\
 6 &  \nodata              & 2496   &  12.011 \\
 7 &  \nodata              & 4800   & 6.246  \\
 8 &  \nodata              & 8640  & 3.470   \\
 9 & Z99         & 4700  & 6.379 \\
 10 & B3-VLA    & 2695   & 11.124 \\
 11 & \nodata               & 4850  & 6.181  \\
 12 & \nodata               & 10500   & 2.855  \\
 13 & SN80/SN81/SN82      & 1580--14750 & 18.974--2.032 \\
$\vdots$ &    $\vdots$   & $\vdots$  \\
 17 & \nodata      & 1580--14750 & 18.974--2.032  \\
18 & WENSS     & 326  & 91.961   \\
19 & Texas     & 365   &   82.135  \\
20 & GB6       & 4850   & 6.181 \\
21 & NORTH6CM  & 4850  & 6.181
\enddata
\label{tab:Ielements}
\end{deluxetable*}

\begin{deluxetable*}{ccccccc}
\tablewidth{0pt}
\tablecaption{Selection of a Depol\_Physics classification, using Depol\_Type and the $c_{i}$ coefficients}
\tablehead{
\colhead{Depol\_Type}  & \colhead{Equation \#}  & \colhead{$c_{1}$} & \colhead{$c_{2}$} & \colhead{$c_{3}$}  & \colhead{$c_{4}$} & \colhead{Depol\_Physics} }
\startdata
gauss & \ref{gaussian} &\nodata   & $\le7.5$  & \nodata & \nodata & burn  \\
gauss & \ref{gaussian} &\nodata   & $>7.5$ & \nodata & \nodata & peaked  \\
power & \ref{powerlaw} & \nodata   & $>0$  & \nodata & \nodata & increase  \\
power & \ref{powerlaw} &\nodata   & $\le0$ & \nodata & \nodata & tribble  \\
gau+t & \ref{gauss+const} &\nodata   & \nodata & \nodata & \nodata & mantovani 
\enddata
\label{tab:models}
\end{deluxetable*}

\begin{deluxetable*}{cccccccc}
\tablewidth{0pt}
\tablecaption{Units of the $c_{i}$ and $d_{i}$ coefficients as per the selected equation. Coefficients $c_{i}$ correspond to the polarized SED, while coefficients $d_{i}$ correspond to the total intensity SED.}
\tablehead{
\colhead{Equation \#}  & \colhead{$c_{1}$} & \colhead{$c_{2}$} & \colhead{$c_{3}$}  & \colhead{$c_{4}$} & \colhead{$d_{1}$} & \colhead{$d_{2}$} & \colhead{$d_{3}$} }
\startdata
\ref{gaussian} & per cent (\%)  & centimetres (cm)  & centimetres (cm) & N/A & \nodata & \nodata & \nodata \\
\ref{powerlaw} & $\log_{10}\left[\text{per cent (\%)}\right]$ & unitless & N/A & N/A & \nodata & \nodata & \nodata \\
\ref{gauss+const} & per cent (\%)  & centimetres (cm)  & centimetres (cm) & per cent (\%) & \nodata & \nodata & \nodata \\
\ref{powerlawI} & \nodata & \nodata & \nodata & \nodata & millijansky (mJy) & unitless & N/A \\
\ref{powerlawIcurved} & \nodata & \nodata & \nodata & \nodata & millijansky (mJy) & unitless & unitless
\enddata
\label{tab:modelsunits}
\end{deluxetable*}

\newpage
\clearpage
\clearpage

\begin{thebibliography}

\bibitem[Abazajian et al.(2009)]{2009ApJS..182..543A}
Abazajian, K.~N., Adelman--McCarthy, J.~K., Ag\"ueros, M.~A., et al., 2009, \apjs, 182, 543.

\bibitem[Akritas \& Bershady(1996)]{1996ApJ...470..706A} 
Akritas, M.~G., \& Bershady, M.~A., 1996, \apj, 470, 706. 

\bibitem[Arshakian \& Beck(2011)]{2011MNRAS.418.2336A} 
Arshakian, T.~G., \& Beck, R., 2011, \mnras, 418, 2336. 

\bibitem[Battye et al.(2011)]{2011MNRAS.413..132B} 
Battye, R.~A., Browne, I.~W.~A., Peel, M.~W., Jackson, N.~J., \& Dickinson, C., 2011, \mnras, 413, 132. 

\bibitem[Beck et al.(2003)]{2003A&A...411...99B}
Beck, R., Shukurov, A., Sokoloff, D., Wielebinski, R., 2003, \aap, 411, 99.

\bibitem[Beck \& Gaensler(2004)]{2004NewAR..48.1289B} 
Beck, R., \& Gaensler, B.~M., 2004, New Astronomy Reviews, 48, 1289. 

\bibitem[Beck(2011)]{2011arXiv1111.5802B} 
Beck, R., 2011, in Magnetic Fields in the Universe: From Laboratory and Stars to Primordial Structures, eds. M.~Soida, K.~Otmianowska-Mazur, E.~M.~de Gouveia Dal Pina \& A.~Lazarian, preprint (arXiv:1111.5802).

\bibitem[Beck et al.(2012)]{2012A&A...543A.113B}
Beck, R., Frick, P., Stepanov, R., Sokoloff, D., 2012, \aap, 543, 113.

\bibitem[Beck et al.(2013)]{2013MNRAS.435.3575B} 
Beck, A.~M., Dolag, K., Lesch, H., Kronberg, P.~P., 2013, MNRAS, 435, 3575.

\bibitem[Becker et al.(1991)]{1991ApJS...75....1B} 
Becker, R.~H., White, R.~L., \& Edwards, A.~L., 1991, \apjs, 75, 1. 

\bibitem[Bentley(1975)]{bentley1975} 
Bentley, J. L., 1975, Communications of the Association for Computing Machinery, 18, 509.

\bibitem[Bernet et al.(2008)]{2008Natur.454..302B} 
Bernet, M.~L., Miniati, F., Lilly, S.~J., Kronberg, P.~P., \& Dessauges-Zavadsky, M., 2008, \nat, 454, 302. 

\bibitem[Bernet et al.(2012)]{2012ApJ...761..144B} 
Bernet, M.~L., Miniati, F., \& Lilly, S.~J., 2012, \apj, 761, 144. 

\bibitem[Bock et al.(1999)]{1999AJ....117.1578B} 
Bock, D.~C.~J., Large, M.~I., Sadler, E.~M., 1999, AJ, 117, 1578.

\bibitem[\protect\citeauthoryear{Brentjens \& de Bruyn}{2005}]{2005A&A...441.1217B}
Brentjens, M.~A., de Bruyn, A.~G., 2005, A\&A, 441, 1217.

\bibitem[Budav{\'a}ri \& Szalay(2008)]{2008ApJ...679..301B} 
Budav{\'a}ri, T., \& Szalay, A.~S., 2008, \apj, 679, 301. 

\bibitem[Burn(1966)]{1966MNRAS.133...67B} 
Burn, B.~J., 1966, \mnras, 133, 67.
 
\bibitem[Carretti(2011)]{2011JApA...32..457C} 
Carretti, E., 2011, JApA, 32, 457. 

\bibitem[Carretti et al.(2013)]{2013Natur.493...66C} 
Carretti, E., Crocker, R.~M., Staveley-Smith, L., et al., 2013, \nat, 493, 66. 

\bibitem[Cioffi \& Jones(1980)]{1980AJ.....85..368C} 
Cioffi, D.~F., \& Jones, T.~W., 1980, \aj, 85, 368. 

\bibitem[Clauset et al.(2009)]{2007arXiv0706.1062C} 
Clauset, A., Rohilla Shalizi, C., \& Newman, M.~E.~J., 2009, Society for Industrial and Applied Mathematics Review, 51, 661. 

\bibitem[Condon(1992)]{1992ARA&A..30..575C} 
Condon, J.~J., 1992, \araa, 30, 575. 

\bibitem[Condon et al.(1998)]{1998AJ....115.1693C} 
Condon, J.~J., Cotton, W.~D., Greisen, E.~W., et al., 1998, \aj, 115, 1693. 

\bibitem[Conway et al.(1974)]{1974MNRAS.168..137C} 
Conway, R.~G., Haves, P., Kronberg, P.~P., et al., 1974, \mnras, 168, 137. 

\bibitem[Dallal \& Wilkinson(1986)]{dallal-wilkinson1986} 
Dallal, G.~E., Wilkinson, L., 1986, Journal of the American Statistical Association, 40, 294.

\bibitem[Douglas et al.(1996)]{1996AJ....111.1945D} 
Douglas, J.~N., Bash, F.~N., Bozyan, F.~A., Torrence, G.~W., \& Wolfe, C., 1996, \aj, 111, 1945. 

\bibitem[Eichendorf \& Reinhardt(1979)]{1979Ap&SS..61..153E} 
Eichendorf, W., \& Reinhardt, M., 1979, \apss, 61, 153. 

\bibitem[Fanti et al.(2004)]{2004A&A...427..465F} 
Fanti, C., Branchesi, M., Cotton, W.~D., et al., 2004, \aap, 427, 465. 

\bibitem[Farnes et al.(2014)]{FARNESETAL} 
Farnes, J.~S., Green, D.~A., Kantharia, N.~G., 2014, \mnras, 437, 3236.

\bibitem[Farnsworth et al.(2011)]{2011AJ....141..191F} 
Farnsworth, D., Rudnick, L., \& Brown, S., 2011, \aj, 141, 191. 

\bibitem[Feigelson \& Babu(1992)]{1992ApJ...397...55F} 
Feigelson, E.~D., \& Babu, G.~J., 1992, \apj, 397, 55. 

\bibitem[Fisher(1922)]{fisher1922}
Fisher, R.~A., 1922, Phil. Trans. R. Soc. Lond., 222, 309. 

\bibitem[Frick et al.(2011)]{2011MNRAS.414.2540F}
Frick, P., Sokoloff, D., Stepanov, R., Beck, R., 2011, \mnras, 414, 2540.

\bibitem[Gaensler et al.(2004)]{2004NewAR..48.1003G} 
Gaensler, B.~M., Beck, R., \& Feretti, L., 2004, New Astronomy Reviews., 48, 1003. 

\bibitem[Gaensler et al.(2005)]{2005AAS...20713703G} 
Gaensler, B.~M., Beck, R., \& Feretti, L., 2005, BAAS, 37, 137.03. 

\bibitem[Garn et al.(2009)]{2009MNRAS.397.1101G} 
Garn, T., Green, D.~A., Riley, J.~M., \& Alexander, P., 2009, \mnras, 397, 1101.

\bibitem[Garrington et al.(1988)]{1988Natur.331..147G} 
Garrington, S.~T., Leahy, J.~P., Conway, R.~G., \& Laing, R.~A., 1988, \nat, 331, 147. 

\bibitem[Garrington et al.(1991)]{1991MNRAS.250..198G}
Garrington, S.~T., Conway, R.~G., 1991, \mnras, 250, 198.

\bibitem[Gie{\ss}\"{u}bel et al.(2013)]{2013A&A...559A..27G}
Gie{\ss}\"{u}bel, R., Heald, G., Beck, R., Arshakian, T.~G., 2013, \aap, 559, 27.

\bibitem[Goldstein \& Reed(1984)]{1984ApJ...283..540G} 
Goldstein, S.~J., Jr., \& Reed, J.~A., 1984, \apj, 283, 540.

\bibitem[Goodlet(2004)]{2004PhDT.........6G} 
Goodlet, J.~A., 2004, Ph.D.~Thesis, University of Southampton.

\bibitem[Grant et al.(2010)]{2010ApJ...714.1689G} 
Grant, J.~K., Taylor, A.~R., Stil, J.~M., et al., 2010, \apj, 714, 1689.

\bibitem[Gregory et al.(1996)]{1996ApJS..103..427G} 
Gregory, P.~C., Scott, W.~K., Douglas, K., \& Condon, J.~J., 1996, \apjs, 103, 427. 

\bibitem[Hales et al.(2011)]{2011AAS...21714232H} 
Hales, C.~A., Gaensler, B.~M., Norris, R.~P., \& Middelberg, E., 2011, BAAS, 43, 142.32. 

\bibitem[Hales(2013)]{halesmnras13} 
Hales, C.~A., 2013, Ph.D.~Thesis, University of Sydney.

\bibitem[Hammond et al.(2012)]{2012arXiv1209.1438H} 
Hammond, A.~M., Robishaw, T., \& Gaensler, B.~M., 2012, preprint (arXiv:1209.1438v3). 

\bibitem[Homan et al.(2002)]{2002ApJ...568...99H} 
Homan, D.~C., Ojha, R., Wardle, J.~F.~C., et al., 2002, \apj, 568, 99. 

\bibitem[Homan(2012)]{2012ApJ...747L..24H} 
Homan, D.~C., 2012, \apjl, 747, L24. 

\bibitem[Hovatta et al.(2012)]{2012AJ....144..105H} 
Hovatta, T., Lister, M.~L., Aller, M.~F., et al., 2012, \aj, 144, 105. 

\bibitem[Isobe et al.(1990)]{1990ApJ...364..104I} 
Isobe, T., Feigelson, E.~D., Akritas, M.~G., \& Babu, G.~J., 1990, \apj, 364, 104. 

\bibitem[Ivezi{\'c} et al.(2002)]{2002AJ....124.2364I} 
Ivezi{\'c}, {\v Z}., Menou, K., Knapp, G.~R., et al., 2002, \aj, 124, 2364. 

\bibitem[Jackson et al.(2010)]{2010MNRAS.401.1388J} 
Jackson, N., Browne, I.~W.~A., Battye, R.~A., Gabuzda, D., \& Taylor, A.~C., 2010, \mnras, 401, 1388. 

\bibitem[Jansson \& Farrar(2012)]{2012ApJ...757...14J} 
Jansson, R., Farrar, G.~R., 2012, \apj, 757, 14. 

\bibitem[Johnston et al.(2008)]{2008ExA....22..151J} 
Johnston, S., Taylor, R., Bailes, M., et al., 2008, Exp. Astron., 22, 151. 

\bibitem[Kass \& Raftery(1995)]{kassraftery1995}
Kass, R.~E., Raftery, A.~E., 1995, Journal of the American Statistical Association, 90, 430.

\bibitem[Kelly(2007)]{2007ApJ...665.1489K} 
Kelly, B.~C., 2007, \apj, 665, 1489. 

\bibitem[Kimball \& Ivezi{\'c}(2008)]{2008AJ....136..684K} 
Kimball, A.~E., \& Ivezi{\'c}, {\v Z}., 2008, \aj, 136, 684. 

\bibitem[Klamer et al.(2006)]{2006MNRAS.371..852K} 
Klamer, I.~J., Ekers, R.~D., et al., 2006, \mnras, 371, 852. 

\bibitem[Klein et al.(2003)]{2003A&A...406..579K} 
Klein, U., Mack, K.~H., Gregorini, L., \& Vigotti, M., 2003, \aap, 406, 579. 

\bibitem[Konigl(1982)]{1982IAUS...97..363K} 
Konigl, A., 1982, `Relativistic jets as radio and X-ray sources', Extragalactic radio sources; Proceedings of the Symposium, Albuquerque, NM, Reidel Publishing Co., p.~363. Edited by Dordrecht, D.

\bibitem[Kronberg et al.(2008)]{2008ApJ...676...70K} 
Kronberg, P.~P., Bernet, M.~L., Miniati, F., et al., 2008, \apj, 676, 70. 

\bibitem[Laing(1988)]{1988Natur.331..149L}
Laing, R.~A., 1988, \nat, 331, 149.

\bibitem[Lilliefors(1967)]{lilliefors1967} 
Lilliefors, H.~W., 1967, Journal of the American Statistical Association, 62, 399.

\bibitem[Liu \& Pooley(1991)]{1991MNRAS.249..343L}
Liu, R., Pooley, G., 1991, \mnras, 249, 343.

\bibitem[Longair(2011)]{2011hea..book.....L} 
Longair, M.~S., 2011, High Energy Astrophysics, Cambridge, UK: Cambridge University Press.

\bibitem[Mantovani et al.(2009)]{2009A&A...502...61M} 
Mantovani, F., Mack, K.~H., Montenegro-Montes, F.~M., Rossetti, A., \& Kraus, A., 2009, \aap, 502, 61. 

\bibitem[Mao et al.(2014)]{2014arXiv1401.1875M} 
Mao, S.~A., Banfield, J., Gaensler, B., et al., 2014, preprint (arXiv:1401.1875).

\bibitem[Massardi et al.(2013)]{2013arXiv1309.2527M} 
Massardi, M., Burke-Spolaor, S.~G., Murphy, T., et al., 2013, MNRAS, 436, 2915. 

\bibitem[Mesa et al.(2002)]{2002A&A...396..463M} 
Mesa, D., Baccigalupi, C., De Zotti, G., et al., 2002, \aap, 396, 463.

\bibitem[Murphy et al.(2010)]{2010MNRAS.402.2403M} 
Murphy, T., Sadler, E.~M., Ekers, R.~D., et al., 2010, \mnras, 402, 2403. 

\bibitem[Noutsos et al.(2008)]{2008MNRAS.386.1881N} 
Noutsos, A., Johnston, S., Kramer, M., \& Karastergiou, A., 2008, \mnras, 386, 1881. 

\bibitem[O'Sullivan et al.(2008)]{shanepaper}
O'Sullivan S., Stil J., Taylor A. R., Ricci R., Grant J. K., Shorten K., 2008, in POS, Proc. 9th Eur. VLBI Network Symp. for Radio Astron., p.~107.

\bibitem[O'Sullivan et al.(2012)]{2012MNRAS.421.3300O} 
O'Sullivan, S.~P., Brown, S., Robishaw, T., et al., 2012, \mnras, 421, 3300. 

\bibitem[Oppermann et al.(2012)]{2012A&A...542A..93O} 
Oppermann, N., et al., 2012, A\&A, 542, 93. 

\bibitem[Pacholczyk \& Swihart(1967)]{1967ApJ...150..647P}
Pacholczyk, A.~G., \& Swihart, T.~L., 1967, ApJ, 150, 647.

\bibitem[Perley et al.(2009)]{2009IEEEP..97.1448P} 
Perley, R., Napier, P., Jackson, J., et al., 2009, IEEE Proceedings, 97, 1448. 

\bibitem[Planck Collaboration et al.(2013)]{2013arXiv1303.5076P} 
Planck Collaboration, Ade, P.~A.~R., Aghanim, N., et al., 2013, preprint (arXiv:1303.5076). 

\bibitem[Raftery(1995)]{raftery1995}
Raftery, A.~E., 1995, Sociol. Methodol., 25, 111.

\bibitem[Reich et al.(2000)]{2000A&A...363..141R} 
Reich, W., F{\"u}rst, E., Reich, P., et al., 2000, \aap, 363, 141.

\bibitem[Rengelink et al.(1997)]{1997A&AS..124..259R} 
Rengelink, R.~B., Tang, Y., de Bruyn, A.~G., et al., 1997, \aaps, 124, 259. 

\bibitem[Ricci et al.(2013)]{ricci2013} 
Ricci, R., et al., 2013, MNRAS, 435, 2793.

\bibitem[Rossetti et al.(2008)]{2008A&A...487..865R} 
Rossetti, A., Dallacasa, D., Fanti, C., Fanti, R., \& Mack, K.~H., 2008, \aap, 487, 865. 

\bibitem[Roy et al.(2005)]{2005MNRAS.360.1305R} 
Roy, S., Rao, A.~P., \& Subrahmanyan, R., 2005, \mnras, 360, 1305. 

\bibitem[Rudnick \& Owen(2014)]{2014arXiv1402.3637R}
Rudnick, L., \& Owen, F., 2014, preprint (arXiv:1402.3637).

\bibitem[Sajina et al.(2011)]{2011ApJ...732...45S} 
Sajina, A., Partridge, B., Evans, T., et al., 2011, \apj, 732, 45. 

\bibitem[Sazonov(1970)]{1970R&QE...13..162S} 
Sazonov, V.~N., 1970, Radiophysics and Quantum Electronics, 13, 162. 

\bibitem[Schwarz(1978)]{schwarz1978}
Schwarz, G., 1978, Ann. Statist., 6, 461.

\bibitem[Simard-Normandin et al.(1980)]{1980A&AS...40..319S} 
Simard-Normandin, M., Kronberg, P.~P., \& Neidhoefer, J., 1980, \aaps, 40, 319. 

\bibitem[Simard-Normandin et al.(1981a)]{1981ApJS...45...97S} 
Simard-Normandin, M., Kronberg, P.~P., \& Button, S., 1981, \apjs, 45, 97. 

\bibitem[Simard-Normandin et al.(1981b)]{1981A&AS...43...19S} 
Simard-Normandin, M., Kronberg, P.~P., \& Neidhoefer, J., 1981, \aaps, 43, 19. 

\bibitem[Simard-Normandin et al.(1982)]{1982A&AS...48..137S} 
Simard-Normandin, M., Kronberg, P.~P., \& Button, S., 1982, \aaps, 48, 137. 

\bibitem[Simmons \& Stewart(1985)]{1985A&A...142..100S}
Simmons, J.~F.~L., Stewart, B.~G., 1985, \aap, 142, 100.

\bibitem[Sokoloff et al.(1998)]{1998MNRAS.299..189S} 
Sokoloff, D.~D., Bykov, A.~A., Shukurov, A., et al., 1998, \mnras, 299, 189. 

\bibitem[Spiegelhalter et al.(2002)]{spiegelhalter2002}
Spiegelhalter, D.~J., Best, N.~G., Carlin, B.~P., van der Linde, A., 2002, Journal of the Royal Statistical Society: Series B (Statistical Methodology), 64, 583.

\bibitem[Sprent(1969)]{sprent1969}
Sprent, P., 1969, `Models in regression and related topics', Methuen, London.

\bibitem[Stil et al.(2007)]{2007mru..confE..69S} 
Stil, J., Taylor, A.~R., Krause, M., Beck, R., 2007, `Polarization of mJy radio sources', From Planets to Dark Energy: the Modern Radio Universe, Published online at SISSA, Proceedings of Science, p.~69. Edited by Beswick R. et al.

\bibitem[Strom \& Jaegers(1988)]{1988A&A...194...79S} 
Strom, R.~G., \& Jaegers, W.~J., 1988, \aap, 194, 79. 

\bibitem[Tabara \& Inoue(1980)]{1980A&AS...39..379T} 
Tabara, H., \& Inoue, M., 1980, \aaps, 39, 379. 

\bibitem[Taylor et al.(2009)]{2009ApJ...702.1230T} 
Taylor, A.~R., Stil, J.~M., \& Sunstrum, C., 2009, \apj, 702, 1230. 

\bibitem[Taylor \& Salter(2010)]{2010ASPC..438..402T} 
Taylor, A. R., \& Salter, C. J., ‘GALFACTS: The G-ALFA Continuum Transit Survey’, The Dynamic Interstellar Medium: A Celebration of the Canadian Galactic Plane Survey. Proceedings of the conference, held at the Naramata Centre, Naramata, British Columbia, Canada, June 6–10, 2010. Edited by R. Kothes, T. L. Landecker, and A. G. Willis. San Francisco: Astronomical Society of the Pacific, 2010, p.402.

\bibitem[Tingay et al.(2003)]{2003PASJ...55..351T} 
Tingay, S.~J., Jauncey, D.~L., King, E.~A., et al., 2003, \pasj, 55, 351. 

\bibitem[Tingay et al.(2013)]{2013PASA...30....7T} 
Tingay, S.~J., Goeke, R., Bowman, J.~D., et al., 2013, PASA, 30, 7. 

\bibitem[Tribble(1991)]{1991MNRAS.250..726T} 
Tribble, P.~C., 1991, \mnras, 250, 726.
 
\bibitem[van Haarlem et al.(2013)]{2013A&A...556A...2V}
van Haarlem, M.~P., et al., 2013, A\&A, 556, 2.

\bibitem[Warton et al.(2006)]{warton2006}
Warton, D.~I., Wright, I.~J., Falster, D.~S., Westoby, M., 2006, Biological Reviews, 81, 259.

\bibitem[Whittam et al.(2013)]{2013MNRAS.429.2080W} 
Whittam, I.~H., Riley, J.~M., Green, D.~A., et al., 2013, \mnras, 429, 2080. 

\bibitem[Wilson et al.(2011)]{2011MNRAS.416..832W} 
  Wilson W.~E., Ferris R.~H., Axtens P., et al., 2011, MNRAS, 416, 832.

\bibitem[Zukowski et al.(1999)]{1999A&AS..135..571Z} 
Zukowski, E.~L.~H., Kronberg, P.~P., Forkert, T., \& Wielebinski, R., 1999, \aaps, 135, 571.

\end{thebibliography}
\end{document}